\documentclass[amsmath,amssymb,aps,prx,twocolumn,groupedaddress,superscriptaddress,citeautoscript,longbibliography]{revtex4-1}

\usepackage{ulem}
\usepackage{graphicx}
\usepackage{multirow}
\usepackage{bm}
\usepackage{xcolor}
\usepackage{hyperref}
\usepackage{dsfont}
\usepackage{enumerate} 
\usepackage{slashed}
\usepackage{tikz}
\usetikzlibrary{tikzmark}

\newcommand{\dr}{\partial}
\newcommand{\nn}{\nonumber}
\newcommand{\bb}[1]{\boldsymbol{#1}}

\newcommand{\braket}[1]{\langle{#1}\rangle}
\newcommand{\bra}[1]{\langle{#1}\rvert}
\newcommand{\ket}[1]{\lvert{#1}\rangle}

\def\ba#1\ea{\begin{align}#1\end{align}}

\begin{document}
\title{Fragile Topology Protected by Inversion Symmetry: Diagnosis, Bulk-Boundary Correspondence, and Wilson Loop}
\author{Yoonseok \surname{Hwang}}
\affiliation{Department of Physics and Astronomy, Seoul National University, Seoul 08826, Korea}
\affiliation{Center for Correlated Electron Systems, Institute for Basic Science (IBS), Seoul 08826, Korea}
\affiliation{Center for Theoretical Physics (CTP), Seoul National University, Seoul 08826, Korea}
\author{Junyeong \surname{Ahn}}
\affiliation{Department of Physics and Astronomy, Seoul National University, Seoul 08826, Korea}
\affiliation{Center for Correlated Electron Systems, Institute for Basic Science (IBS), Seoul 08826, Korea}
\affiliation{Center for Theoretical Physics (CTP), Seoul National University, Seoul 08826, Korea}
\author{Bohm-Jung \surname{Yang}}
\email{bjyang@snu.ac.kr}
\affiliation{Department of Physics and Astronomy, Seoul National University, Seoul 08826, Korea}
\affiliation{Center for Correlated Electron Systems, Institute for Basic Science (IBS), Seoul 08826, Korea}
\affiliation{Center for Theoretical Physics (CTP), Seoul National University, Seoul 08826, Korea}

\date{\today}

\begin{abstract}
We study the bulk and boundary properties of fragile topological insulators (TIs) protected by inversion symmetry, mostly focusing on the class A of the Altland-Zirnbauer classification. First, we propose an efficient method for diagnosing fragile band topology by using the symmetry data in momentum space. Using this method, we show that among all the possible parity configurations of inversion-symmetric insulators, at least 17$\%$ of them have fragile topology in two dimensions while fragile TIs are less than 3$\%$ percent in three dimensions.
Second, we study the bulk-boundary correspondence of fragile TIs protected by inversion symmetry. In particular, we generalize the notion of $d$-dimensional ($d$D) $k$th-order TIs, which is normally defined for $0<k\le d$, to the cases with $k>d$, and show that they all have fragile topology. In terms of the Dirac Hamiltonian, a $d$D $k$th-order TI has $(k-1)$ boundary mass terms. We show that a minimal fragile TI with the filling anomaly can be considered as the $d$D $(d+1)$th-order TI, and all the other $d$D $k$th-order TIs with $k>(d+1)$ can be constructed by stacking $d$D $(d+1)$th-order TIs.
Although $d$D $(d+1)$th-order TIs have no in-gap states, the boundary mass terms carry an odd winding number along the boundary, which induces localized charges on the boundary at the positions where the boundary mass terms change abruptly. In the cases with $k>(d+1)$, we show that the net parity of the system with boundaries can distinguish topological insulators and trivial insulators. Also, by studying the Wilson loop and nested Wilson loop spectra, we show that all the spectral windings of the Wilson loop and nested Wilson loop should be unwound to resolve the Wannier obstruction of fragile TIs. By counting the minimal number of bands required to unwind the spectral winding of the Wilson loop and nested Wilson loop, we determine the minimal number of bands to resolve the Wannier obstruction, which is consistent with the prediction from our diagnosis method of fragile topology.
Finally, we show that a $(d+1)$D $(k-1)$th-order TI can be obtained by an adiabatic pumping of $d$D $k$th-order TI, which generalizes the previous study of the $2$D third-order TI.
\end{abstract}

\maketitle

% \tableofcontents

\section{Introduction \label{sec: Intro}}
Understanding the role of crystalline symmetries in the topological properties of materials is a central topic in recent studies of condensed matter.
Soon after the discovery of topological insulators protected by time-reversal symmetry~\cite{kane2005z, hasan2010colloquium}, topological insulators (TIs) and topological superconductors (TSCs) have been systematically classified into 10-fold Altland-Zirnbauer (AZ) classes~\cite{altland1997nonstandard}, based on non-spatial symmetries including time-reversal, particle-hole, and chiral symmetries~\cite{schnyder2008classification, kitaev2009periodic, ryu2010topological}. In this classification, TIs are characterized by nontrivial topological invariants and have a Wannier obstruction that is stable against adding trivial bands. Also $d$-dimensional ($d$D) TIs possess stable gapless states at the $(d-1)$D boundaries, which is called the bulk-boundary correspondence.

On the other hand, the topological properties of crystalline solids can also be protected by space-group symmetries, which are realized in topological crystalline insulators (TCIs)~\cite{fu2011topological, hsieh2012topological}.
Compared to the TIs and TSCs classified based on 10-fold AZ classes, TCIs have extremely rich structures because of the large variety of crystalline symmetries forming $230$ space groups and $1651$ magnetic space groups~\cite{kruthoff2017topological, bradlyn2017topological, po2017symmetry, watanabe2018structure, song2018quantitative, thorngren2018gauging, shiozaki2018atiyah}. For instance, recent studies of TCIs have revealed two intriguing properties that are absent in the 10-fold classification of TIs and TSCs, that is, the fragile topology and the higher-order bulk-boundary correspondence.

Fragile topology is one interesting characteristics of some TCIs discovered in recent theoretical studies~\cite{po2018fragile, cano2018topology, bradlyn2019disconnected, bouhon2018wilson, wieder2018axion, po2019faithful, else2019fragile, ahn2019failure, liu2019shift, song2019fragile, kooi2019classification, wieder2019strong}.
In general, TCIs are not adiabatically deformable to atomic insulators without breaking symmetries, that is, they have an obstruction to constructing exponentially localized symmetric Wannier states~\cite{thonhauser2006insulator, soluyanov2011Wannier, marzari2012maximally}.
However, the Wannier obstruction of fragile TIs can be resolved by adding appropriately chosen trivial bands. This property of fragile TIs is clearly distinct from the properties of TCIs with stable topology whose band topology is not affected by additional trivial bands. Because of this, TCIs with stable topology can be classified by $K$-theory~\cite{freed2013twisted, shiozaki2014topology, shiozaki2016topology, shiozaki2017topological, kruthoff2017topological, stehouwer2018classification}, where the equivalence class of insulators is examined when adding trivial bands are allowed. This, at the same time, implies that $K$-theory is not an appropriate tool to study fragile TIs.
In general, two different approaches can be considered to examine the Wannier obstruction of fragile TIs. One is to investigate the symmetry representation of exponentially localized Wannier states~\cite{po2018fragile,po2019faithful,else2019fragile,ahn2019failure,liu2019shift,kooi2019classification,wieder2019strong}, and the other is to study the winding pattern of the Wilson loop spectrum~\cite{cano2018topology,bradlyn2019disconnected,bouhon2018wilson,wieder2018axion,ahn2019failure,kooi2019classification,wieder2019strong}. While both approaches have been applied successfully to characterize the fragile topology of some specific models, the general characterization of fragile TIs based on them has not been achieved yet.

Another intriguing property of some TCIs is that they do not follow the conventional bulk-boundary correspondence. Recently, it has been found that some TCIs have gapless states only in some subspace of the boundary such as hinges or corners~\cite{benalcazar2017quantized, benalcazar2017electric, schindler2018higher1, langbehn2017reflection, song2017d, fang2017rotation, ezawa2018higher, khalaf2018symmetry, khalaf2018higher, van2018higher, geier2018second, matsugatani2018connecting, wang2018higher, schindler2018higher2, ahn2019symmetry, cualuguaru2019higher, trifunovic2019higher,liu2017novel,liu2019helical,cornfeld2019classification,kooi2019classification,song2018topological,okuma2019topological,shiozaki2018generalized,shiozaki2019classification,benalcazar2019quantization,schindler2019fractional}.
In general, $d$D topological phases with gapless states in $(d-k)$D boundaries are called $d$D $k$th-order topological phases. In particular, when $1<k\le d$, they are called higher-order topological phases and their bulk-boundary correspondence has recently been well-established~\cite{khalaf2018symmetry,khalaf2018higher,geier2018second,trifunovic2019higher}.

On the other hand, the nature of boundary states of fragile TIs is not clearly understood yet. It is generally expected that the bulk-boundary correspondence of fragile TIs belongs to a rather trivial category. This is because fragile TIs cannot have a peculiar spectral feature at the boundary distinct from that of atomic insulators, since the boundary energy spectrum is insensitive to adding trivial bands below the bulk gap.
This is also consistent with the fact that topological phases without gapless boundary states are either fragile topological or atomic~\cite{liu2019shift, trifunovic2019higher}.
Nevertheless, some characteristic boundary features of fragile TIs have been reported~\cite{wieder2018axion, wang2018higher}.
For instance, it was shown that, in a two-dimensional (2D) fragile TI protected by inversion symmetry, fractional corner charges appear at the boundary when the geometry of the system preserves inversion symmetry~\cite{wieder2018axion, wang2018higher}.
Moreover, many atomic insulators protected by rotation symmetry were shown to host fractional corner charges in $2$D, which is similar to the case of $2$D second-order TIs~\cite{benalcazar2017quantized, benalcazar2017electric, song2017d, benalcazar2019quantization, van2018higher, schindler2019fractional}.
The origin of corner charges in fragile topological and atomic insulators has been attributed to the filling anomaly~\cite{benalcazar2019quantization}, which denotes an obstruction to fulfill the electron filling for charge neutrality when the relevant crystalline symmetry is preserved~\cite{wieder2018axion, song2017d, wang2018higher, benalcazar2019quantization, schindler2019fractional}.
To understand the generic boundary properties of fragile TIs and atomic insulators, it is highly desirable to establish their bulk-boundary correspondence beyond that of stable topological insulators.

In this work, we fully characterize the bulk and boundary properties of fragile TIs protected by inversion symmetry. First, we establish the bulk-boundary correspondence of fragile TIs protected by inversion symmetry. For this purpose, we generalize the notion of the $d$D $k$th-order TIs to the cases with $k>d$, and show that the $d$D $(d+1)$th-order TIs have fragile topology and exhibit the filling anomaly with the associated boundary charge accumulation. On the other hand, $d$D $k$th-order TIs with $k>d+1$ have a featureless boundary even though they have fragile topology.
Second, we propose an efficient scheme to diagnose fragile band topology by using the symmetry data in momentum space. Using this method, we show that among all the possible parity configurations of inversion-symmetric insulators, at least 17$\%$ of them have fragile topology in $2$D while only 3$\%$ are fragile TIs in $3$D.
Finally, we show that the fragile topology of $d$D $k$th-order TI with $k>d$ is manifested in the relative winding of the Wilson loop and nested Wilson loop spectra.
Here, we mainly focus on the class A of AZ classification that does not have time-reversal, particle-hole, chiral symmetries.
However, we anticipate that the idea we propose here can be generalized to other AZ classes.
In particular, the Dirac Hamiltonians of generalized higher-order topological phases are discussed in Appendix~\ref{app: LCandAZ}.

To generalize the notion of the higher-order topology to the cases with $k>d$, we use the Dirac Hamiltonian approach that has been developed recently to understand the higher-order topology with $1<k\le d$~\cite{khalaf2018higher}.
In Ref.~\onlinecite{khalaf2018higher}, it is shown that when a Dirac Hamiltonian has $(k-1)$ mass terms odd under inversion in addition to the mass term even under inversion, the Dirac Hamiltonian describes a $k$th-order topological phase.
The mass term even under inversion is the bulk mass term $\mathcal{M}$ while the other mass terms odd under inversion are boundary mass terms $m_{a,\bb{r}}$ ($a=1,\dots,k-1$).
Then, the energy gap at the boundary $E_g=2\sqrt{\sum_{a=1}^{k-1}m^2_{a,\bb{r}}}$ vanishes in the $(d-k)$D subspace of the boundary, because the boundary mass terms vanish there due to the constraint $m_{a,\bb{r}}=-m_{a,-\bb{r}}$ imposed by inversion symmetry.

Using the Dirac Hamiltonian formalism, it is straightforward to extend the notion of higher-order topology to the cases with $k>d$~\cite{cualuguaru2019higher, okuma2019topological, shiozaki2019classification}.
For instance, $d$D $(d+1)$th-order TIs have $d$ boundary mass terms, so the boundary is fully gapped in general.
In this case, even though the boundary is gapped, inversion symmetry requires that the $(d+1)$ mass terms, including $d$ boundary mass terms and the bulk mass term, form a skyrmionlike structure in real space, and have an odd winding number along the boundary.
In Refs.~\cite{okuma2019topological} and \cite{shiozaki2019classification}, it was shown that 3D fourth-order Dirac Hamiltonian in $\mathbb{R}^3$ is equivalent to a bound state at the origin. However, the relevant bulk-boundary correspondence was not discussed. 
In this paper, we show that the winding number of boundary mass terms gives rise to the filling anomaly using the Goldstone-Wilczek formula~\cite{goldstone1981fractional}.
A related idea has already been proposed for $d=2$ in Ref.~\onlinecite{wieder2018axion} but our theory is generally valid in arbitrary spatial dimensions.
We also clarify the spatial location of extra charge accumulation or depletion.
According to the Goldstone-Wilczek formula, since the extra charge density is determined by the mass winding density, the extra charges are localized in the region where the mass terms change rapidly.
Since the bulk mass term changes rapidly at the boundary, the induced charges are naturally located at the boundary. The actual location of the extra charge is the position on the boundary where the boundary mass terms vary abruptly. Because of this reason, when the geometry of the system has sharp corners, the extra charges appear there. On the other hand, when the geometry of the system has only smooth boundary, the extra charges are spread over the whole boundary, as confirmed by numerical calculations.

Under the periodic boundary condition in which the fragile band topology is defined, $d$D $(d+1)$th-order TIs can be constructed in the following way. We start from an insulator with $2^{d-1}$ occupied bands in which all the occupied bands have positive parities at all inversion-invariant momenta.
Let us call such an insulator as a trivial-parity insulator.
Then, replacing all the positive-parity states at an odd number of inversion-invariant momenta by negative-parity states gives a $d$D $(d+1)$th-order TIs. In particular, when a $d$D $(d+1)$th-order TI has negative-parity states only at a single inversion-invariant momentum, the transition between the trivial-parity insulator and the $d$D $(d+1)$th-order TI can naturally be described by a Dirac Hamiltonian.
In this construction, $d$D $k$th-order TIs with $k>d$ have the fragile band topology.
In particular, a minimal fragile TI with the filling anomaly can be considered as a $d$D $(d+1)$th-order TI.
One can construct a $d$D $k$th-order TI with $k>(d+1)$ by stacking $d$D $(d+1)$th-order TIs.
In general, a $d$D $k$th-order TI with $k>(d+1)$ does not have any nontrivial feature at the boundary although it has fragile topology.
Nevertheless, we show that the trivial-parity insulator and $d$D $k$th-order TI can be distinguished by the net parity $\mathcal{I}=N_--N_+$~\cite{thorngren2018gauging,song2017topological,huang2017building} of a finite-size system with open boundary, defined by the difference between the number of occupied states with odd parity $N_-$ and the number of occupied states with even parity $N_+$.
In fact, the net parity $\mathcal{I}$ defined for the given geometry of the finite-size system is identical to one of bulk topological invariants for inversion-symmetric fragile TIs and atomic insulators.
This is consistent with the layer construction proposed in Refs.~\onlinecite{trifunovic2019higher,song2017topological,huang2017building,shiozaki2018generalized,song2018topological}.

The properties of fragile TIs discussed above are rigorously confirmed by analyzing the symmetry data, that is, the inversion parities at inversion-invariant momenta in the Brillouin zone.
In fact, in recent studies, it is found that parities can be organized in the form of symmetry indicators that distinguish stable topological phases from fragile topological or atomic insulators~\cite{po2017symmetry}. However, this method cannot analyze fragile TIs.
Here, we establish a simple criterion that can be used to distinguish fragile TIs from atomic insulators efficiently.
Using this method, we estimate the number of the parity configurations corresponding to inversion-symmetric stable TIs, fragile TIs, and atomic insulators, respectively, in both $2$D and $3$D.

To investigate the fragile topology further, we systematically examine how many trivial bands are required to turn a fragile TI to an atomic insulator.
Using our criterion for fragile topology, we show that for the $d$D $(d+1)$th-order TI which has $2^{d-1}$ occupied bands, $(2^{d-1}-1)$ bands are needed to trivialize the fragile topology.
We explain why $(2^{d-1}-1)$ bands are necessary for trivialization by using the Wilson loop method.
The idea is that all the spectral winding of Wilson loops~\cite{yu2011equivalent,alexandradinata2014wilson,alexandradinata2016topological} and (generalized) nested Wilson loops~\cite{benalcazar2017quantized, benalcazar2017electric} have to be unwound to completely resolve the Wannier obstruction in fragile topological phases.
Explicitly, we show that the total number of bands needed to trivialize the $d$D $(d+1)$th-order TIs can be decomposed into $2^{d-1}-1=2^{d-2}+2^{d-3} + \cdots + 1$, where $2^{d-2-l}$ is the number of bands needed to remove the spectral winding of the $l$th nested Wilson loop ($l=0,1,2,\dots$). Here, the zeroth nested Wilson loop denotes the conventional Wilson loop.
We also generalize this idea to $d$D $k$th-order TIs with $k>(d+1)$ which also have fragile topology.

The rest of this paper is organized as follows.
We start Sec.~\ref{sec: SSHIntro} with the Su-Schrieffer-Heeger (SSH) model in order to introduce the concepts of $d$D $(d+1)$th-order TIs and its higher-order generalizations.
Although there is no fragile TI in one dimension because there is no Wannier obstruction in one dimension~\cite{marzari2012maximally, alexandradinata2014wilson,  read2017compactly, alexandradinata2019crystallographic}, all the ideas related with the bulk-boundary correspondence can be generalized straightforwardly to higher dimensions.
In Sec.~\ref{sec: Methods}, we introduce the Dirac Hamiltonians of $d$D $k$th-order TIs for arbitrary $d$ and $k$.
Then, we show that the $d$ boundary mass terms of $d$D $(d+1)$th-order TI have a nontrivial winding number on the boundary in the presence of inversion symmetry.
The winding number of boundary mass terms is directly connected to the induced current which dictates the nontrivial charge accumulation at the boundary.
In Sec.~\ref{sec: Classification}, we construct a mapping between a parity configuration in the Brillouin zone and the inversion-symmetric Wannier states.
With the mapping, we obtain $(2^d+1)$ invariants for the class A inversion-symmetric insulators in $d$ dimensions.
From this, we develop an efficient method for diagnosing the atomic, fragile topological, and stable topological phases and classify all the possible parity configurations in two and three dimensions.
Also, we introduce the net parity of a finite-size system with open boundary and the condition for the presence of the filling anomaly.
In Sec.~\ref{sec: FragileReal}, we demonstrate our theoretical prediction through numerical calculations of tight-binding models with open boundary condition.
In Sec.~\ref{sec: FragileMomentum}, we show that fragile topological phases protected by inversion symmetry are characterized by the relative winding~\cite{alexandradinata2014wilson} in the (nested) Wilson loop spectrum, which is fragile against adding trivial bands.
Using the $3$D fourth-order TI as a concrete example, we count the number of trivial bands needed to unwind the relative winding in the Wilson loop and the nested Wilson loop spectra and show that this matches with the result expected from our diagnosis method.
Then, we generalize this counting to the case of $d$D $k$th-order TIs.
In Sec.~\ref{sec: Pumping}, we study the general pumping process of inversion-symmetric insulators.
Specifically, we construct $(d+1)$D $(k-1)$th-order TI by a pumping of $d$D $k$th-order TI, which generalizes the result in Ref.~\onlinecite{wieder2018axion} for $d=2$ and $k=3$.
Finally, we conclude in Sec.~\ref{sec: Discussions}.

\section{Symmetry data and generalized higher-order topology in one-dimension \label{sec: SSHIntro}}
Here we introduce the notion of the generalized higher-order topology in one-dimensional (1D) systems by considering the Su-Schrieffer-Heeger (SSH) model~\cite{su1979solitons} and its multiple copies.
Let us note that the meaning of ``topological phases'' is a bit subtle in 1D, since only atomic and obstructed atomic insulators~\cite{bradlyn2017topological} exist in 1D~\cite{marzari2012maximally, alexandradinata2014wilson,  read2017compactly, alexandradinata2019crystallographic}, in contrast to higher dimensional systems where topological phases mean the phases with either fragile or stable band topology. For convenience, however, here we use the term ``topological phases" to denote the obstructed atomic insulators in $1$D systems.

\subsection{Second-order topology in 1D \label{subsec: SSH(1,2)}}
Let us begin with the $1$D SSH model Hamiltonian with inversion symmetry,
\ba
\label{eq: SSH}
H_{\rm SSH}(k) = \sin k \, \sigma_x + (t+\cos k) \, \sigma_z,
\ea
which is invariant under inversion symmetry $I=\sigma_z$ as well as chiral symmetry $S=\sigma_y$, that is, 
\ba
I H_{\rm SSH}(k) I^{-1} &= H_{\rm SSH}(-k), \nn \\
S H_{\rm SSH}(k) S^{-1} &= -H_{\rm SSH}(k).
\ea
The energy spectrum is given by $E(k)=\pm \sqrt{\sin^2 k+(t+\cos k)^2}$, which is gapped when $|t| \ne 1$. When $t=1$ ($t=-1$), the band gap closes at $k=\pi$ ($k=0$).

%%%%%%%%%%%%%%%%%%%%%%%%%%%%%%%%%%%%%%%%%%%%%%%%%%%%%%%%%%%%%%%%
{
	\setlength{\tabcolsep}{12pt}
	\begin{table}[b!]
		\caption{The parity of the occupied state at inversion-invariant momenta of inversion-symmetric atomic insulators. $s$ and $p$ denote the even- and odd-parity states, respectively. $A$ and $B$ are Wyckoff positions of $1$D inversion-symmetric lattices shown in Fig.~\ref{fig: FAnomaly}(a).}
		\begin{tabular}{c|c|c}
			\hline
			\hline
			Orbital & $k=0$ & $k=\pi$ \\ \hline
			$s(A)$ & $+$ & $+$ \\ \hline
			$p(A)$ & $-$ & $-$ \\ \hline
			$s(B)$ & $+$ & $-$ \\ \hline
			$p(B)$ & $-$ & $+$ \\ \hline
			\hline
		\end{tabular}
		\label{table: 1dRep}
	\end{table}
}
%%%%%%%%%%%%%%%%%%%%%%%%%%%%%%%%%%%%%%%%%%%%%%%%%%%%%%%%%%%%%%%%

The insulating phases are distinguished by a topological invariant, which is nothing but the winding number of the two-component vector $(t+\cos k,\sin k)$ around the origin $(0,0)$, so that the insulating phase is topological when $|t|<1$ while it is trivial when $|t|>1$.

\begin{figure}[t!]
	\centering
	\includegraphics[width=0.49\textwidth]{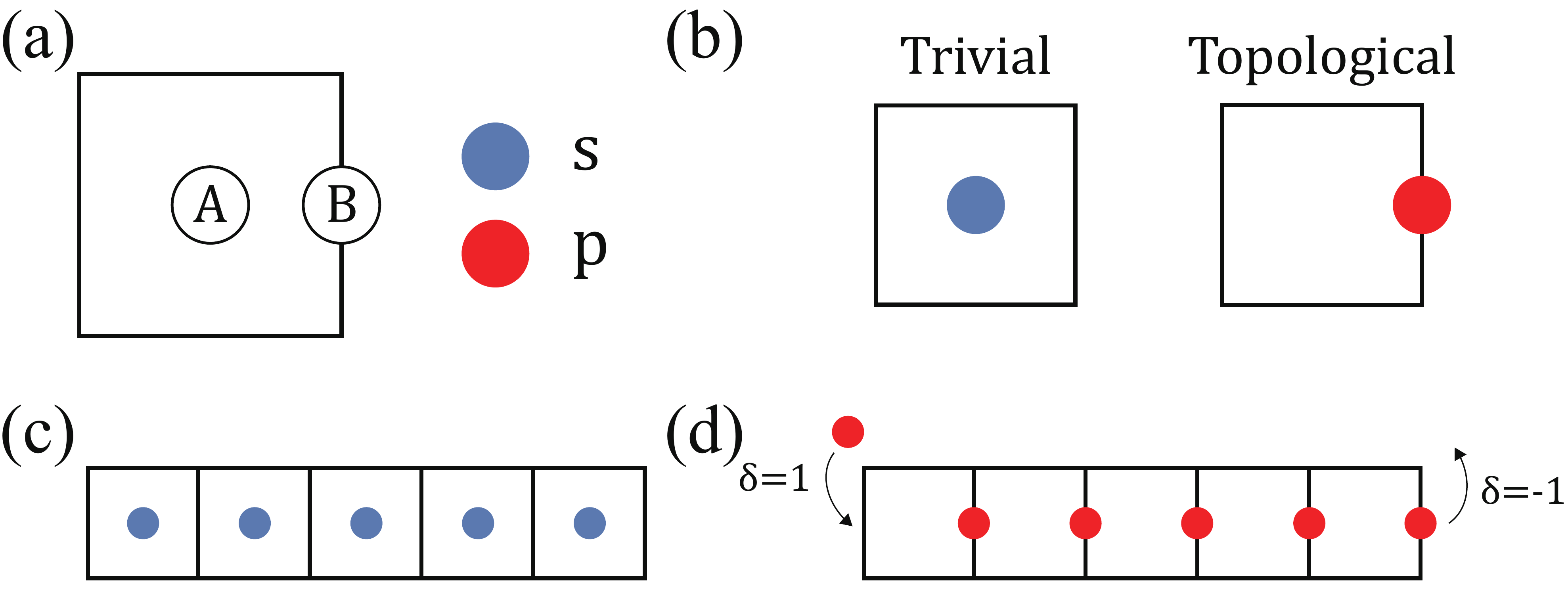}
	\caption{
		The Wannier center of $1$D SSH model and filling anomaly.
		(a) Wyckoff positions $A$ and $B$ of a $1$D inversion-symmetric lattice. $s$ and $p$ orbitals are denoted by the blue and red circles, respectively.
		(b) The parity configurations of trivial and topological phases are equivalent to those of the atomic insulators $s(A)$ and $p(B)$, respectively.
		(c) A finite-size trivial insulator with open boundaries, which is inversion symmetric.
		(d) A finite-size topological phase with open boundaries cannot be made inversion symmetric. In order to have inversion-symmetric ground state, one extra electron must be added ($\delta=1$) or removed ($\delta=-1$) to the half-filled system, which is called the filling anomaly~\cite{benalcazar2019quantization}.}
	\label{fig: FAnomaly}
\end{figure}

The topological and trivial insulators can also be distinguished by the parity configuration of occupied states at inversion-invariant momenta.
In the trivial phase, the parity of the occupied band at $k=0$ ($p_0$) and that at $k=\pi$ ($p_{\pi}$) is $(p_0,p_{\pi})=(-,-)$ when $t>1$ while $(p_0,p_{\pi})=(+,+)$ when $t<-1$. In the topological phase with $|t|<1$, $(p_0,p_{\pi})=(-,+)$.
Table~\ref{table: 1dRep} shows the parity configurations of atomic insulators with inversion symmetry, which shows that the parity configuration of the topological insulator is equivalent to that of the atomic insulator with a $p$ orbital at the Wyckoff position $B$. On the other hand, the parity configuration of the trivial insulator is equivalent to that of the atomic insulator with a $s$ ($p$) orbital at the Wyckoff position $A$ when $t<-1$ ($t>1$) [see Fig.~\ref{fig: FAnomaly}(b)].

%%%%%%%%%%%%%%%%%%%%%%%%%%%%%%%%%%%%%%%%%%%%%%%%%%%%%%%%%%%%%%%%
%%%%%%%%%%%%%%%%%%%%%%%%%%%%%%%%%%%%%%%%%%%%%%%%%%%%%%%%%%%%%%%%

%%%%%%%%%%%%%%%%%%%%%%%%%%%%%%%%%%%%%%%%%%%%%%%%%%%%%%%%%%%%%%%%
\begin{figure}[t!]
	\centering
	\includegraphics[width=0.49\textwidth]{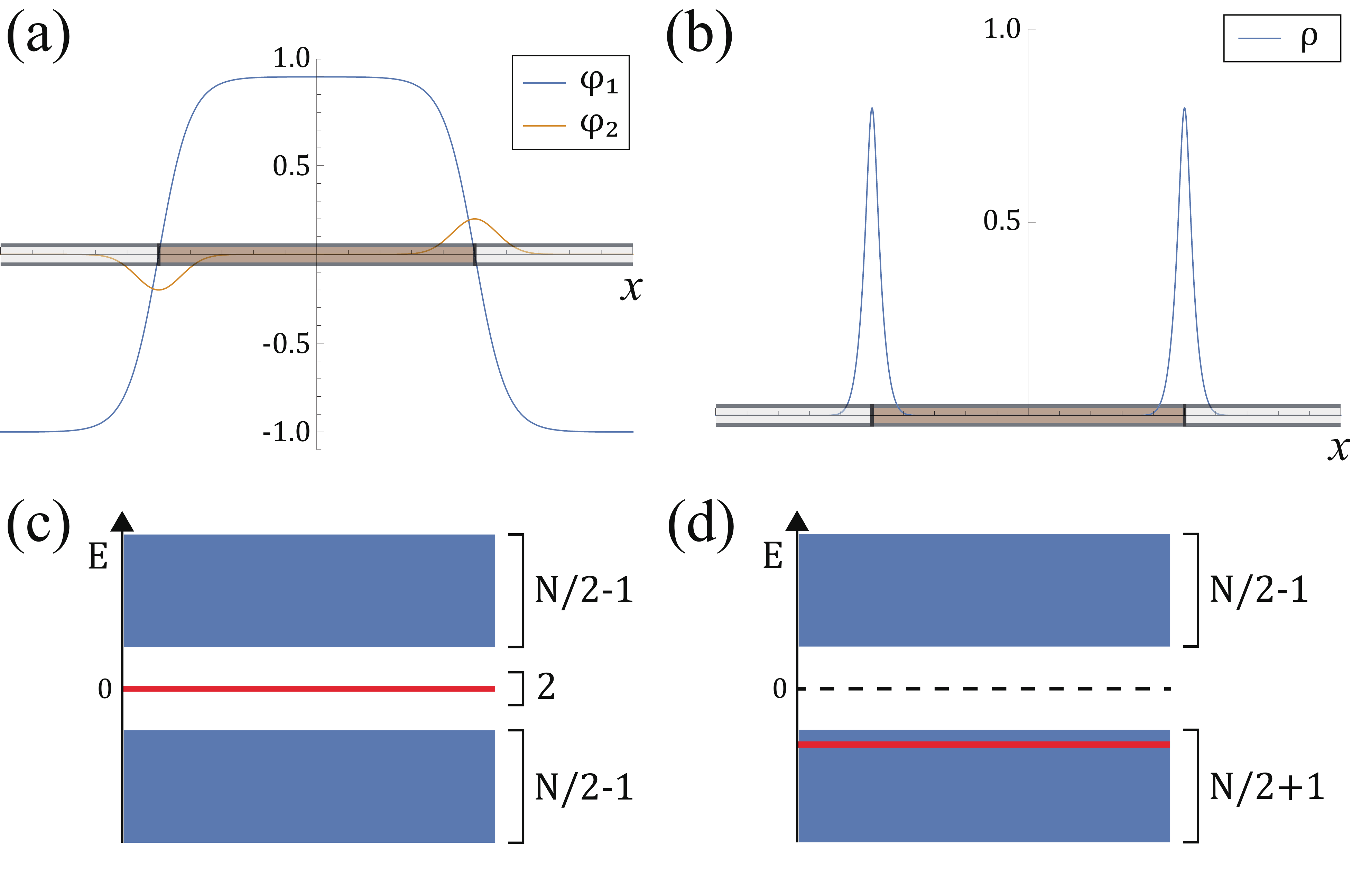}
	\caption{
		Mass profiles of the $1$D SSH model and the relevant charge density and energy spectrum.
		(a) A heterostructure composed of a $1$D topological insulator at $|x|<L/2$ sandwiched by trivial insulators at $|x|>L/2$. The domain walls between the trivial and topological insulators described in Eqs.~\eqref{eq: 1d1stSSH} and~\eqref{eq: 1d2ndSSH} are implemented by using $\varphi_1(x)=-1-\tanh(\frac{x-L/2}{l_1})+\tanh(\frac{x+L/2}{l_1})$ and $\varphi_2(x)=0.2 \big( \exp [-(\frac{x-L/2}{l_2})^2 ] - \exp [-(\frac{x+L/2}{l_2})^2 ] \big)$. $l_1=l_2=\frac{L}{10}$ are used. The domain of the topological insulator is depicted by a brown rod. Two ends of the rod are located at $x=\pm \frac{L}{2}$.
		(b) The accumulated charge density $\rho$ when the second mass term $\varphi_2(x)$ is added. $\rho(x)$ is localized only in the region where the bulk mass $\varphi_1(x)$ changes its sign.
		(c) The energy spectrum of $\mathcal{H}_{(1,1)}$ in Eq.~\eqref{eq: 1d1stSSH}. Two zero-energy states localized at domain walls are degenerate due to inversion symmetry.
		(d) The energy spectrum of $\mathcal{H}_{(1,2)}$ in Eq.~\eqref{eq: 1d2ndSSH} with the boundary mass term $\varphi_2(x)$ that breaks chiral symmetry but preserves inversion symmetry. The pair of domain wall states are merged into the bulk spectrum. There is a single hole in the valence band at half-filling indicating the filling anomaly.}
	\label{fig: SSH}
\end{figure}
%%%%%%%%%%%%%%%%%%%%%%%%%%%%%%%%%%%%%%%%%%%%%%%%%%%%%%%%%%%%%%%%

To examine the surface states, let us construct a heterostructure composed of a topological insulator in the region with $|x|<L/2$ and trivial insulators on the sides with $|x|>L/2$. Such a heterostructure can also be described by the Hamiltonian in Eq.~\eqref{eq: SSH} when $t$ depends on the spatial coordinate $x$. For simplicity, we assume that $-1<t(x)<1$ for $|x|<L/2$ and $t(x)<-1$ for $|x|>L/2$. Since the band gap closes at $k=0$ at the critical point with $t=-1$, one can consider the following low-energy Dirac Hamiltonian:
\ba
\label{eq: 1d1stSSH}
\mathcal{H}_{(1,1)}= k \sigma_x+\varphi_1(x) \sigma_z,
\ea
where $\varphi_1(x)\equiv 1+t(x)$ that plays the role of a bulk mass term. Then, $\varphi_1(x)>0$ for $|x|<L/2$ and $\varphi_1(x)<0$ for $|x|>L/2$. $\varphi_1(x)=\varphi_1(-x)$ due to inversion symmetry. Let us note that the symbol $\mathcal{H}_{(d,k)}$ indicates the $d$D Dirac Hamiltonian with $k$ anticommuting mass terms.

When the length $L$ is sufficiently large, there are two zero-energy states at two edges (or two domain walls), where the bulk mass term $\varphi_1(x)$ changes its sign [see Fig.~\ref{fig: SSH}(c)].
Because of inversion symmetry, the two edge states should be either occupied or unoccupied simultaneously.
Thus, an inversion-symmetric ground state cannot satisfy the exact half-filling condition, which is described in Figs.~\ref{fig: FAnomaly}(c) and~\ref{fig: FAnomaly}(d).
This phenomenon is known as the {\it filling anomaly}~\cite{benalcazar2019quantization}.

Let us now break chiral symmetry by adding an {inversion-symmetric second mass term $\varphi_2(x)$ to $\mathcal{H}_{(1,1)}$, which gives
\ba
\label{eq: 1d2ndSSH}
\mathcal{H}_{(1,2)} = k \sigma_x+\varphi_1(x)\sigma_z+\varphi_2(x)\sigma_y,
\ea
where $\varphi_2(x)=-\varphi_2(-x)$.
We assume that $\varphi_2(x)$ is finite only around the domain wall and call this type of mass terms as ``boundary'' mass terms.
The Dirac Hamiltonian with $(k-1)$ boundary mass terms describes a $k$th-order topological insulator.
In this respect, $\mathcal{H}_{(1,2)}$ means the Dirac Hamiltonian of $1$D second-order TI.
We generalize the notion of $d$D $k$th-order TI for arbitrary $k$ in Sec.~\ref{subsec: Dirac Hamiltonian}.

The boundary mass term $\varphi_2$ makes the energy of the edges states shifted from zero. However, since inversion symmetry is maintained, the two edge states are still degenerate, and thus the filling anomaly remains.
In particular, the profile of $\varphi_2$ depicted in Fig.~\ref{fig: SSH}(a) lowers the energy of the edge states. When the edge states are occupied simultaneously, the system has one extra charge relative to the half-filling, and the resulting phase is said to have the filling anomaly with $\delta=1$ [see Fig.~\ref{fig: SSH}(d)].

The extra charge density accumulated at a domain wall shown in Fig.~\ref{fig: SSH}(b) is given by the Goldstone-Wilczek formula~\cite{goldstone1981fractional}
\ba
\label{eq: GWformula}
\rho(x)
&=\frac{1}{2\pi} \epsilon^{a b} \frac{\varphi_a \dr_{x} \varphi_b}{\varphi^2} \nn \\
&=\frac{1}{2\pi} \dr_{x} \tan^{-1}\Big(\frac{\varphi_2}{\varphi_1}\Big).
\ea
It is worth noting that Eq.~\eqref{eq: GWformula} shows that the extra charge is accumulated mostly at the domain wall, even when the edge spectrum merges into the bulk spectrum. This is because the bulk mass term $\varphi_1$ varies only near the domain wall where the accumulated charge density is located. This property is shared by any $d$D $(d+1)$th-order inversion-symmetric TIs as shown in Sec.~\ref{subsec: bbcorres}.

The total amount of the extra charge $Q=\int dx \rho(x)$ is given by the winding number of $(\varphi_1,\varphi_2)$.
Note that we can compactify $R^1$ to $S^1$ since $(\varphi_1,\varphi_2) \rightarrow (-1,0)$ at $x\rightarrow \pm \infty$, and thus the integer winding number can be defined. 
The boundary conditions on $\varphi_1$ and $\varphi_2$ imposed by inversion symmetry constrain $Q$ to be 1.
This result is consistent with the picture based on the shifting of the energy level of the edge states by adding $\varphi_2$.
When the bulk mass term $\varphi_1(x)$ changes its sign near each domain wall, $\tan^{-1}\Big(\frac{\varphi_2}{\varphi_1}\Big)$ is abruptly changed by $\pi$. Therefore, a half-integral charge should be accumulated at each domain wall.

\subsection{Third- and fourth-order topology \label{subsec: SSH(1,4)}}
The idea of higher-order topology can be generalized further even to the cases with $k>(d+1)$. To illustrate the idea, let us first consider two copies of the SSH model. The relevant Hamiltonian is obtained by doubling $H_{\rm SSH}(k)$ in an inversion-symmetric way as
\ba
H(k)
&= \tau_x \otimes H_{\rm SSH}(k) \nn \\
&= \sin k \, \tau_x \sigma_x + (t+\cos k) \, \tau_x \sigma_z.
\ea
The corresponding Dirac Hamiltonian linearized near $k=0$ is
\ba
\label{eq: DoubleSSH}
\mathcal{H}= k \, \tau_x \sigma_x+\varphi_1(x) \, \tau_x \sigma_z,
\ea
where $\varphi_1(x)>0$ for $|x|<L/2$ and $\varphi_1(x)<0$ for $|x|>L/2$, which again describes a heterostructure with two domain walls at $x=\pm L/2$.
The Hamiltonian $\mathcal{H}$ is invariant under inversion symmetry $I=\tau_x \sigma_z$ and chiral symmetry $S=\tau_z \sigma_0$. 
When chiral symmetry $S=\tau_z \sigma_0$ is preserved, two boundary mass terms $\varphi_2(x) \tau_x \sigma_y + \varphi_3(x) \tau_y \sigma_0$ are allowed. Adding them to $\mathcal{H}$ gives
\ba
\label{eq: 1d3rdSSH}
\mathcal{H}_{(1,3)}=&k \, \tau_x \sigma_x + \varphi_1(x) \, \tau_x \sigma_z + \nn \\
+&\varphi_2(x) \, \tau_x \sigma_y + \varphi_3(x) \, \tau_y \sigma_0,
\ea
where $\varphi_{2,3}(x)=-\varphi_{2,3}(-x)$. The resulting Dirac Hamiltonian describes a $1$D third-order TI belonging to class AIII.

There is one remaining boundary mass term $\varphi_4(x) \tau_z \sigma_0$ that breaks $S$ but keeps $I$. By adding this term to $\mathcal{H}_{(1,3)}$, we obtain the following Dirac Hamiltonian:
\ba
\label{eq: 1d4thSSH}
\mathcal{H}_{(1,4)}=&k \, \tau_x \sigma_x + \varphi_1(x) \, \tau_x \sigma_z + \varphi_2(x) \, \tau_x \sigma_y \nn \\
&+ \varphi_3(x) \, \tau_y \sigma_0 + \varphi_4(x) \, \tau_z \sigma_0,
\ea
which describes a $1$D fourth-order TI belonging to class A.

%%%%%%%%%%%%%%%%%%%%%%%%%%%%%%%%%%%%%%%%%%%%%%%%%%%%
\begin{figure}[t!]
	\centering
	\includegraphics[width=0.49\textwidth]{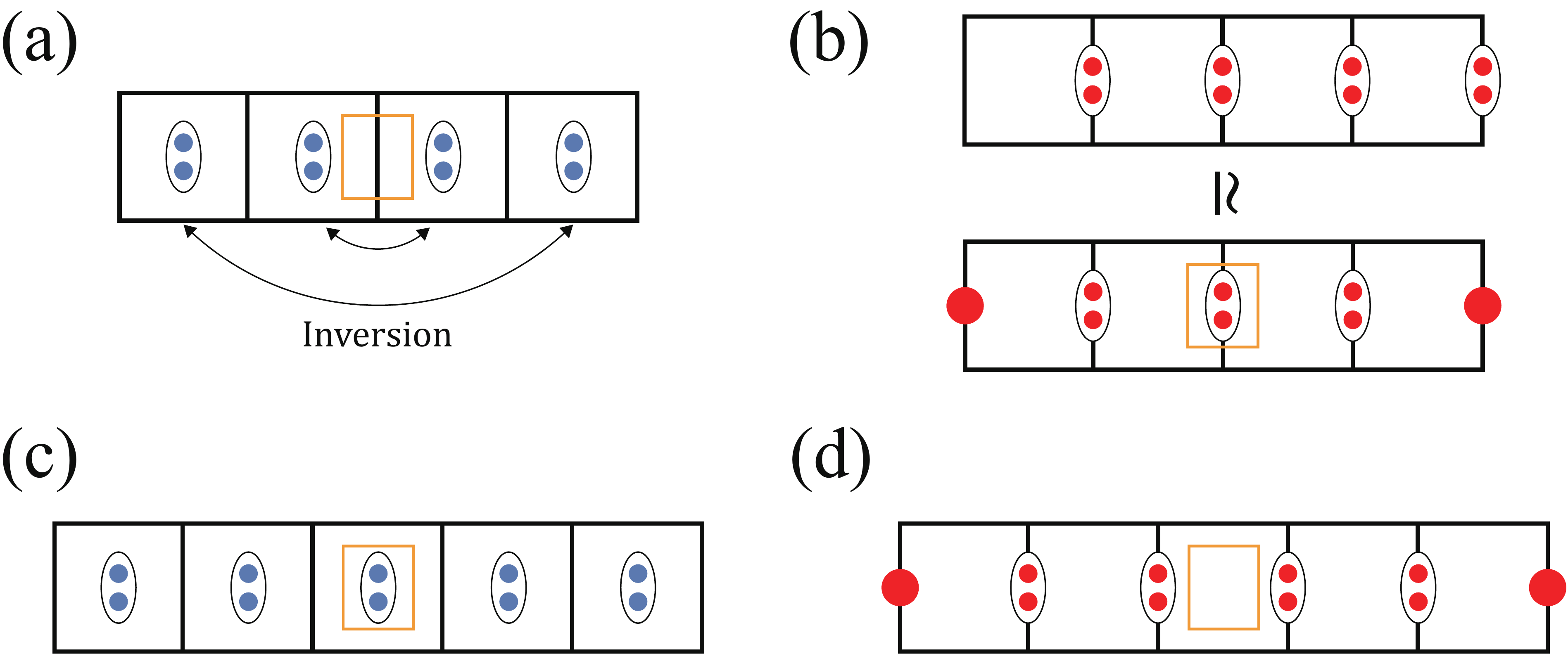}
	\caption{
		The Wannier centers of $1$D third- and fourth-order TIs and net parities.
		(a) Two copies of trivial insulators of a finite-size with even $N_{\rm cell}$. All unit cells form inversion-symmetric pairs giving an equal number of even-parity and odd-parity states, and thus $\mathcal{I}=0$.
		(b) Two copies of trivial insulators of a finite-size with even $N_{\rm cell}$. The upper configuration is rearranged into the lower one which is inversion symmetric. All $p$ orbitals except the two at the center within the orange box form the equal number of bonding and antibonding states, and thus $\mathcal{I}=2$.
		(c) Similar to (a) but with odd $N_{\rm cell}$. All $s$ orbitals except the two at the center form inversion-symmetric pairs, and thus $\mathcal{I}=-2$.
		(d) Similar to (b) but with odd $N_{\rm cell}$. All orbitals form inversion-symmetric pairs, and thus $\mathcal{I}=0$.
		In all cases, $\mathcal{I}\equiv N_--N_+$ is determined by the difference in the number of $p$ and $s$ orbitals at the center within the orange box.}
	\label{fig: 1d4thSSH}
\end{figure}
%%%%%%%%%%%%%%%%%%%%%%%%%%%%%%%%%%%%%%%%%%%%%%%%%%%%

%%%%%%%%%%%%%%%%%%%%%%%%%%%%%%%%%%%%%%%%%%%%%%%%%%%%
{
	\setlength{\tabcolsep}{8pt}
	\begin{table}[b!]
		\caption{The relation between $N_+$, $N_-$, $N_{\rm cell}$, and $\mathcal{I}$ for doubled SSH models of finite-size, which corresponds to the $1$D third and fourth-order topological insulator described by $\mathcal{H}_{(1,3)}$ and $\mathcal{H}_{(1,4)}$.}
		\begin{tabular}{c|c|c|c|r}
			\hline
			\hline
			$N_{\rm cell}$ & Phase & $N_{+}$ & $N_{-}$ & $\mathcal{I}$ \quad \\ \hline
			\hline
			\multirow{2}{*}{Even} & Trivial & $N_{\rm cell}$ & $N_{\rm cell}$ & $0$ \quad \\
			\cline{2-5} & Topological & $N_{\rm cell}-1$ & $N_{\rm cell}+1$ & $2$ \quad \\ \hline
			\multirow{2}{*}{Odd} & Trivial & $N_{\rm cell}+1$ & $N_{\rm cell}-1$ & $-2$ \quad \\
			\cline{2-5} & Topological & $N_{\rm cell}$ & $N_{\rm cell}$ & $0$ \quad \\ \hline
			\hline
		\end{tabular}
		\label{table: 1d4thParity}
	\end{table}
}
%%%%%%%%%%%%%%%%%%%%%%%%%%%%%%%%%%%%%%%%%%%%%%%%%%%%%

For both $\mathcal{H}_{(1,3)}$ and $\mathcal{H}_{(1,4)}$, there are neither the filling anomaly nor zero-energy states.
However, the relative topology between the trivial and topological phases can still be revealed by real-space wave function.

To demonstrate the idea, let us consider a finite-size doubled SSH model composed of $N_{\rm cell}$ unit cells in which the total number of energy eigenstates is $4N_{\rm cell}$. In the case of a trivial insulator, two $s$ orbitals are located at the center of each unit cell, whereas, in the case of a topological insulator, two $p$ orbitals are located at a boundary of each unit cell. We first consider a trivial insulator with an even integer $N_{\rm cell}$ shown in Fig.~\ref{fig: 1d4thSSH}(a). Since a unit cell is related with another unit cell via inversion symmetry, their bonding and antibonding combinations generate equal number of even- and odd-parity states. That is, $N_+=N_-=N_{\rm cell}$ where $N_+$ ($N_-$) denotes the number of occupied bands with even (odd) parity. Similarly, when $N_{\rm cell}$ is odd as shown in Fig.~\ref{fig: 1d4thSSH}(c), all the unit cells except the one at the center form inversion-symmetric pairs, and contribute equally to $N_+$ and $N_-$. However, the central unit cell containing two $s$ orbitals at the inversion center contributes 2 only to $N_+$, which leads to $N_+=N_{\rm cell}+1$ and $N_-=N_{\rm cell}-1$.

Similar analysis can be done for topological insulators. Since there is no filling anomaly, a finite-size doubled SSH model can have inversion-symmetric atomic configurations as shown in Figs.~\ref{fig: 1d4thSSH}(b) and~\ref{fig: 1d4thSSH}(d). In contrast to the trivial insulator in which orbitals are located at the center of each unit cell, the orbitals of a topological insulator are located at the edges of each unit cell. When $N_{\rm cell}$ is even, $p$ orbitals at an edge are related with another $p$ orbitals via inversion symmetry, and their bonding and antibonding combinations contribute equally to $N_+$ and $N_-$. Since the remaining two $p$ orbitals at the inversion center contribute 2 only to $N_-$, we have $N_+=N_{\rm cell}-1$ and $N_-=N_{\rm cell}+1$. On the other hand, when $N_{\rm cell}$ is odd, all $p$ orbitals are related by inversion symmetry and thus $N_+=N_-=N_{\rm cell}$. The dependence of $N_\pm$ on the parity of $N_{\rm cell}$ is summarized in Table~\ref{table: 1d4thParity}.

The four distinct cases considered above can be distinguished in terms of the net parity $\mathcal{I}$~\cite{thorngren2018gauging,song2017topological,huang2017building} defined as (with an additional $-1$ in our convention)
\ba
\label{eq: NetParity}
\mathcal{I}\equiv N_--N_+.
\ea
Since a pair of orbitals related by inversion symmetry contribute equally to $N_+$ and $N_-$, they do not contribute to $\mathcal{I}$. Thus, it is determined by the difference in the number of $p$ and $s$ orbitals localized at the inversion center of the finite-size system.
When $N_{\rm cell}$ is even, one can easily find $\mathcal{I}=0$ ($\mathcal{I}=2$) for a trivial (topological) insulator. On the other hand, when $N_{\rm cell}$ is odd, we obtain $\mathcal{I}=-2$ ($\mathcal{I}=0$) for a trivial (topological) insulator [see Fig.~\ref{fig: 1d4thSSH}].

%%%%%%%%%%%%%%%%%%%%%%%%%%%%%%%%%%%%%%%%%%%%%%%%%%%%%
\begin{figure}[t!]
	\centering
	\includegraphics[width=0.49\textwidth]{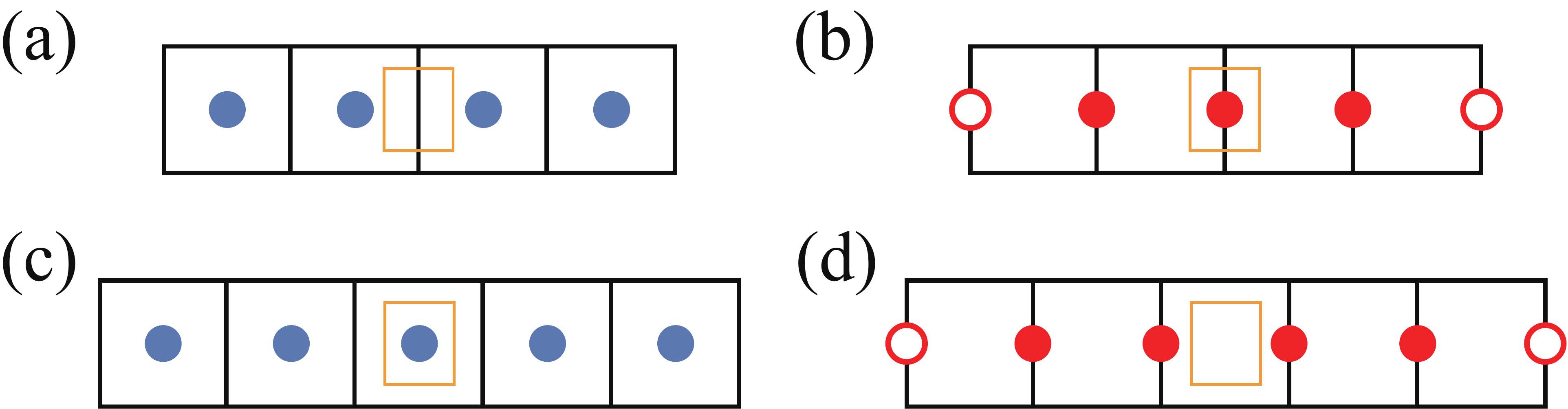}
	\caption{
		The Wannier centers of $1$D second-order TI and net parities.
		(a) Atomic configuration for a finite-size trivial insulator relevant to the $1$D second-order TI when $N_{\rm cell}$ is even. All $s$ orbitals form inversion-symmetric pairs, and thus $\mathcal{I}=0$.
		(b) Atomic configuration for a finite-size topological insulator relevant to the $1$D second-order TI when $N_{\rm cell}$ is even. The red empty circles at the boundaries must be occupied or unoccupied simultaneously because of the filling anomaly. All $p$ orbitals except the one at the center form inversion-symmetric pairs, and thus $\mathcal{I}=1$.
		(c) Similar to (a) but when $N_{\rm cell}$ is odd. All $s$ orbitals except the one at the center form inversion-symmetric pairs, and thus $\mathcal{I}=-1$.
		(d) Similar to (b) but when $N_{\rm cell}$ is odd. All $p$ orbitals form inversion-symmetric pairs, and thus $\mathcal{I}=0$.}
	\label{fig: 1d2ndSSH}
\end{figure}
%%%%%%%%%%%%%%%%%%%%%%%%%%%%%%%%%%%%%%%%%%%%%%%%%%%%%

%%%%%%%%%%%%%%%%%%%%%%%%%%%%%%%%%%%%%%%%%%%%%%%%%%%%%
{
	\setlength{\tabcolsep}{3pt}
	\renewcommand{\arraystretch}{1.4}
	\begin{table}[b!]
		\caption{The relation between $N_+$, $N_-$, $N_\pm$, and $\mathcal{I}$ for the $1$D second-order TI described by $\mathcal{H}_{(1,2)}$ when the extra charge $\delta$ is added due to the filling anomaly.}
		\begin{tabular}{c|c|c|c|r}
			\hline
			\hline
			$N_{\rm cell}$ & Phase & $N_{+}$ & $N_{-}$ & $\mathcal{I}$ \quad \\ \hline
			\hline
			\multirow{2}{*}{Even} & Trivial & $\frac{1}{2}N_{\rm cell}$ & $\frac{1}{2}N_{\rm cell}$ & $0$ \quad \\
			\cline{2-5} & Topological & $\frac{1}{2}(N_{\rm cell}+\delta-1)$ & $\frac{1}{2}(N_{\rm cell}+\delta+1)$ & $1$ \quad \\ \hline
			\multirow{2}{*}{Odd} & Trivial & $\frac{1}{2}(N_{\rm cell}+1)$ & $\frac{1}{2}(N_{\rm cell}-1)$ & $-1$ \quad \\
			\cline{2-5} & Topological & $\frac{1}{2}(N_{\rm cell}+\delta)$ & $\frac{1}{2}(N_{\rm cell}+\delta)$ & $0$ \quad \\ \hline
			\hline
		\end{tabular}
		\label{table: 1d2ndParity}
	\end{table}
}
%%%%%%%%%%%%%%%%%%%%%%%%%%%%%%%%%%%%%%%%%%%%%%%%%%%%%

A similar analysis can also be applied to the $1$D second-order TI in which the filling anomaly exists. The result is summarized in Fig.~\ref{fig: 1d2ndSSH} and Table~\ref{table: 1d2ndParity}. In fact, $\mathcal{I}$ can serve as a bulk topological invariant for general inversion-symmetric insulators without gapless boundary states, which is discussed in Sec.~\ref{subsec: NetParity}.

\section{Generalized higher-order topology in $d>1$ dimensions \label{sec: Methods}}

\subsection{Dirac Hamiltonian approach \label{subsec: Dirac Hamiltonian}}
Here we extend the notion of $d$D $k$th-order TI with $k>d$ to general dimensions $d>1$. We focus on the symmetry class A of the AZ symmetry classification. For this purpose, we use the Dirac Hamiltonian describing the transition between two topologically distinct phases.
The Dirac Hamiltonians of inversion-symmetric TI/TSCs have been exhaustively studied in Ref.~\onlinecite{khalaf2018higher} that we refer to construct the Dirac Hamiltonian with generalized higher-order topology.

In general, a Dirac Hamiltonian is expressed in terms of $2^n \times 2^n$ $\gamma$ matrices which satisfy the Clifford algebra $\{\gamma_i, \gamma_j\}=2\delta_{ij}$ for $i,j=1,\dots,2n+1$ as
\ba
\label{eq: Dirac}
\mathcal{H}_0(\bb{k}) = \sum_{i=1}^d k_i \gamma_i + \lambda \mathcal{M},
\ea
where $\{\mathcal{M},\gamma_i\} = 0$ and $\mathcal{M}^2=\mathds{1}_{2^n}$.
Here, $\mathds{1}_{2^n}$ denotes the $2^n \times 2^n$ identical matrix.
In Eq.~\eqref{eq: Dirac}, $\mathcal{M}$ is the bulk mass term because the energy eigenvalues are given by $E(\bb{k})=\pm\sqrt{\bb{k}^2+\lambda^2}$.
Changing the sign of the bulk mass term $\mathcal{M}$ flips all the inversion parities at $\bb{k}=0$. We choose $\lambda>0$ ($\lambda<0$) to mean the topological (trivial) insulator.
Here, we consider the cases when the inversion symmetry operator is identical to the bulk mass term, i.e., $I=\mathcal{M}$.
When we choose $\mathcal{M}=\gamma_{2n+1}$, there remain $(2n-d)$ additional mass terms, $m_1 \gamma_{d+1}+\cdots+m_{2n-d} \gamma_{2n}$, which anticommute with $\mathcal{M}$, and thus break inversion symmetry when $m_{i=1,\dots,2n-d}$ are constant.

Now we consider a topological insulator with boundaries whose global shape is inversion symmetric, and allow the position $\bb{r}$ dependence of both the bulk mass term $\lambda(\bb{r}) \mathcal{M}$ and the boundary mass terms $m_{a,\bb{r}} \gamma_{d+a}$ with $a=1,\dots,2n-d$. The corresponding inversion-symmetric Dirac Hamiltonian including the boundary mass terms is given by
\ba
\label{eq: GeneralDirac}
\mathcal{H}(\bb{k},\bb{r}) = \sum_{i=1}^d k_i \gamma_i + \lambda(\bb{r}) \mathcal{M} + \sum_{a=1}^{2n-d} m_{a,\bb{r}} M_{a},
\ea
where $M_a=\gamma_{d+a}$ and $\lambda(\bb{r})>0$ [$\lambda(\bb{r})<0$] inside (outside) the topological insulator. $\mathcal{H}(\bb{k},\bb{r})$ satisfies
\ba
I \mathcal{H}(\bb{k},\bb{r}) I^{-1}=\mathcal{H}(-\bb{k},-\bb{r}),
\ea
in which $m_{a,\bb{r}}=-m_{a,-\bb{r}}$. 

The order $k$ of $\mathcal{H}(\bb{k},\bb{r})$ is defined as follows.
At the boundary $\bb{r}$ with $\lambda(\bb{r})=0$, the energy gap is given by $2 m_s(\bb{r})$ where $m_s(\bb{r})=\sqrt{\sum_{a=1}^{2n-d} m_{a,\bb{r}}^2}$.
When $k \le d$ where $k=2n-d+1$, all boundary mass terms $m_{a,\bb{r}}$ vanish simultaneously on $(d-k)$D domain due to the condition $m_{a,\bb{r}}=-m_{a,-\bb{r}}$ imposed by inversion symmetry.
In this case, the order is given by $k$, which is the conventional way of defining the order of a topological insulator.
On the other hand, when $k > d$, the boundary states are fully gapped, and thus the order $k$ cannot be well defined.
In spite of this, however, we generalize the definition of the order $k$ so that $(k-1)$ is identical to the number of boundary mass terms. Namely, when a $d$D Dirac Hamiltonian has $(k-1)$ boundary mass terms, the relevant TI is a $d$D $k$th-order TI as defined in Sec.~\ref{sec: Intro}.
Let us emphasize once again that a $d$D $k$th-order TI in periodic systems can be well defined by the Dirac Hamiltonian when the relevant trivial insulator is the trivial-parity insulator, so that the sign reversal of the bulk mass term mediates the transition between the trivial-parity insulator and the $d$D $k$th-order TI.

Now we consider the Dirac Hamiltonian for $d$D $k$th-order TI with $k>d$.
Since a $d$D $k$th-order TI exists only when $d+k=2l+1$ where $l=1, 2,\dots$,
the minimal order $k$ of a $d$D TI without gapless boundary states described by a Dirac Hamiltonian is $k=d+1$.
The Dirac Hamiltonian of $d$D $(d+1)$th-order TI is 
\ba
\label{eq: Dirac(d,d)}
\mathcal{H}(\bb{k}, \bb{r}) = \sum_{i=1}^d k_i \gamma_i + \lambda(\bb{r}) \mathcal{M} + \sum_{a=1}^{d} m_{a,\bb{r}} M_{a}.
\ea

Then, $d$D $k$th-order TI with $k>(d+1)$ can be obtained by superposing $2^{\frac{k-d-1}{2}}$ copies of $d$D $(d+1)$th-order TI. For instance, in 1D, we have obtained the $1$D fourth-order TI by doubling the $1$D second-order TI.
Interestingly, all $d$D $k$th-order TIs with $k>d$ turn out to be a fragile phase as shown in Sec.~\ref{sec: FragileReal}.
The properties of the Dirac Hamiltonian of $d$D $(d+1)$th-order TI in Eq.~\eqref{eq: Dirac(d,d)} are discussed in detail in Secs.~\ref{subsec: masshomotopy} and~\ref{subsec: bbcorres}.

\subsection{Homotopy of boundary mass terms \label{subsec: masshomotopy}} 
The Dirac Hamiltonian for the boundary theory $h(\bb{k},\bb{r})$ is obtained via the boundary projection of the bulk Hamiltonian~\cite{khalaf2018higher, khalaf2018symmetry, geier2018second, trifunovic2019higher} 
\ba
h(\bb{k},\bb{r}) = P_+(\bb{r}) \mathcal{H}(\bb{k},\bb{r}) P_+(\bb{r}),
\ea
where $P_+(\bb{r})=\frac{1}{2}[\mathds{1}-i (\bb{n_r}\cdot \bb{\gamma}) \mathcal{M}]$ denotes the projection operator to the boundary where $\bb{n}_{\bb{r}}$ is the surface normal vector at the position $\bb{r}$ and $\bb{\gamma}=(\gamma_1, \dots, \gamma_d)$.
Thus, the boundary theory of $d$D $(d+1)$th-order TI is given by
\ba
\label{eq: DiracBoundary(d,d)}
h(\bb{k}, \bb{r}) = \bb{k}_S \cdot \tilde{\bb{\gamma}} + \sum_{a=1}^{d} m_{a,\bb{r}} \tilde{M}_{a},
\ea
where $\bb{k}_S = \bb{k} - (\bb{k}\cdot \bb{n_r})\bb{n_r}$ denotes the boundary momentum and $\tilde{\gamma}_i$ and $\tilde{M}_a$ are boundary-projected gamma matrices and mass matrices.
The details about the boundary projection are summarized in Appendix~\ref{app: Proj}.
The boundary states of Eq.~\eqref{eq: DiracBoundary(d,d)} are gapped everywhere because of $d$ distinct boundary mass terms.
Since there is no gapless boundary state, the system seems to be topologically trivial.
However, we show that the nontrivial higher-order band topology is manifested in the nontrivial winding structure of the boundary mass terms in real space. With a unit vector $\bb{\varphi}(\bb{r})$ defined in terms of the boundary mass terms $\vec{m}_{\bb{r}}=(m_{1,\bb{r}},\dots,m_{d,\bb{r}})$,
\ba
\bb{\varphi}(\bb{r})=\frac{\vec{m}_{\bb{r}}}{|\vec{m}_{\bb{r}}|},
\ea
one can consider a mapping from the boundary position $\bb{r}$ to the unit vector $\bb{\varphi}(\bb{r})$.
Note that $\bb{\varphi}(\bb{r})$ is well defined since the boundary gap $m_s(\bb{r})=|\vec{m}_{\bb{r}}|$ is non-zero everywhere on the boundary. Also, $\bb{\varphi}(\bb{r})$ satisfies $\bb{\varphi}(\bb{r}) \cdot \bb{\varphi}(\bb{r})=1$.
Since both the boundary and $\bb{\varphi}(\bb{r})$ are homotopic to the $(d-1)$D sphere $S^{d-1}$, one can define the homotopy group $\pi_{d-1}(S^{d-1})=\mathbb{Z}$.
The winding numbers $w_{d}$ relevant to $\pi_d(S^d)=\mathbb{Z}$ are given by
\ba
\label{eq: masswinding}
w_1&=\int d\theta \, \frac{1}{2\pi} \dr_{\theta} \tan^{-1} \Big( \frac{\varphi_2(\theta)}{\varphi_1(\theta)} \Big), \\
w_2&=\int d\theta d\phi \, \frac{1}{4\pi} \bb{\varphi}(\theta,\phi) \cdot \dr_{\theta} \bb{\varphi}(\theta,\phi) \times \dr_{\phi} \bb{\varphi}(\theta,\phi),
\ea
and similar expressions in higher dimensions.
When the system is inversion symmetric, the unit vector $\bb{\varphi}(\bb{r})$ should satisfy
\ba
\label{eq: massconstraint}
\bb{\varphi}(\bb{r}) = - \bb{\varphi}(-\bb{r}).
\ea
With this configuration, the winding number is constrained to be an odd integer~\cite{outerelo2009mapping}.
Let us note that even when a phase transition at the boundary occurs, the winding number modulo $2$ remains invariant due to Eq.~\eqref{eq: massconstraint}.
This shows that the winding number $w_d$ of boundary mass terms characterizes the nontrivial bulk topology.
In the following section, Sec.~\ref{subsec: bbcorres}, the connection between the winding number $w_d$ and the bulk-boundary correspondence of $d$D $(d+1)$th-order TI is discussed in detail.

\subsection{Induced U(1) current and bulk-boundary correspondence \label{subsec: bbcorres}}
Consider the Dirac Hamiltonian for a $d$D $(d+1)$th-order TI given in Eq.~\eqref{eq: Dirac(d,d)},
\ba
\mathcal{H}_{(d,d+1)} = -\sum_{i=1}^d i \gamma_i \dr_i + \sum_{a=1}^{d+1} m_{a,\bb{r}} M_a,
\ea
where $m_{d+1,\bb{r}}=\lambda(\bb{r})$ and $M_{d+1}=\mathcal{M}$.
This can be interpreted as the Hamiltonian describing a massless Dirac fermion coupled to $(d+1)$ scalar fields $M_a$.
Then, the induced electric current $J_{d}^\mu$ is given by 
\ba
\label{eq: top_current}
J_{d}^\mu =& \frac{s_d}{d! \, {\rm Area}(S^d)} \epsilon^{\mu \mu_1 \dots \mu_d} \epsilon^{a_1 \dots a_{d+1}} \nn \\
& \times \Phi_{a_1} \dr_{\mu_1} \Phi_{a_2} \dr_{\mu_2} \dots \dr_{\mu_d} \Phi_{a_{d+1}},
\ea
where $\mu,\mu_i=0,1,\dots,d$, $\dr_0=\dr_t$, $\Phi_a=\frac{m_{a,\bb{r}}}{m_{\bb{r}}}$, $m_{\bb{r}}^2=\sum_{a=1}^{d+1} m_{a,\bb{r}}^2$, ${\rm Area}(S^d)=\frac{2 \pi^{(d+1)/2}}{\gamma(\frac{d+1}{2})}$, and $\gamma(d)$ is the gamma function~\cite{abanov2000theta}.
Here, the sign of $s_d=\pm 1$ depends on the representation of gamma matrices where $\text{Tr}[\gamma_1,\dots,\gamma_{2d+1}]=2^d i^d$.
In our convention for gamma matrices, $s_d=-(-1)^{\frac{d(d-1)}{2}}$.
The derivation of Eq.~\eqref{eq: top_current} is given in Appendix~\ref{app: Current}.
Let us note that $\dr_\mu J_{d}^\mu=0$. Moreover, since $\Phi_a \rightarrow \{0,\dots,0,-1\}$ as $\bb{r}\rightarrow \infty$, $R^d$ can be compactified to $S^d$.
Then, one can regard $\int_{R^d} J_d$ as a winding number on $S^d$.

To compute the induced current explicitly, let us consider a finite-size $d$D $(d+1)$th-order TI with the boundary between the topological phase and the vacuum (or a trivial insulator).
Then, the $d$ boundary mass terms wind along the boundary while the bulk mass term $\lambda(\bb{r})\mathcal{M}$ varies rapidly across the boundary.
According to the expression of $J^\mu_d$ in Eq.~\eqref{eq: top_current}, charges are accumulated in the region where the mass terms vary rapidly, that is, only near the boundary.
Note that $J^\mu_1$ is also used in Sec.~\ref{subsec: SSH(1,2)} to explain the $1$D second-order TI [see Eq.~\eqref{eq: GWformula}].
Since $J^\mu_d$ is non-zero only near the boundary, the current density at the boundary $j^\mu_{d-1}$ can be determined by integrating $J^\mu_d$ along $r$ taking into account $\dr_r m_{a,\bb{r}} \ll \dr_r m_{d+1,\bb{r}}$ for $a=1,\dots,d$ as 
\ba
\label{eq: boundarycurrent}
j^\mu_{d-1}
=& \int_{0}^{\infty} dr \, J^\mu_{d} \nn \\
\simeq& (-1)^d \frac{s_d}{(d-1)! \, {\rm Area}(S^{d-1})} \epsilon^{\mu \mu_1 \dots \mu_{d-1}} \epsilon^{a_1 \dots a_d} \nn \\
& \times \varphi_{a_1} \dr_{\mu_1} \varphi_{a_2} \dr_{\mu_2} \dots \dr_{\mu_{d-1}} \varphi_{a_d},
\ea
where $\varphi_a=\frac{m_{a,\bb{r}}}{|\vec{m}_{\bb{r}}|}$ and $\vec{m}_{\bb{r}}=(m_{1,\bb{r}},\dots,m_{d,\bb{r}})$.
One can also derive Eq.~\eqref{eq: boundarycurrent} rigorously in the following way.
When the domain of the TI is sufficiently large, the boundary theory of $d$D $(d+1)$th-order TI given in Eq.~\eqref{eq: DiracBoundary(d,d)}, can be regarded as the Dirac Hamiltonian defined in $R^{d-1}$ with $d$ mass terms.
Then, the induced current $J^\mu_{d-1}$ is obtained by the boundary projection, and its sign is fixed by comparison with Eq.~\eqref{eq: boundarycurrent}. 
We get $j_{d-1}^\mu$ at the $(d-1)$D boundary of $d$D $(d+1)$th-order TI after compactifying $R^{d-1}$ to $S^{d-1}$ (see Appendix~\ref{app: Current}).

Equation~\eqref{eq: boundarycurrent} shows that the total charge of boundary states $\int_{S^{d-1}} j^0_{d-1}$, which is nothing but the anomalous charge $\delta$ due to the filling anomaly, is determined by the winding of boundary mass terms as
\ba
\delta = \int_{S^{d-1}} j^0_{d-1} =-(-1)^{\frac{d(d+1)}{2}} w_{d-1}.
\ea
We note that the condition $\varphi_{a}(\bb{r})=-\varphi_{a}(-\bb{r})$ imposed by inversion symmetry restricts the winding to be an odd integer.
The anomalous charge $\delta$ modulo two characterizes the bulk-boundary correspondence of $d$D $(d+1)$th-order TIs since any inversion-preserving process can change the boundary charge only by even integers unless it closes the bulk gap.
The charge densities predicted by $j^0_{d-1}$ are as follows: 
\ba
\label{eq: current1d}
j_{0}^0
&=\frac{1}{2},
\ea
\ba
\label{eq: current2d}
j_{1}^0
&=\frac{1}{2\pi} \epsilon^{0 \mu} \epsilon^{a b} \varphi_a \dr_{\mu} \varphi_b 
=\frac{1}{2\pi} \dr_{\theta} \tan^{-1}\Big(\frac{\varphi_2}{\varphi_1}\Big),
\ea
\ba
\label{eq: current3d}
j_{2}^0
&=-\frac{1}{8\pi} \epsilon^{0 \mu \nu} \epsilon^{a b c} \varphi_a \dr_{\mu} \varphi_b \dr_{\nu} \varphi_c \nn \\
&=-\frac{1}{4\pi} \bb{\varphi} \cdot \dr_{\theta} \bb{\varphi} \times \dr_{\phi} \bb{\varphi}.
\ea

It is worth noting that the induced currents shown above were also discussed in Refs.~\onlinecite{goldstone1981fractional, chamon2008electron, chamon2008irrational, ryu2009masses, yao2010topological, fukui2019dirac} in various contexts.
In all the previous studies, however, all the mass terms are considered as either bulk order parameters or bulk pumping parameters in time direction.
In contrast to this, in our case, all $\varphi(\bb{r})$ are defined along the boundary. Thus, the physical implication of the induced currents in our work is different from that in the previous studies.
Before closing this section, let us remark on the cases when the boundary projection is not well defined or the boundary is not smooth. In general, the boundary projection is not well-defined when the boundary mass terms are comparable to or larger than the bulk mass term. Nevertheless, as long as the total amount of boundary charge $\delta$ (modulo two) is preserved under any bulk gap-preserving and inversion-symmetric perturbation, our discussion on the bulk-boundary correspondence is still valid.
When the boundary is not smooth, it can be treated as an extreme curvature limit of the system with smooth boundary. In Sec.~\ref{subsec: Model}, we numerically show that our predictions of boundary charge are still valid even when the boundary mass terms are comparable to the bulk mass term and/or the boundary is not smooth. In addition to the total amount of the boundary charge, the distribution of the boundary charge density is also consistent with that of the winding number density.

\section{Symmetry data and classification of inversion-symmetric insulators \label{sec: Classification}}

In this section, by using symmetry data, we systematically classify inversion-symmetric insulators into three classes: atomic insulators, fragile topological insulators, and stable topological insulators. In particular, we provide a convenient scheme to diagnose fragile topology, and estimate the number of parity configurations belonging to each class.
Also, we discuss the net parity of a finite-size system with open boundary and the condition for the presence of the filling anomaly.

In a $d$D inversion-symmetric system, there are $2^d$ distinct Wyckoff positions $W_i$ and inversion-invariant momenta $\bb{k}^{{\rm inv}}_i$ which are left invariant under inversion symmetry:
\ba
\Big\{ W_{i=1,\dots,2^d} \Big\} =\Big\{ \sum_{l=1}^d x_l \bb{a}_l | \forall x_l=0,\frac{1}{2} \Big\}, \\
\Big\{ \bb{k}^{\rm inv}_{i=1,\dots,2^d} \Big\} = \Big\{ \sum_{l=1}^d y_l \bb{G}_l | \forall y_l=0,\pi \Big\},
\ea
where $\bb{a}$ and $\bb{G}$ are the lattice and reciprocal vectors, respectively, which satisfy $\bb{a}_i \cdot \bb{G}_j = 2\pi \delta_{ij}$.
We label $W_i$ and $\bb{k}^{{\rm inv}}_i$ as follows.
\ba
W_{i=1+\sum_{j=1}^d 2^{j-1} \sigma_j} &= \frac{1}{2} (\sigma_1, \sigma_2, \dots \sigma_d) \\
\bb{k}^{\rm inv}_{i=1+\sum_{j=1}^d 2^{j-1} \sigma_j} &= \pi (\sigma_1, \sigma_2, \dots \sigma_d)
\ea
where $\sigma=0$, $1$.
More explicitly,
\ba
& W_1=(0,0,\dots,0), \, W_2=(\frac{1}{2},0,\dots,0), \nn \\
& W_3=(0,\frac{1}{2},\dots,0), \, W_4=(\frac{1}{2},\frac{1}{2},\dots,0), \nn \\
& \dots, W_{2^d}=(\frac{1}{2},\frac{1}{2},\dots,\frac{1}{2}),
\ea
and similarly,
\ba
& \bb{k}^{\rm inv}_1=(0,0,\dots,0), \, \bb{k}^{\rm inv}_2=(\pi,0,\dots,0), \nn \\
& \bb{k}^{\rm inv}_3=(0,\pi,\dots,0), \, \bb{k}^{\rm inv}_4=(\pi,\pi,\dots,0), \nn \\
& \dots, \bb{k}^{\rm inv}_{2^d}=(\pi,\pi,\dots,\pi).
\ea

\subsection{Mapping between Wannier states and parity configurations \label{subsec: WannierMapping}}
Any given parity configuration $\mathcal{B}$ can be mapped to linear combinations of inversion-symmetric Wannier states $w_\xi(W)$.
Here, $w_\xi(W)$ denotes a Wannier state with parity $\xi$ localized at the Wyckoff position $W$.
At the momentum $\bb{k}^{{\rm inv}}$ in the Brillouin zone, a state with parity $\xi \exp (2 i \bb{k}^{{\rm inv}} \cdot W)$ can be induced by $w_{\xi}(W)$. 
For example, $s$ and $p$ orbitals localized at $W_1$ correspond to $w_{+}(W_1)$ and $w_{-}(W_1)$, respectively.
Let us note that it is sufficient to consider symmetric Wyckoff positions $W_i$ only.
When a orbital is located at generic point $W \ne -W$, its partner must also be located at $-W$ because of inversion symmetry.
Then, the symmetric and antisymmetric combinations of the two states give $s$ and $p$ orbitals at a certain symmetric Wyckoff position, say $W_i$. It is worth noting that this $s$ and $p$ orbital pair located at $W_i$ can be deformed adiabatically to another Wyckoff position, say $W_j$, without breaking inversion symmetry.
Such an adiabatic deformation of Wannier states can be referred as ``Wannier gauge redundancy.''

For a given Wannier state representation $\bigoplus_{W,\xi} \, \mu_{W,\xi} w_{\xi}(W)$, there is a corresponding parity configuration $\mathcal{B}$ in the Brillouin zone. The equivalence relation between them~\cite{song2017d, fang2017rotation, cano2018topology, bradlyn2019disconnected, po2019faithful, benalcazar2019quantization, wieder2018axion, ahn2019failure, song2019fragile} is denoted as
\ba
\label{eq: EquivRel}
\mathcal{B} \Leftrightarrow \bigoplus_{W, \xi} \, \mu_{W, \xi} w_{\xi}(W).
\ea
Here, $\mu_{W,\xi}$ is a rational number~\cite{po2017symmetry}, i.e., $\mu_{W,\xi} \in \mathbb{Q}$, and dictates how many $w_{\xi}(W)$ are superposed in a given atomic configuration. 
For later convenience, we denote a parity configuration of $N_{\mathcal{B}}$ bands in $d$D that satisfies $n_-(\gamma=\bb{k}^{{\rm inv}}_1)=n_+(\bb{k}^{{\rm inv}} \ne \gamma)=N_{\mathcal{B}}$ as $\mathcal{B}_{\gamma}(d,N_{\mathcal{B}})$. Namely, all occupied bands have odd parity at the $\gamma$ point whereas they have even parity at the other $\bb{k}^{{\rm inv}}$ in $\mathcal{B}_{\gamma}(d,N_{\mathcal{B}})$.

The equivalence relation Eq.~\eqref{eq: EquivRel} is not one to one since there is an ambiguity in choosing Wannier states, that is, the Wannier gauge redundancy:
\ba
\label{eq: WGaugeRedun}
\bigoplus_{\xi=\pm} w_{\xi}(W) = \bigoplus_{\xi=\pm} w_{\xi}(W').
\ea
This follows from the fact that $w_{+1}(W) \oplus w_{-1}(W)$ gives $n_{+}(\bb{k}^{{\rm inv}})=n_{-}(\bb{k}^{{\rm inv}})=1$ for any $W$ and $\bb{k}^{{\rm inv}}$.

According to the recent studies~\cite{po2017symmetry, bradlyn2017topological, po2018fragile}, a parity configuration $\mathcal{B}$ can be classified into three distinct classes depending on the type of $\mu_{W, \xi} \in \mathbb{Q}$.
(i) $\mathcal{B}$ is diagnosed to be a stable topological phase when the set of $\mu_{W, \xi}$ includes a fraction. Adding trivial bands corresponds to $\Delta \mu_{W, \xi} \in \mathbb{Z}$, and thus it does not change the fractional part of $\mu_{W}$.
(ii) $\mathcal{B}$ is diagnosed to be a fragile topological phase when the set of $\mu_{W, \xi}$ includes at least one negative integer in any fixed Wannier gauge.
(iii) $\mathcal{B}$ is diagnosed to be an (obstructed) atomic phase when all $\mu_{W, \xi}$ are integers greater than or equal to $0$ in at least one fixed Wannier gauge.

For example, 
(i) $\mathcal{B}_{\gamma}(2, 1)$ is equivalent to $\ominus \frac{1}{2} p(W_1) \oplus \frac{1}{2} p(W_2) \oplus \frac{1}{2} p(W_3) \oplus \frac{1}{2} p(W_4)$.
Because of the fractional coefficients of $p$ orbitals, the phase is diagnosed to be a stable topological phase. In fact, according to the recent symmetry indicator analysis, this phase is characterized by an odd Chern number~\cite{fang2012bulk, po2017symmetry}, which indicates its stable band topology.
(ii) $\mathcal{B}_{\gamma}(2, 2)$ is equivalent to $\ominus p(W_1) \oplus p(W_2) \oplus p(W_3) \oplus p(W_4)$.
The negative coefficient of $p(W_1)$ cannot be removed in any Wannier gauge, and thus the phase is diagnosed to be a fragile topological phase.
(iii) $\mathcal{B} \Leftrightarrow s(W_1) \ominus s(W_2) \oplus 2 p(W_1)$ has a negative coefficient of $s(W_2)$, and seems to be a fragile topological phase.
However, $\mathcal{B} \Leftrightarrow p(W_1) \oplus p(W_2)$ up to the Wannier gauge redundancy $s(W_1) \oplus p(W_1)=s(W_2) \oplus p(W_2)$, thus $\mathcal{B}$ is diagnosed to be an atomic phase.

Now, let us compare the real-space data $\{w_{\xi}(W)\}$ and the momentum-space data $\{n_{\xi}(\bb{k}^{{\rm inv}})\}$.
There are $2^{d+1}$ possible types of $w_{\xi}(W)$, counting the number of Wyckoff positions with two different parities. But, only $(2^d+1)$ of them are independent due to the Wannier gauge redundancy shown in Eq.~\eqref{eq: WGaugeRedun}.
Namely, the Wannier gauge redundancy $s(W_1) \oplus p(W_1)=s(W_2) \oplus p(W_2)=\dots=s(W_{2^d}) \oplus p(W_{2^d})$ give $(2^d-1)$ constraint equations.
Taking into account this, one can write $(2^d+1)$ numbers invariant under the Wannier gauge redundancy as
\ba
\label{eq: inversionindices}
\{n_{\rm occ}, \nu_{W_1}, \nu_{W_2}, \dots, \nu_{W_{2^d}}\},
\ea
where $\nu_W=\mu_{W,-}-\mu_{W,+}$.
Note that $\mu_{W,-}$ and $\mu_{W,+}$ change by the same amount under the Wannier gauge redundancy, thus, $\nu_W$ can serve as a topological invariant.
This approach was introduced in Ref.~\onlinecite{van2018higher} for rotation-symmetric ($C_{n=2,3,4,6}$) insulators and $\nu_W$ indices were found explicitly. 
However, the possibility to interpret $\nu_W$ indices as the indicators for fragile topological phases was not appreciated.
We show that inversion-symmetric insulators can be completely classified in terms of the indices in Eq.~\eqref{eq: inversionindices}, which can also be used for the diagnosis of fragile topological phases.

%%%%%%%%%%%%%%%%%%%%%%%%%%%%%%%%%%%%%%%%%%%%%%%%%%%%%%%%%%%
\begin{figure*}[ht]
	\centering
	\includegraphics[width=0.98\textwidth]{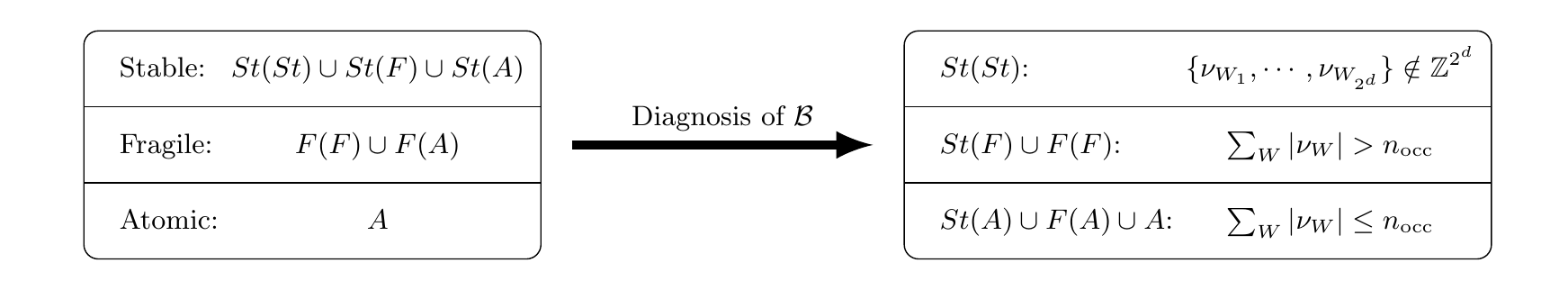}
	\caption{Diagnosis of band topology. Left: all sets of possible band topology are divided into three sets, $[\Omega_{i=1,2,3}]=\cup_{j \le i} \Omega_i(\Omega_j)$ where $\Omega_1$, $\Omega_{2}$, and $\Omega_{3}$ denote the atomic, fragile topological, and stable topological phases, respectively. Notation $\Omega_i(\Omega_j)$ means that the phase is diagnosed to be $\Omega_j$ only with the symmetry data while the phase is actually $\Omega_i$. Right: these phases are rearranged into three sets, $\cup_{j \ge i} \Omega_i(\Omega_j)$, and these are diagnosed in terms of $\nu_W$ indices.}
	\label{fig: Classification}
\end{figure*}
%%%%%%%%%%%%%%%%%%%%%%%%%%%%%%%%%%%%%%%%%%%%%%%%%%%%%%%%%%

On the other hand, the parity configuration $\mathcal{B}$ is characterized by $2^{d+1}$ integers,
\ba
\{n_{+}(\bb{k}^{{\rm inv}}_1),n_{-}(\bb{k}^{{\rm inv}}_1),\dots,n_{+}(\bb{k}^{{\rm inv}}_{2^d}),n_{-}(\bb{k}^{{\rm inv}}_{2^d})\}.
\ea
In an insulating phase, the number of occupied states $n_{\rm occ}$ is fixed to be $n_{\rm occ} = n_{+}(\bb{k}^{{\rm inv}})+n_{-}(\bb{k}^{{\rm inv}})$ for all $\bb{k}^{{\rm inv}}$.
This is called the compatibility relation~\cite{kruthoff2017topological, po2017symmetry, bradlyn2017topological}.
Thus, $\mathcal{B}$ can be characterized by $(2^d+1)$ independent integers.
One canonical choice of these $(2^d+1)$ integers is
\ba
\{n_{\rm occ},n_{-}(\bb{k}^{{\rm inv}}_1)-n_{+}(\bb{k}^{{\rm inv}}_1),\dots,n_{-}(\bb{k}^{{\rm inv}}_{2^d})-n_{+}(\bb{k}^{{\rm inv}}_{2^d})\},
\ea
including $n_{-}(\bb{k}^{{\rm inv}}_i)-n_{+}(\bb{k}^{{\rm inv}}_i)$ with $i=1,\dots,2^d$.
Observing that $w_{\xi}(W)$ has the parity $\xi \exp (2 i \bb{k}^{{\rm inv}} \cdot W)$ at $\bb{k}^{{\rm inv}}$, one can find a mapping between $\{\nu_W\}$ and $\{n_{-}(\bb{k}^{{\rm inv}})-n_{+}(\bb{k}^{{\rm inv}})\}$ as
\ba
\label{eq: Indices}
\begin{pmatrix}
	\nu_{W_1} \\
	\nu_{W_2} \\
	\nu_{W_3} \\
	\vdots \\
	\nu_{W_{2^d}}
\end{pmatrix}
=A_d
\begin{pmatrix}
	n_{-}(\bb{k}^{{\rm inv}}_1) - n_{+}(\bb{k}^{{\rm inv}}_1) \\
	n_{-}(\bb{k}^{{\rm inv}}_2) - n_{+}(\bb{k}^{{\rm inv}}_2) \\
	n_{-}(\bb{k}^{{\rm inv}}_3) - n_{+}(\bb{k}^{{\rm inv}}_3) \\
	\vdots \\
	n_{-}(\bb{k}^{{\rm inv}}_{2^d}) - n_{+}(\bb{k}^{{\rm inv}}_{2^d})
\end{pmatrix},
\ea
where the $2^d \times 2^d$ matrix $A_d$ maps the real data $\{\nu_W\}$ and the momentum-space data $\{n_{-}(\bb{k}^{{\rm inv}})-n_{+}(\bb{k}^{{\rm inv}})\}$, and its components are given by
\ba
\label{eq: AProp}
\big[ A_d \big]_{ij}= \frac{1}{2^d} \exp(2 i \bb{k}^{{\rm inv}}_i \cdot W_j).
\ea
Also, $A_d$ satisfies the following relations:
\ba
A_d^{-1} = 2^d A_d, \quad A_d^{T} = A_d, \quad \sum_{i=1}^{2^d} \big[ A_d \big]_{ij} = \delta_{j,1}.
\ea
For example, $A_{d=2}$ is given by
\ba
A_2=
\frac{1}{4}
\begin{pmatrix}
	+1 & +1 & +1 & +1 \\
	+1 & -1 & +1 & -1 \\
	+1 & +1 & -1 & -1 \\
	+1 & -1 & -1 & +1
\end{pmatrix},
\ea
and the expressions for $\nu_{W_{i=1,2,3,4}}$ are the same as those in Ref.~\onlinecite{van2018higher}.
From the fact that $n_{\pm}(\bb{k}^{{\rm inv}})$ must be a non-negative integer, the possible sets of $\{\nu_{W}\}$ are restricted.
The most fundamental restriction is that $2^d \big| \sum_{j=1}^{2^d} [A_d]_{ij} \nu_{W_j} \big|$ must be an integer less than or equal to $n_{\rm occ}$.
Below, we list some useful restrictions and inequalities derived from this:
\begin{gather}
|\nu_{W_i}| \le n_{\rm occ}, \, |\nu_{W_i} \pm \nu_{W_{j \ne i}}| \le n_{\rm occ}, \, \big|\sum_{W} \nu_{W} \big| \le n_{\rm occ} \nn \\
2^{d-1} \nu_{W_{i}} \in \mathbb{Z}, \quad 2^{d-2} (\nu_{W_{i}} \pm \nu_{W_{j}}) \in \mathbb{Z}.
\label{eq: nuRestr}
\end{gather}
For example,
\ba
\label{eq: df}
\{\nu_{W_1},\dots,\nu_{W_4}\}=\{\frac{1}{2},0,-\frac{1}{2},0\} \text{ or } \{-1,0,1,0\}
\ea
cannot be realized when $n_{\rm occ}=1$ in 2D.
The first set $\{\frac{1}{2},0,-\frac{1}{2},0\}$ is forbidden due to $2^{d-2} (\nu_{W_{i}} \pm \nu_{W_{j \ne i}}) \in \mathbb{Z}$ while the second set $\{-1,0,1,0\}$ is forbidden because of $|\nu_{W_i} \pm \nu_{W_j}| \le n_{\rm occ}$.
The properties of $A_d$ and $\nu_W$, and the derivation of the equations from Eq.~\eqref{eq: Indices} to Eq.~\eqref{eq: nuRestr} are shown in Appendix~\ref{app: DerivProp}.

\subsection{Diagnosis of fragile topology \label{subsec: Diagnosis}}
Now, we provide an efficient scheme for diagnosing the band topology in terms of $\{n_{\rm occ}, \nu_{W_1}, \dots, \nu_{W_{2^d}}\}$.
Any given inversion-symmetric phase is an element of $[\Omega]$ where $\Omega=A,~F,~St$.
Here, $[A]$, $[F]$, and $[St]$ denote a set of atomic, fragile topological, stable topological phases, respectively.
Each set can be partitioned into mutually exclusive subsets: $[St]=St(St) \cup St(F) \cup St(A)$, $[F]=F(F) \cup F(A)$, and $[A]=A(A)$.
The notation $\Omega_i(\Omega_j)$ means that the given inversion-symmetric phase belongs to $\Omega_i$, but its parity configuration $\mathcal{B}$ predicts it to be in the set $\Omega_j$. We denote $A(A)=A$ for simplicity.

The diagnosis of stable topological phases has been well-established in $2$D and $3$D in terms of symmetry indicators~\cite{po2017symmetry,watanabe2018structure}. However, an efficient method to diagnose fragile phases has not been established. Here, we propose one efficient scheme.
The procedure for diagnosing inversion-symmetric insulators proceeds as follows.
(i) $St(St)$ is diagnosed by a criterion $\{\nu_{W_1},\dots,\nu_{W_{2^d}}\} \notin \mathbb{Z}^{2^d}$ where $\mathbb{Z}^{2^d}$ indicates the set of $2^d$ integers.
(ii) A fragile phase can be diagnosed by a criterion $\sum_{W} |\nu_W| > n_{\rm occ}$, which can be obtained by the fact that an atomic phase must satisfy $\sum_{W} |\nu_W| \le n_{\rm occ}$.
Thus, a phase which satisfies $\sum_{W} |\nu_W| > n_{\rm occ}$ belongs to $St(F) \cup F(F)$.
(iii) A phase which satisfies $\sum_{W} |\nu_W| \le n_{\rm occ}$ cannot be distinguished from atomic phases, so it belongs to $St(A) \cup F(A) \cup A$.
This two-step diagnosis procedure is summarized in the following diagram, and also in Fig.~\ref{fig: Classification}.

%%%%%%%%%%%%%%%%%%%%%%%%%%%%%%%%%%%%%%%%%%%%%%%%%%%%%%%%%%%%%%%%
\begin{center}
	\fbox{
		\centering
		\parbox{0.34\textwidth}
		{\centering
			$\{\nu_{W_1},\dots,\nu_{W_{2^d}}\} \notin \mathbb{Z}^{2^d}$ $\Rightarrow$ $St(St)$}
	}
	\\
	$\Downarrow$ \\
	\fbox{
		\centering
		\parbox{0.34\textwidth}
		{\centering
			$\sum_{W} |\nu_W| > n_{\rm occ}$ $\Rightarrow$ $St(F) \cup F(F)$}
	}
	\\
	$\Downarrow$ \\
	\fbox{
		\centering
		\parbox{0.34\textwidth}
		{\centering
			$\sum_{W} |\nu_W| \le n_{\rm occ}$ $\Rightarrow$ $St(A) \cup F(A) \cup A$}
	}.
\end{center}
%%%%%%%%%%%%%%%%%%%%%%%%%%%%%%%%%%%%%%%%%%%%%%%%%%%%%%%%%%%%%%%%

Let us emphasize that our diagnosis procedure goes beyond the diagnosis method proposed previously for stable topological phases~\cite{po2017symmetry, song2018quantitative} in that our approach can diagnose even fragile topological phases.
The criterion for fragile topological and atomic phases is derived in Appendix~\ref{app: DerivDiag}.
It is worth noting that our classification of inversion-protected fragile TIs is complete. The band topology of class A insulators is fully characterized by inversion parities at high-symmetry points and all the Chern numbers defined in the Brillouin zone~\cite{kruthoff2017topological, po2017symmetry, shiozaki2018atiyah}. Since the Chern numbers are stable topological invariants, fragile TIs and atomic insulators can be fully characterized by using only inversion-parities. This implies that $F(A)$ is an empty set and $[F]=F(F)$.

%%%%%%%%%%%%%%%%%%%%%%%%%%%%%%%%%%%%%%%%%%%%%%%%%%%%%%%%%%%%%%%%
\begin{table}[b!]
	\caption{Results of the parity configuration diagnosis in 2D. When $n_{\rm occ}$ is sufficiently large, $N(A)$, $N(F)$, and $N(St)$ approach to $\frac{1}{3} N_{\rm tot}$, $\frac{1}{6} N_{\rm tot}$, and $\frac{1}{2} N_{\rm tot}$. At least, two-third of parity configurations are topologically nontrivial.
	}
	\begin{tabular}{c|c|c|c|c}
		\hline
		\hline
		$n_{\rm occ}$ & $N_{\rm tot}$ & $N(A)$ & $N(F)$ & $N(St)$ \\ \hline
		$1$ & $16$ & $8$ & $0$ & $8$ \\ \hline
		$2$ & $81$ & $33$ & $8$ & $40$ \\ \hline
		$3$ & $256$ & $96$ & $32$ & $128$ \\ \hline
		$4$ & $625$ & $225$ & $88$ & $312$ \\ \hline
		$5$ & $1296$ & $456$ & $192$ & $648$ \\ \hline
		$6$ & $2401$ & $833$ & $368$ & $1200$ \\ \hline
		$7$ & $4096$ & $1408$ & $640$ & $2048$ \\ \hline
		$8$ & $6561$ & $2241$ & $1040$ & $3280$ \\ \hline
		$9$ & $10000$ & $3400$ & $1600$ & $5000$ \\ \hline
		$10$ & $14641$ & $4961$ & $2360$ & $7320$ \\ \hline
		$n_{\rm occ} \rightarrow \infty$ & $(n_{\rm occ}+1)^4$ & $\frac{1}{3} N_{\rm tot}$ & $\frac{1}{6} N_{\rm tot}$ & $\frac{1}{2} N_{\rm tot}$ \\ \hline
		\hline
	\end{tabular}
	\label{table: Distribution2d}
\end{table}
%%%%%%%%%%%%%%%%%%%%%%%%%%%%%%%%%%%%%%%%%%%%%%%%%%%%%%%%%%%%%%%%

%%%%%%%%%%%%%%%%%%%%%%%%%%%%%%%%%%%%%%%%%%%%%%%%%%%%%%%%%%%%%%%%
\begin{table*}[t!]
	\caption{Results of the parity configuration diagnosis in 3D. When $n_{\rm occ}$ is sufficiently large, $N(A)$ is $\frac{1}{315} N_{\rm tot}$. As for $N(F)$ and $N(St)$, although their exact numbers in the $n_{\rm occ}\rightarrow\infty$ limit are not known, considering their $n_{\rm occ}$ dependence, we conjecture that $N(F)=\frac{283}{10080} N_{\rm tot}$ and $N(St)=\frac{31}{32} N_{\rm tot}$ in the large $n_{\rm occ}$ limit. In contrast to $2$D, the majority of parity configurations are stable topological phases. Here, $z_4=-4 \nu_{W_1}$ denotes the $\mathbb{Z}_4$ symmetry indicator \cite{po2017symmetry,khalaf2018symmetry, song2018quantitative} for inversion symmetry. The relation between symmetry indicators and $\nu_W$ indices is discussed in Appendix~\ref{app: RelationtoSI}. Let us also note that stable topological phases are Weyl semimetal when $z_4$ is odd~\cite{turner2012quantized, hughes2011inversion}.}
	\begin{tabular}{c|c|c|c|c|c|c|c}
		\hline
		\hline
		\multirow{2}{*}{$n_{\rm occ}$} & \multirow{2}{*}{$N_{\rm tot}$} & \multirow{2}{*}{$N(A)$} & \multirow{2}{*}{$N(F)$} & \multicolumn{4}{c}{$N(St)$} \\
		\cline{5-8} & & & & $z_4=0$ & $z_4=1$ & $z_4=2$ & $z_4=3$ \\ \hline
		$1$ & $256$ & $16$ & $0$ & $56$ & $64$ & $56$ & $64$ \\
		$2$ & $6561$ & $129$ & $112$ & $1400$ & $1640$ & $1640$ & $1640$ \\
		$3$ & $65536$ & $704$ & $1344$ & $14336$ & $16384$ & $16384$ & $16384$ \\
		$4$ & $390625$ & $2945$ & $9536$ & $85176$ & $97656$ & $97656$ & $97656$ \\
		$5$ & $1679616$ & $10128$ & $42368$ & $367416$ & $419904$ & $419896$ & $419904$ \\
		$6$ & $5764801$ & $29953$ & $151248$ & $1260000$ & $1441200$ & $1441200$ & $1441200$ \\
		$7$ & $16777216$ & $78592$ & $445696$ & $3670016$ & $4194304$ & $4194304$ & $4194304$ \\
		$8$ & $43046721$ & $187137$ & $1160944$ & $9413600$ & $10761680$ & $10761680$ & $10761680$ \\
		$9$ & $100000000$ & $411280$ & $2713728$ & $21875000$ & $25000000$ & $24999992$ & $25000000$ \\
		$10$ & $214358881$ & $845185$ & $5859936$ & $46884600$ & $53589720$ & $53589720$ & $53589720$ \\ \hline
		\hline
	\end{tabular}
	\label{table: Distribution3d}
\end{table*}
%%%%%%%%%%%%%%%%%%%%%%%%%%%%%%%%%%%%%%%%%%%%%%%%%%%%%%%%%%%%%%%%

\subsection{Results of diagnosis \label{subsec: Result2d3d}}
We tabulate how many parity configurations are diagnosed to be atomic, fragile topological, and stable topological phases.
We first fix the number of occupied bands $n_{\rm occ}$.
Then, we apply the diagnosis procedure to all possible parity configurations, and count $N(A)$, $N(F)$, and $N(St)$, which are the number of parity configurations diagnosed to be atomic, fragile topological, and stable topological phases, respectively.
The sum of $N(A)$, $N(F)$, and $N(St)$ must be $N_{\rm tot}$.
Here, $N_{\rm tot}$ is the number of all possible parity configurations and is given by $(n_{\rm occ}+1)^{2^d}$.
Let us note that when the ordering of energy bands is neglected, there can be $0, 1,\dots, n_{\rm occ}$ states with odd parity at each inversion-invariant momentum, so there are $(n_{\rm occ}+1)^{2^d}$ possible parity configurations.

In 2D, as shown in Table~\ref{table: Distribution2d}, $N(A)$, $N(F)$, and $N(St)$ are given by $\frac{1}{3} N_{\rm tot}$, $\frac{1}{6} N_{\rm tot}$, and $\frac{1}{2} N_{\rm tot}$, respectively, when $n_{\rm occ}$ is sufficiently large. Namely, at least, one-sixth (one-half) of parity configurations are predicted to be fragile (stable) topological phases.
On the other hand, in 3D, most of parity configurations are diagnosed to be stable topological phases as shown in Table~\ref{table: Distribution3d}.
When $n_{\rm occ}$ is sufficiently large, we find that $N(A)$ is given by $\frac{1}{315} N_{\rm tot}$ but it is difficult to determine the exact values of $N(F)$ and $N(St)$ in the $n_{\rm occ}\rightarrow\infty$ limit. However, from the values of $N(A)$, $N(F)$, and $N(St)$ for various $n_{\rm occ}$, we conjecture that $N(F)=\frac{283}{10080} N_{\rm tot}$ and $N(St)=\frac{31}{32} N_{\rm tot}$, which implies that only about 3$\%$ of parity configurations are diagnosed to be fragile topological phases.

The implications of Tables~\ref{table: Distribution2d} and~\ref{table: Distribution3d} are as follows.
First, there is no fragile TI with $n_{\rm occ}=1$ in any dimensions~\cite{alexandradinata2018no}.
This follows from the second inequality in Eq.~\eqref{eq: nuRestr}, $|\nu_{W_i} \pm \nu_{W_j}| \le n_{\rm occ}$.
Second, a layer-stacking of $2$D fragile TIs with negligible interlayer coupling gives rise to fragile TIs in higher dimensions.
For instance, in $2$D, there are eight fragile TIs with $n_{\rm occ}=2$.
The $\nu_W$ indices of these fragile TIs are given by $\{\mp 1,\pm 1,\pm 1,\pm 1\}$ and its permutations.
With these representative $\nu_W$ indices, the fragile TIs with $n_{\rm occ}=2$ in $3$D can be constructed in the following way.
Let us consider $\{\nu_W\}=\{-1,1,1,1,0,0,0,0\}$ as an example of indices for a $3$D insulator.
These indices correspond to a $3$D insulator obtained by a stacking of 2D layers with $\{\nu_W\}=\{-1,1,1,1\}$, which satisfy $\bb{d} \cdot (x,y,z) = 0$ (mod 1) where $\bb{d}=(0,0,1)$ and $(x,y,z)$ denotes the position on the $2$D layers.
Since there are $14$ independent $2$D layers that satisfy $\bb{d} \cdot (x,y,z) = 0$ or $\frac{1}{2}$ (mod $1$) where $\bb{d}=(1,0,0),(0,1,0),(0,0,1),(1,1,0),(1,0,1),(0,1,1),(1,1,1)$, we have $8 \times 14 = 112$ fragile TIs with $n_{\rm occ}=2$ in $3$D.
Also, we note that the $1344$ fragile TIs with $n_{\rm occ}=3$ in $3$D can be obtained by adding a trivial band to $112$ fragile TIs with $n_{\rm occ}=2$.
There are 16 ways to add a trivial band, which is one of $s(W_i)$ and $p(W_i)$ for $i=1,\dots,8$. However, four of them trivialize a 3D fragile TI with $n_{\rm occ}=2$.
Accordingly, we have $112 \times 12 =1344$ fragile TIs with $n_{\rm occ}=3$ in 3D.
From the layer-stacking picture, we conclude that the minimal number of occupied bands to have the fragile topology is two in any dimensions.
This process for obtaining a higher-dimensional fragile TI by layer stacking $2$D fragile TIs can be described by using a pumping process in momentum space as discussed in Sec.~\ref{sec: Pumping}.
Third, fragile topology can be trivialized under the doubling of the unit cell.
For instance, let us consider the doubling of the unit cell in all spatial directions.
In this case, all the symmetric Wyckoff positions except $W_{2^d}$ form inversion-symmetric pairs at generic positions, thus, the resulting $\nu_W$ is composed of $2^d$ $\nu_{W_{2^d}}$ of the original system, that is, $\nu_{W_i}=\nu_{W_{2^d}}$ for $i=1,\dots,2^d$.
If we repeat the same doubling process, the number of occupied bands increases while $\sum_W |\nu_W| = 2^d |\nu_{W_{2^d}}|$ is fixed.
Then, the result of the diagnosis for fragile TI, $\sum_W |\nu_W| > n_{\rm occ}$, can change to $\sum_W |\nu_W| \le n_{\rm occ}$.
Finally, when $n_{-}(\bb{k}^{\rm inv}_i)$ for $i=1,\dots 2^d$ are less than $2^{d-1}$, at least one $\nu_W$ index among $\nu_{W_{2,3,\dots,2^d}}$ is zero, i.e., $\prod_{i=2}^{2^d} \nu_{W_i}=0$.
We have verified that this is true for arbitrary number of occupied bands in $2$D and $3$D, and we expect that this is also true in any dimension $d>3$.

\subsection{Filling anomaly in general dimensions \label{subsec: FillingAnomaly}}
In this section, we propose a condition for the presence or absence of the filling anomaly, independent of the geometry of finite-size systems.
First, we observe that, for a given finite-size system, the number of the Wannier centers at the boundary modulo two, $N_{\dr}$, is invariant under an attaching or a detaching of unit cells to the boundary in an inversion-symmetric way.
This is depicted in Fig.~\ref{fig: GeneralFilling}(a).
Second, there are $2^d$ building blocks for the finite-size systems in $d$D.
Any finite-size inversion-symmetric insulators can be constructed by attaching unit cells at the boundary of these building blocks.
For example, in 2D, there are four building blocks composed of $1 \times 1$, $2 \times 1$, $1 \times 2$, and $2 \times 2$ unit cells [see Fig.~\ref{fig: GeneralFilling}(b)].
For the building block made of the $1 \times 1$ unit cell, $N_{\dr}$ is given by $\nu_{W_2}+\nu_{W_3}+\nu_{W_4}$ modulo $2$.
However, in the case of the building block composed of $1 \times 2$ unit cells, $N_{\dr}$ is given by $2 \nu_{W_2}+\nu_{W_3}+2 \nu_{W_4}$ modulo $2$.
Since the filling anomaly exists when $N_\dr$ is odd and $N_\dr$ depends on the building blocks, the presence or absence of the filling anomaly depends on the type of these building blocks for given $\nu_{W_{2,3,4}}$.
Investigating all the building blocks in $d$D, we conclude that $\nu_{W_i}$ for $i=2,3,\dots,2^d$ must be odd integers (even integers) at the same time to have (not to have) the filling anomaly independent of the type of the building blocks.
Physically, this condition is equivalent to the vanishing polarization, $\bb{P}=\sum_{W} W \nu_W=(0,0)$ in $2$D.
In the rest of this paper, we focus on the cases where the filling anomaly is well defined independent of the type of the building blocks.

%%%%%%%%%%%%%%%%%%%%%%%%%%%%%%%%%%%%%%%%%%%%%%%%%%%%%%%%%%%
\begin{figure}[b!]
	\centering
	\includegraphics[width=0.49\textwidth]{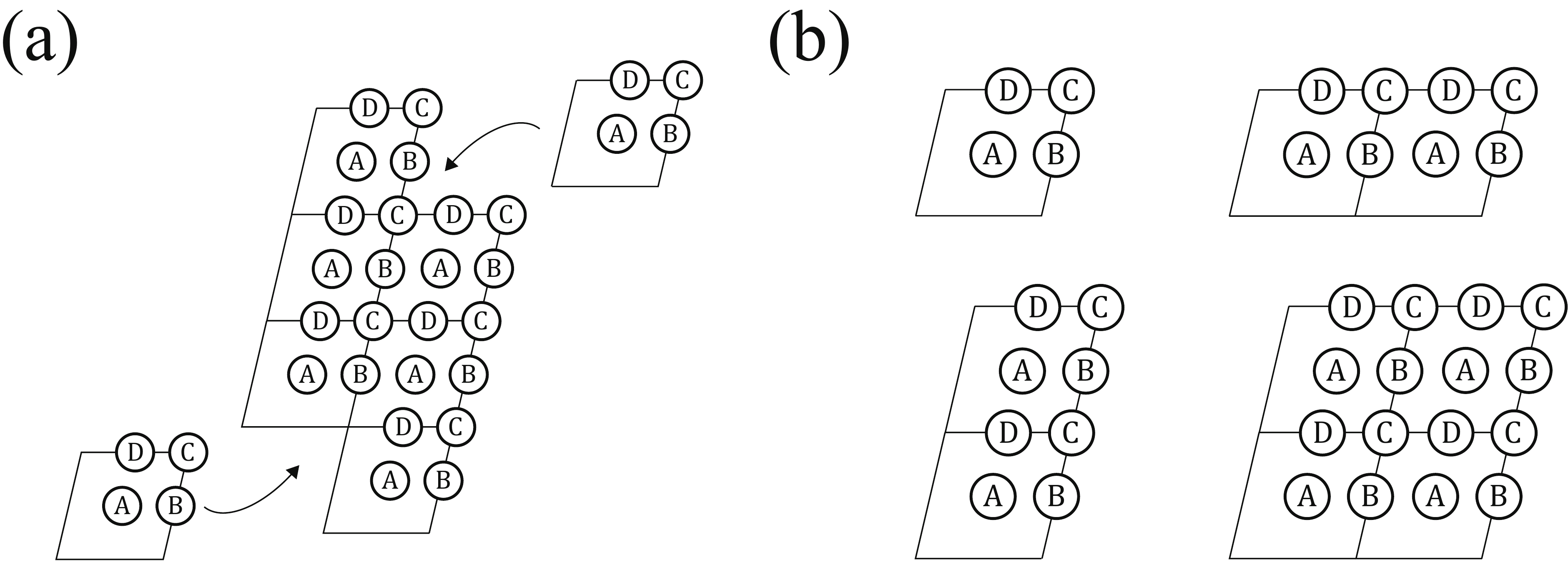}
	\caption{Filling anomaly and the building blocks of finite-size systems. Attaching or detaching the unit cells to the boundary of the finite-size inversion-symmetric insulators does not change the number of Wannier centers at the boundary modulo two, $N_{\dr}$. Four Wyckoff positions $W_{1,2,3,4}$ are denoted as $A$, $B$, $D$, and $C$, respectively. (a) A process changing $N_{\dr}$ by $\sum_{\xi} (\mu_{B,\xi}+\mu_{C,\xi}+\mu_{D,\xi})-\sum_{\xi}(\mu_{B,\xi}+3\mu_{C,\xi}+\mu_{D,\xi})=-2\sum_{\xi} \mu_{C,\xi}=0$ modulo $2$. Note that $\nu_W=\mu_{W,-}-\mu_{W,+}=\sum_{\xi} \mu_{W,\xi}$ modulo $2$.
	(b) For any finite-size system, the corresponding building block is determined uniquely. In $2$D, there are four building blocks. The building block corresponding to the geometry depicted in (a) is $2 \times 2$ unit cells. The presence (absence) of the filling anomaly is well defined only when $\nu_{B,C,D}$ are odd (even) integers simultaneously.
	}
	\label{fig: GeneralFilling}
\end{figure}
%%%%%%%%%%%%%%%%%%%%%%%%%%%%%%%%%%%%%%%%%%%%%%%%%%%%%%%%%%

The condition for the presence of the filling anomaly puts a constraint to the number of occupied bands.
As at least one $\nu_W$ index among $\nu_{W_{2,3,\dots,2^d}}$ is zero when $n_{-}(\bb{k}^{\rm inv}_i)$ for $i=1,\dots, 2^d$ are less than $2^{d-1}$, the filling anomaly occurs only when $n_{\rm occ} \ge 2^{d-1}$.
For atomic insulators, the number of occupied bands must be greater than or equal to $(2^d-1)$ in order to have all $\nu_{W_{2,3,\dots,2^d}}$ as odd integers.
Thus, there are $2^{2^d-1}$ minimal atomic insulators with the filling anomaly, and they have $\{\nu_W\}=\{0,\pm 1,\pm 1,\pm 1\}$ in $2$D and $\{0,\pm 1,\pm 1,\pm 1,\pm 1,\pm 1,\pm 1,\pm1\}$ in $3$D. Here, all $\pm$ are independent.
In contrast to atomic insulators, fragile TIs can have the filling anomaly when $n_{\rm occ} \ge 2^{d-1}$.
In other words, all insulators with the filling anomaly are fragile TIs when $2^{d-1} \le n_{\rm occ} \le 2^d-2$.

\subsection{Net parity in open boundary \label{subsec: NetParity}}
In this section, we discuss the net parity~\cite{trifunovic2019higher,song2017topological,huang2017building} of finite-size inversion-symmetric insulators without gapless boundary states, which generalizes the discussion in Sec.~\ref{subsec: SSH(1,4)}.
For any inversion-symmetric insulator in $d$D which does not host gapless boundary states, the energy spectra of occupied and unoccupied states are well defined.
In this case, the net parity of a finite-size system $\mathcal{I}=N_--N_+$ in Eq.~\eqref{eq: NetParity}, characterizes the bulk band topology~\cite{trifunovic2019higher,song2017topological,huang2017building}. Here, $N_\pm$ denotes the number of occupied states with parity $\pm$.
Now, we argue that $\mathcal{I}$ does not change if we add localized states at the boundary.
For instance, when an electron is added to the boundary, its inversion partner also has to be added to preserve inversion symmetry as shown in Fig.~\ref{fig: NetParity}(a).
Then, the two electrons form a symmetric and an antisymmetric states.
In this process, both $N_+$ and $N_-$ increase by one, so $\mathcal{I}$ is invariant.
The only way to change $\mathcal{I}$ is to add electrons at the inversion center.
However, this cannot be done in an inversion-symmetric way.
Therefore, $\mathcal{I}$ is the bulk topological invariant of inversion-symmetric systems that does not change as long as the bulk gap is preserved.

%%%%%%%%%%%%%%%%%%%%%%%%%%%%%%%%%%%%%%%%%%%%
\begin{figure}[t!]
	\centering
	\includegraphics[width=0.49\textwidth]{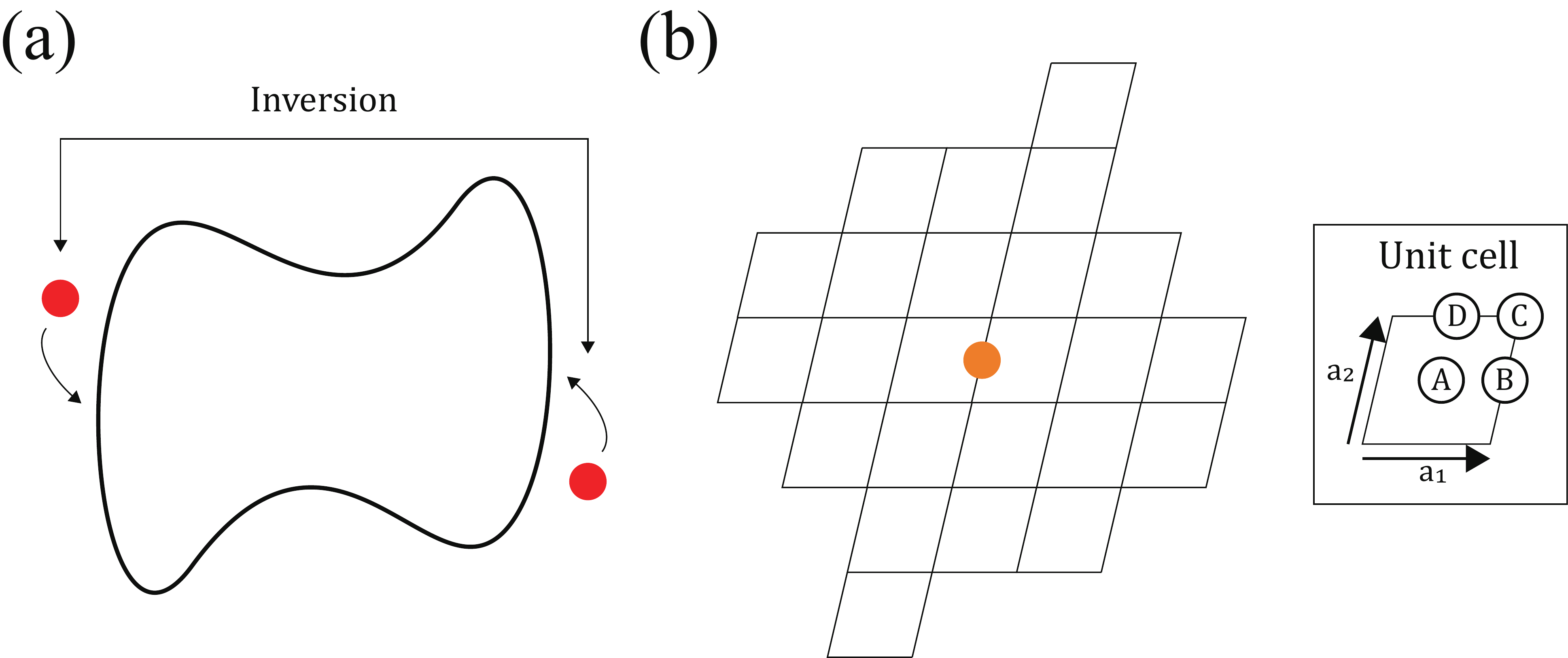}
	\caption{Schematic figures describing how the net parity is defined for inversion-symmetric insulators without gapless boundary states. (a) Localized degrees of freedom, red dots, added to the boundary in an inversion-symmetric way. Their symmetric and antisymmetric combinations generate $s$ and $p$ orbitals, and thus $\mathcal{I}$ does not change. (b) A finite-size system with an open boundary in which the Wyckoff position $B$ is at the inversion center, the orange dot. In this case, the net parity $\mathcal{I}$ is given by $\nu_B$.}
	\label{fig: NetParity}
\end{figure}
%%%%%%%%%%%%%%%%%%%%%%%%%%%%%%%%%%%%%%%%%%%%%%%

As discussed in Sec.~\ref{subsec: SSH(1,4)}, the net parity $\mathcal{I}$ is determined by the difference in the number of $p$ and $s$ orbitals at the center of the system with an open boundary, $\Delta N_{\rm center}$.
When the system with an open boundary is inversion symmetric, its center must be one of the Wyckoff position $W^*$.
Hence, $\Delta N_{\rm center}$ is given by $\nu_{W^*}$, and the net parity $\mathcal{I}$ is manifested as one of $\nu_W$ indices as shown in Fig.~\ref{fig: NetParity}(b).

\section{$d$D $(d+1)$th-order TI as minimal fragile topological insulator with filling anomaly \label{sec: FragileReal}}

In this section, we propose a minimal Dirac Hamiltonian for a fragile TI with filling anomaly.
We assume that the number of occupied bands is $N_{\mathcal{B}}$, and all the occupied bands at the $\gamma$ point have negative parity while they have positive parity at all the other inversion-invariant momenta. Then, the corresponding parity configuration is $\mathcal{B}_\gamma(d,N_{\mathcal{B}})$.
Since such a parity configuration can be generated by a band inversion at $\gamma$, starting from a trivial-parity insulator having only positive parity states at inversion-invariant momenta, the state corresponding to $\mathcal{B}_\gamma(d,N_{\mathcal{B}})$ can be described by a Dirac Hamiltonian.

The minimal dimension of the Dirac Hamiltonian can be determined by our diagnosis method of fragile topology as follows.
First, we compute $2^d$ inversion indices $\{\nu_{W_1},\dots,\nu_{W_{2^d}}\}$ by using Eq.~\eqref{eq: Indices}. Namely,
\ba
\begin{pmatrix}
	\nu_{W_1} \\
	\nu_{W_2} \\
	\nu_{W_3} \\
	\vdots \\
	\nu_{W_{2^d}}
\end{pmatrix}
&=A_d
\begin{pmatrix}
	N_{\mathcal{B}} \\
	-N_{\mathcal{B}} \\
	-N_{\mathcal{B}} \\
	\vdots \\
	-N_{\mathcal{B}}
\end{pmatrix},
\ea
which gives
\ba
\label{eq: BZIndices}
\{\nu_{W_1},\dots,\nu_{W_{2^d}}\}=2^{1-d} N_{\mathcal{B}} \{1-2^{d-1},1,1,\dots, 1\}.
\ea

%%%%%%%%%%%%%%%%%%%%%%%%%%%%%%%%%%%%%%%%%%%%%%%%%%%%%%%%%%%
\begin{figure}[b!]
	\centering
	\includegraphics[width=0.49\textwidth]{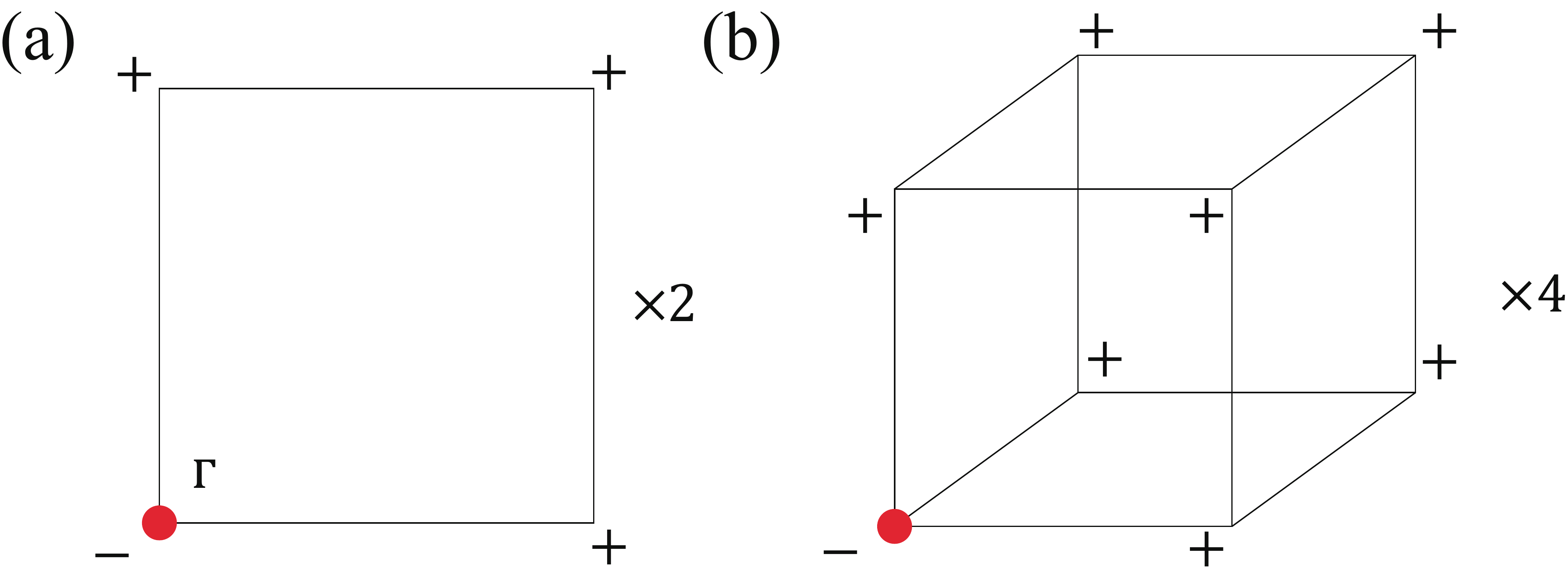}
	\caption{Parity configurations for $d$D $(d+1)$th-order TIs. (a) $\mathcal{B}_\gamma(2,2)$ for a $2$D third-order TI. (b) $\mathcal{B}_\gamma(3,4)$ for a $3$D fourth-order TI.
		}
	\label{fig: ParityConfig}
\end{figure}
%%%%%%%%%%%%%%%%%%%%%%%%%%%%%%%%%%%%%%%%%%%%%%%%%%%%%%%%%%

To be a fragile topological phase, $\{\nu_W\}$ must be a set of integers and satisfy $\sum_W |\nu_{W}|>N_{\mathcal{B}}$.
These conditions constrain $N_{\mathcal{B}}$ to be an integer multiple of $2^{d-1}$ in any dimensions greater than one.
Thus, the minimal number of occupied bands for the fragile phase is $N_{\mathcal{B}}=2^{d-1}$.
Importantly, all $\nu_{W_i}$ for $i=2,\dots,2^d$ are odd integers when $N_{\mathcal{B}}=2^{d-1}$, and thus the filling anomaly exists. This is consistent with the result discussed in Sec.~\ref{subsec: FillingAnomaly} where it is shown that the filling anomaly can occur when $n_{\rm occ} \ge 2^{d-1}$.
Also, we can find the equivalent representation for Wannier states corresponding to $\mathcal{B}_{\gamma}(d,2^{d-1})$ by using $\nu_W=\mu_{W,-}-\mu_{W,+}$ such as
\ba
\label{eq: FrgWannier}
\mathcal{B}_{\gamma}(d,2^{d-1}) \Leftrightarrow \bigoplus_{i=1}^{2^d} \, p(W_i) \ominus 2^{d-1} \, p(W_1).
\ea
Let us note three implications of Eqs.~\eqref{eq: BZIndices} and~\eqref{eq: FrgWannier}.
First, the Wannier obstruction can be lifted after adding $(2^{d-1}-1)$ $p(W_1)$ orbitals.
Second, $\mathcal{B}_{\gamma}(d, N_{\mathcal{B}})$ with $N_{\mathcal{B}}<2^{d-1}$ must have stable band topology since $\nu_W$ must be a fractional number.
Third, an integer multiple of $\mathcal{B}_{\gamma}(d, 2^{d-1})$ retains the fragile topology since multiplying Eq.~\eqref{eq: BZIndices} by an integer does not change the relevant inequality, i.e., $\sum_W |\nu_W| > n_{\rm occ}$.

In Sec.~\ref{subsec: Dirac Hamiltonian}, it is shown that the Dirac Hamiltonian for a class A $d$D $k$th-order TI is expressed by $2^\frac{d+k-1}{2} \times 2^\frac{d+k-1}{2}$ gamma matrices.
Hence, the Dirac Hamiltonian corresponding to $\mathcal{B}_\gamma(d,2^{d-1})$ can describe a $d$D $(d+1)$th-order TI.
For example, the Dirac Hamiltonian relevant to $\mathcal{B}_{\gamma}(2,2)$ describes a $2$D third-order TI, and similarly, the Dirac Hamiltonian relevant to $\mathcal{B}_{\gamma}(3,4)$ describes a $3$D fourth-order TI. The corresponding parity configurations are shown in Fig.~\ref{fig: ParityConfig}. When the number of occupied states is greater than $2^{d-1}$, the corresponding Dirac Hamiltonian describes a $d$D $k$th-order TI with $k>(d+1)$.
As discussed above, all $d$D $k$th-order TIs with $k>d$ have fragile band topology.

In Secs.~\ref{subsec: Model} and~\ref{subsec: Numerics}, we examine the charge accumulation in the $d$D $(d+1)$th-order TI and establish the bulk-boundary correspondence, which can be viewed as $2$D and $3$D generalizations of the idea discussed in Sec.~\ref{subsec: SSH(1,2)}.

Before we proceed, let us comment on the case where the occupied states with negative parity are distributed over the multiple inversion-invariant momenta. 
In this case, the minimal number of occupied bands to have fragile topology is two in any dimensions which is smaller than $2^{d-1}$ for $d>2$ (see Sec.~\ref{subsec: Result2d3d}).
However, these insulators cannot have the filling anomaly unless the number of occupied bands is greater than or equal to $2^{d-1}$ as discussed in Sec.~\ref{subsec: FillingAnomaly}.
In Appendix~\ref{app: SumFormula}, we discuss how an odd number of band inversions corresponding to the $d$D $(d+1)$th-order Dirac Hamiltonian generates the filling anomaly.

\subsection{Tight-binding model \label{subsec: Model}}

First, we define $4 \times 4$ gamma matrices $\gamma_{n=1,...,5}$ and $8 \times 8$ gamma matrices $\check{\gamma}_{m=1,...,7}$ in order to construct the tight-binding models for class A $2$D third-order and $3$D fourth-order TIs protected by inversion symmetry.
We choose that $\gamma_{n}$ and $\check{\gamma}_{n}$ with an even (odd) $n$ are real (imaginary).
For example, $\{\gamma_1, \dots, \gamma_5\}=\{\tau_x\sigma_x, \tau_x\sigma_y, \tau_x\sigma_z, \tau_y\sigma_0, \tau_z\sigma_0\}$ and $\{\check{\gamma}_1 , \dots, \check{\gamma}_7\}=\{\mu_x\tau_x\sigma_x, \mu_x\tau_x\sigma_y, \mu_x\tau_x\sigma_z, \mu_x\tau_y\sigma_0, \mu_x\tau_z\sigma_0, \mu_y\tau_0\sigma_0, \mu_z\tau_0\sigma_0\}$.

%%%%%%%%%%%%%%%%%%%%%%%%%%%%%%%%%%%%%%%%%%%%%%%%%%%%%%%%%%%%%
\begin{figure*}[ht]
	\centering
	\includegraphics[width=0.98\textwidth]{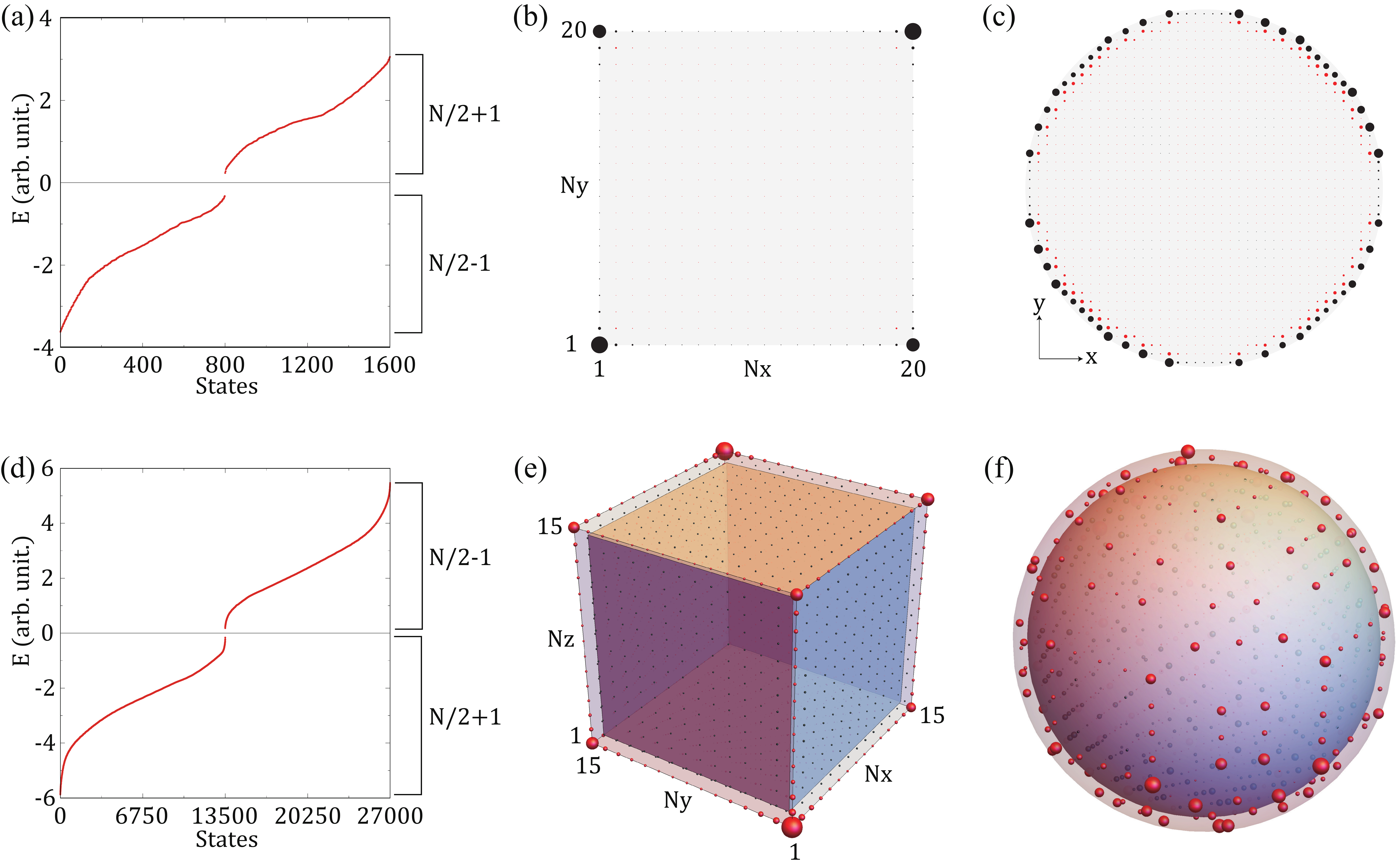}
	\caption{Energy spectra and boundary charge accumulations of finite-size $2$D third-order and $3$D fourth-order TIs. (a) The energy spectrum of the lattice model for $2$D third TIs in Eq.~\eqref{eq: HA2dtb}, in the topological phase ($\lambda=1$). The system has the square geometry with $20 \times 20$ unit cells ($N=1600$). (b) The extra charge accumulation $\Delta \rho(x,y)$ defined in Eq.~\eqref{eq: density2d} in the square geometry with $20 \times 20$ unit cells. (c) The extra charge accumulation $\Delta \rho(x,y)$ in the disk geometry. The radius of the disk ($R$) is equal to the length of $20.5$ unit cells ($R=20.5$). (d) The energy spectrum of the lattice model for $3$D fourth TIs in Eq.~\eqref{eq: HA3dtb}, in the topological phase ($\lambda=1$). The system has the cube geometry with $15 \times 15 \times 15$ unit cells ($N=27000$). (e) The extra charge accumulation $\Delta \rho(x,y,z)$ in Eq.~\eqref{eq: density3d} in the cubic geometry with $15 \times 15 \times 15$ unit cells. (f) The extra charge accumulation $\Delta \rho(x,y,z)$ in the sphere geometry. The radius of the sphere ($R$) is equal to the length of $6.7$ unit cells ($R=6.7$). The orientation of the coordinate system is same as (e). In (b) and (c) and (e) and (f), the size of the dots and spheres dictates the absolute value of the accumulated or depleted charges. A black (red) dot and sphere denotes the depletion (accumulation). To enhance the visibility of the boundary, the cube with the edge length of $14.5$ unit cells and the sphere with the radius of $6.2$ unit cells are depicted in (e) and (f), respectively.}
	\label{fig: FiniteSize}
\end{figure*}
%%%%%%%%%%%%%%%%%%%%%%%%%%%%%%%%%%%%%%%%%%%%%%%%%%%%%%%%%%%%%

Now, we discuss how the boundary mass terms can be introduced in lattice models.
Following~\cite{geier2018second,trifunovic2019higher}, we briefly review the mechanism for generating the boundary mass terms from symmetric perturbations to the bulk Hamiltonian.
An inversion-symmetric Dirac Hamiltonian without any perturbation is
\ba
\mathcal{H}_{0}(\bb{k}) = \sum_{i=1}^d k_i \gamma_i + \lambda \mathcal{M},\\
I \mathcal{H}_{0}(\bb{k}) I^{-1} = \mathcal{H}_{0}(-\bb{k}),
\ea
where $I$ denotes the operator for inversion symmetry.
Inversion-symmetric perturbation $H_1$ is given by
\ba
H_{1} = i \sum_{i=1}^d \sum_{a=1}^d \Delta_{i,a} \, \gamma_i M_a \mathcal{M},
\ea
where $M_a=-I M_a I^{-1}$ with $a=1,\dots,d$.

After projection to the boundary, $H_1$ becomes
\ba
P_+(\bb{r}) \, H_1 \, P_+(\bb{r}) = \sum_{i=1}^d \sum_{a=1}^d \Delta_{i,a} \, \bb{n}_{\bb{r},i} \tilde{M}_a,
\ea
where $\tilde{M}_a= M_a P_+(\bb{r})$.
Then, one can identify the boundary mass terms as $m_{a,\bb{r}}=\sum_{i=1}^d \Delta_{i,a} \bb{n}_{\bb{r}, i}$.
Since $m_{a,\bb{r}}$ is proportional to the surface normal vector $\bb{n}_{\bb{r}}$, $m_{a, \bb{r}}=-m_{a,-\bb{r}}$ is naturally implemented.
Finally, the lattice regularization with the substitution $k_i \rightarrow \sin k_i$ and $\lambda \rightarrow -(d-\lambda-\sum_{i=1}^d \cos k_i)$ leads to the bulk Hamiltonian
\ba
\label{eq: Htb}
H(\bb{k}) = \sum_{i=1}^d \sin k_i \gamma_i - ( d - \lambda - \sum_{i=1}^d \cos k_i ) \mathcal{M} + H_1,
\ea
which naturally contains the boundary mass terms whose explicit form appears after projection to the boundary.
All the tight-binding models used in the paper are constructed by using this method.
For example, the tight-binding models for class A $2$D third-order and $3$D fourth-order TIs are given by
\ba
\label{eq: HA2dtb}
H_{{\rm A} (2,3)} =& \sum_{i=1}^2 \sin k_i \gamma_i - (2 - \lambda - \sum_{i=1}^2 \cos k_i ) \gamma_5 \nn \\
& + i \sum_{i=1}^2 \sum_{a=3}^4 \Delta_{i,a} \, \gamma_i \gamma_a \gamma_5,
\ea
and
\ba
\label{eq: HA3dtb}
H_{{\rm A} (3,4)} =& \sum_{i=1}^3 \sin k_i \check{\gamma}_i - (3 - \lambda - \sum_{i=1}^3 \cos k_i ) \check{\gamma}_7 \nn \\
& + i \sum_{i=1}^3 \sum_{a=4}^6 \check{\Delta}_{i,a} \, \check{\gamma}_i \check{\gamma}_a \check{\gamma}_7.
\ea

For the tight-binding models defined above, since the boundary mass terms are given by $m_{a, \bb{r}}=\sum_{i=1}^{d} \Delta_{i,a} \bb{n}_{\bb{r},i}$, the corresponding winding numbers are determined by the sign of the determinant of $\Delta_{i,a}$.
For example,
\ba
w_1 &= \frac{1}{2 \pi} \oint d\theta \, \big( \varphi_1 \dr_{\theta} \varphi_2 - \varphi_1 \dr_{\theta} \varphi_2 \big) \nn \\
&= \frac{1}{2 \pi} \oint d\theta \, \frac{1}{m_1^2+m_2^2} \big( m_1 \dr_{\theta} m_2 - m_1 \dr_{\theta} m_2 \big) \nn \\
&= \text{Det} (\bb{\Delta}) \, \frac{1}{2\pi} \oint d\theta \, \frac{1}{m_1^2+m_2^2} \nn \\
&= \frac{\text{Det} (\bb{\Delta})}{|\text{Det} (\bb{\Delta})|} \nn \\
&= \text{sgn}(\text{Det} (\bb{\Delta})),
\ea
where we use $m_{a=1,2}=\Delta_{1,a+2} \cos \theta + \Delta_{2,a+2} \sin \theta$ and
\ba
\frac{1}{2\pi} \oint d\theta \frac{1}{A \cos 2\theta + B \sin 2\theta + C} \\
= \text{sgn}(C-A) \frac{1}{\sqrt{C^2-A^2-B^2}}.
\ea
Similarly, we obtain $w_d = \text{sgn}(\text{Det} (\bb{\Delta}))$.
Thus, the $d$D $(d+1)$th-order TI described by the relevant Dirac Hamiltonian is characterized by $w_d= 1$ (mod $2$).

\subsection{Numerical calculations \label{subsec: Numerics}}
In order to verify our theoretical prediction for charge accumulation at the boundary based on the continuum Dirac Hamiltonian, here we calculate the energy spectrum and the eigenstates of $2$D third-order and $3$D fourth-order TIs described by the lattice Hamiltonian $H_{{\rm A}(2,3)}$ and $H_{{\rm A}(3,4)}$, by considering various shapes of open boundaries.
We choose square and disk geometries for $2$D third-order TI, and cube and sphere geometries for $3$D fourth-order TI.
For numerical calculations, we choose $\Delta_{1,3}=0.240$, $\Delta_{1,4}=-0.196$, $\Delta_{2,3}=-0.259$, $\Delta_{2,4}=-0.204$ for the $2$D third-order TI, and $\check{\Delta}_{1,4}=-0.25$, $\check{\Delta}_{1,5}=0.05$, $\check{\Delta}_{1,6}=0.10$, $\check{\Delta}_{2,4}=-0.05$, $\check{\Delta}_{2,5}=0.30$, $\check{\Delta}_{2,6}=-0.10$, $\check{\Delta}_{3,4}=0.05$, $\check{\Delta}_{3,5}=-0.05$, $\check{\Delta}_{3,6}=0.35$ for the $3$D fourth-order TI.
Then $\text{sgn}({\text{Det} (\bb{\Delta}})) = -1$ and $\text{sgn}({\text{Det} (\bb{\check{\Delta}}})) = -1$ in both cases.
Also, we assume that all sublattice sites are located at the unit-cell center for simplicity.

The energy spectra in Figs.~\ref{fig: FiniteSize}(a) and~\ref{fig: FiniteSize}(d) show that the number of occupied bands and that of unoccupied bands are different in the topological phase ($\lambda=1$) whereas they are the same in the trivial phase ($\lambda=-1$).
Figures~\ref{fig: FiniteSize}(a) and~\ref{fig: FiniteSize}(d) correspond to the $2$D third-order TI and the $3$D fourth-order TI, respectively.
Moreover, when the states below (above) the gap are fully occupied (unoccupied) by adding an extra electron or hole to the half-filled system, we find that the extra charge accumulation is localized only at certain subregions of the boundary where the surface and the bulk mass terms change rapidly.
More explicitly, it is found that the extra charge is localized at the corners of the system with a square or cube geometry, while the extra charge is spread over the whole boundary region when the system geometry has a disk or sphere shape, as shown in Fig.~\ref{fig: FiniteSize}.

The difference in the number of occupied and unoccupied states in $d$D $(d+1)$th-order TIs can be understood by examining the change in the energy spectrum of a $d$D $d$th-order TI with chiral symmetry when chiral-symmetry-breaking perturbations are added. 
Specifically, a $d$D $d$th-order TI with chiral symmetry has two zero-energy modes related by inversion. 
Adding chiral-symmetry-breaking perturbation, the zero-energy modes are pushed either upward or downward until they are merged to the unoccupied or occupied states, leading to a $d$D $(d+1)$th-order TI without in-gap states.
As long as inversion symmetry is preserved, the two in-gap states should be degenerate during this process. The filling anomaly requires that an extra electron or hole is added to the half-filled system, and the states below (above) the gap are fully occupied (unoccupied), to preserve inversion symmetry. Then, one can examine the extra charge accumulation relative to the half-filling (see Sec.~\ref{subsec: SSH(1,2)} for the related discussion in $1$D second-order TIs).

%%%%%%%%%%%%%%%%%%%%%%%%%%%%%%%%%%%%%%%%%%%%%%%%%%%%%%%%%%%%%%%%
\begin{table}[b!]
	\caption{The extra charge accumulation in the $2$D third-order TI. $Q^{\rm num.}_{i=1,2,3,4}$ ($Q_{i=1,2,3,4}$ ) indicates the extra charge in the $i$th quadrant computed from the lattice model (from the induced charge of the continuum model). Note that $Q^{\rm num.}_1=Q^{\rm num.}_3$ and $Q^{\rm num.}_2=Q^{\rm num.}_4$ due to inversion symmetry, and similar relations hold for $Q_i$.}
	\begin{tabular}{c|c|c|c}
		\hline
		\hline
		$Q^{\rm num.}_1$ & $Q_1$ & $Q^{\rm num.}_2$ & $Q_2$ \\ \hline
		$-0.282329$ & $-0.284825$ & $-0.217611$ & $-0.215175$ \\ \hline
		\hline
	\end{tabular}
	\label{table: Charge2d}
\end{table}
%%%%%%%%%%%%%%%%%%%%%%%%%%%%%%%%%%%%%%%%%%%%%%%%%%%%%%%%%%%%%%%%

%%%%%%%%%%%%%%%%%%%%%%%%%%%%%%%%%%%%%%%%%%%%%%%%%%%%%%%%%%%%%%%%
\begin{table}[b!]
	\caption{The extra charge accumulation in the $3$D fourth-order TI. $Q^{\rm num.}_{i=1,\dots,8}$ ($Q_{i=1,\dots,8}$) indicates the extra charge in the $i$th octant computed from the lattice model (from the induced charge of the continuum model). Note that $Q^{\rm num.}_1=Q^{\rm num.}_7$, $Q^{\rm num.}_2=Q^{\rm num.}_8$, $Q^{\rm num.}_3=Q^{\rm num.}_5$, and $Q^{\rm num.}_4=Q^{\rm num.}_6$ due to inversion symmetry, and similar relations hold for $Q_i$.}
	\begin{tabular}{c|c|c|c}
		\hline
		\hline
		$Q^{\rm num.}_1$ & $Q_1$ & $Q^{\rm num.}_2$ & $Q_2$ \\ \hline
		$0.116481$ & $0.113613$ & $0.203660$ & $0.223673$ \\ \hline
		\hline
		$Q^{\rm num.}_3$ & $Q_3$ & $Q^{\rm num.}_4$ & $Q_4$ \\ \hline
		$0.089725$ & $0.081586$ & $0.090134$ & $0.081127$ \\ \hline
		\hline
	\end{tabular}
	\label{table: Charge3d}
\end{table}
%%%%%%%%%%%%%%%%%%%%%%%%%%%%%%%%%%%%%%%%%%%%%%%%%%%%%%%%%%%%%%%%

However, to understand the distribution of the extra charge accumulation at the boundary, we need the additional information of the induced current $j^\mu_{d}$ as explained in Sec.~\ref{subsec: bbcorres}.
To confirm the charge accumulation predicted by the induced current in Sec.~\ref{subsec: bbcorres}, we first numerically solve the lattice model for a $2$D third-order TI and compute the extra charge density $\Delta \rho(x,y)$ at the position $(x,y)$ relative to the half-filling ($\rho_0=2$).
In both the square and disk geometries, the extra charge density $\Delta \rho(x,y)$ is defined as
\ba
\label{eq: density2d}
\Delta \rho(x,y) = \sum_{i=1}^{N_{\rm occ}} |\psi_i(x,y)|^2-\rho_0,
\ea
where $\psi_i(x,y)$ is the $i$th energy eigenstate and $N_{\rm occ}$ is the number of occupied states.
Since total extra charge is $-1$ according to $\delta= \int j^0_1=\text{sgn}(\text{Det} (\bb{\Delta}))=-1$, the total sum of $\Delta \rho(x,y)$ is equal to $-1$. Namely,
\ba
\sum_{x=1}^{N_x}\sum_{y=1}^{N_y} \Delta \rho(x,y) = -1,
\ea
where $N_x$ and $N_y$ denote the number of lattice sites in the $x$ and $y$ directions, respectively.

In the case of the square geometry, the extra charge is mostly localized at the corners as shown in Fig.~\ref{fig: FiniteSize}(b).
This is simply because the boundary mass terms vary abruptly near the corners.
To demonstrate the spatial distribution of the extra charge accumulation, we divide the system into four quadrants. Then we compare the total extra charge in each quadrant, $Q^{\rm num.}_{i=1,2,3,4}$ with the total induced charge density $j^0_1$ along the boundary of each quadrant, $Q_i$.
Because of inversion symmetry, $Q^{\rm num.}_1=Q^{\rm num.}_3$ and $Q^{\rm num.}_2=Q^{\rm num.}_4$.
Explicitly, $Q^{\rm num.}_{1,2}$ are defined as
\ba
&Q^{\rm num.}_1= \sum_{\frac{N_x}{2} < x} \sum_{\frac{N_y}{2} < y} \Delta \rho(x,y), \nn \\
&Q^{\rm num.}_2= \sum_{x < \frac{N_x}{2}} \sum_{\frac{N_y}{2} < y} \Delta \rho(x,y).
\ea
According to the numerical data displayed in Table~\ref{table: Charge2d}, the deviation of $Q_i$ from $Q^{\rm num.}_i$ is only about one-percent. This is because when the bulk gap is sufficiently large, our formula for the induced current becomes more accurate due to the negligible contributions of higher-order terms in the derivative expansion. More detailed information about how the induced currents are computed is in Appendix~\ref{app: Current}.
On the other hand, in contrast to the case of the square geometry, the extra charge is spread over the whole boundary in the disk geometry as shown in Fig.~\ref{fig: FiniteSize}(c). This delocalized charge accumulation is also consistent with the profile of $j^0_1(\theta)$ in Eq.~\eqref{eq: current2d}.

Now, let us consider the $3$D fourth-order TI.
Both in the cube and sphere geometries, the extra charge density $\Delta \rho(x,y,z)$ at the position $(x,y,z)$ compared to half-filling ($\rho_0=4$) is defined as
\ba
\label{eq: density3d}
\Delta \rho(x,y,z) = \sum_{i=1}^{N_{\rm occ}} |\psi_i(x,y,z)|^2-\rho_0,
\ea
where $\psi_i(x,y,z)$ is the $i$th energy eigenstate and $N_{\rm occ}$ is the number of occupied states.
Also, the total accumulated charge is $1$ according to $\delta= \int j^0_2=-\text{sgn}(\text{Det} (\bb{\check{\Delta}}))=1$, and thus the total sum of $\Delta \rho(x,y,z)$ is equal to $1$. Namely,
\ba
\sum_{x=1}^{N_x}\sum_{y=1}^{N_y}\sum_{z=1}^{N_z} \Delta \rho(x,y,z) = 1,
\ea
where $N_{i=x,y,z}$ denote the number of lattice sites in the $i$ direction.

In the case of the cube geometry, the extra charge is mostly accumulated at the corners where the boundary mass terms vary rapidly, as shown in Fig.~\ref{fig: FiniteSize}(e). To demonstrate the spatial distribution of the extra charge accumulation, we divide the system into eight octants.
Then, we compare the total extra charge in each octant, $Q^{\rm num.}_{i=1,\dots,8}$ with the total induced charge density $j^0_2$ along the boundary of each octant, $Q_i$.
Inversion symmetry imposes $Q^{\rm num.}_1=Q^{\rm num.}_7$, $Q^{\rm num.}_2=Q^{\rm num.}_8$, $Q^{\rm num.}_3=Q^{\rm num.}_5$, and $Q^{\rm num.}_4=Q^{\rm num.}_6$.
Explicitly, $Q^{\rm num.}_{1,2,3,4}$ are defined as
\ba
&Q^{\rm num.}_1= \sum_{\frac{N_x}{2} < x} \sum_{\frac{N_y}{2} < y} \sum_{\frac{N_z}{2} < z} \Delta \rho(x,y,z), \nn \\
&Q^{\rm num.}_2= \sum_{x < \frac{N_x}{2}} \sum_{\frac{N_y}{2} < y} \sum_{\frac{N_z}{2} < z} \Delta \rho(x,y,z), \nn \\
&Q^{\rm num.}_3= \sum_{x < \frac{N_x}{2}} \sum_{y < \frac{N_y}{2}} \sum_{\frac{N_z}{2} < z} \Delta \rho(x,y,z), \nn \\
&Q^{\rm num.}_4= \sum_{\frac{N_x}{2} < x} \sum_{y < \frac{N_y}{2}} \sum_{\frac{N_z}{2} < z} \Delta \rho(x,y,z).
\ea
%Note that when the cube is incommensurate, $N_{x,y,z}=15$ for example, we divide the unit cell evenly.
The results of numerical calculations are displayed in Table~\ref{table: Charge3d}, which shows that the deviation of $Q_i$ from $Q^{\rm num.}_i$ is about 10$\%$ percent.
This is because our formula for the induced current in Eq.~\eqref{eq: top_current} get corrections from the higher-order terms when the bulk gap is small. When the bulk gap becomes larger, the deviation of $Q_i$ from $Q^{\rm num.}_i$ becomes smaller.
In contrast to the case of the cube geometry, the extra charge is spread over the whole boundary in the sphere geometry as shown in Fig.~\ref{fig: FiniteSize}(f).
This delocalized charge accumulation is also consistent with the profile of $j^0_2(\theta,\phi)$ in Eq.~\eqref{eq: current3d}.
These numerical studies for $2$D third-order and $3$D fourth-order TIs in various open boundaries demonstrate the validity of the theory based on the induced current.

\section{Wilson loop of fragile TI \label{sec: FragileMomentum}}
As we have shown in Secs.~\ref{sec: Classification} and~\ref{sec: FragileReal}, the inversion parities completely characterize inversion-protected fragile topological insulators.
On the other hand, there is an alternative powerful method, called the Wilson loop method~\cite{yu2011equivalent,alexandradinata2014wilson,alexandradinata2016topological,fidkowski2011model,benalcazar2017quantized,benalcazar2017electric}, that can be used to diagnosis the Wannier obstruction.
Here, we demonstrate the consistency between the inversion-parity analysis and the Wilson loop method to complete the analysis of the fragile band topology.

The Wilson loop is an operator defined along a periodic direction in the Brillouin zone, whose spectral winding along the other periodic direction indicates the Wannier obstruction on the 2D subspace in the Brillouin zone~\cite{soluyanov2011Wannier, yu2011equivalent, fidkowski2011model}.
When the Wilson loop spectrum is gapped, the nested Wilson loop can be defined by treating the Wilson loop spectrum like the energy spectrum~\cite{benalcazar2017quantized,benalcazar2017electric}.
As shown below, the spectral winding of the Wilson loop and the nested Wilson loop spectra contain the essential characteristics of fragile TIs in three dimensions.

For completeness, in this section, we investigate the fragile topology of $d$D $(d+1)$th-order TIs by using the Wilson loop method and show that the Wilson loop analysis is consistent with the diagnosis based on the parity eigenvalues.
First, we observe that the Wilson loop spectrum of fragile TIs has a non-zero relative winding~\cite{alexandradinata2014wilson}.
Then, we show that the relative winding of the Wilson loop can be unwound by adding appropriately chosen trivial bands.
In dimensions greater than two, however, the number of additional orbitals required to trivialize the fragile topology does not match with the number of additional orbitals required to unwind the relative winding of the Wilson loop spectrum. We find that their number difference counts the number of orbitals required to unwind the relative winding of the nested Wilson loop spectrum.
Comparing the Wilson loop and nested Wilson loop spectra of $2$D third-order and $3$D fourth-order TIs, we show that the fragile topology of $3$D insulators is characterized by a non-zero relative winding of the nested Wilson loop, and its unwinding requires extra trivial bands to be added.
We show that this idea is generally valid for $d$D $k$th-order TIs with $d \ge 3$ and $k>d$.
In all dimensions $d$ and for all order $k$, the number of trivial bands needed to unwind all nested Wilson loops is consistent with the number of bands needed to trivialize the fragile topology determined by parity eigenvalues, which is indicated in Eq.~\eqref{eq: FrgWannier}.

\subsection{Fragile band topology and the relative winding number \label{sec: Wilsonloopwinding}}
%%%%%%%%%%%%%%%%%%%%%%%%%%%%%%%%%%%%%%%%%%%%%%%%%%%%%
\begin{figure}[b!]
	\centering
	\includegraphics[width=0.49\textwidth]{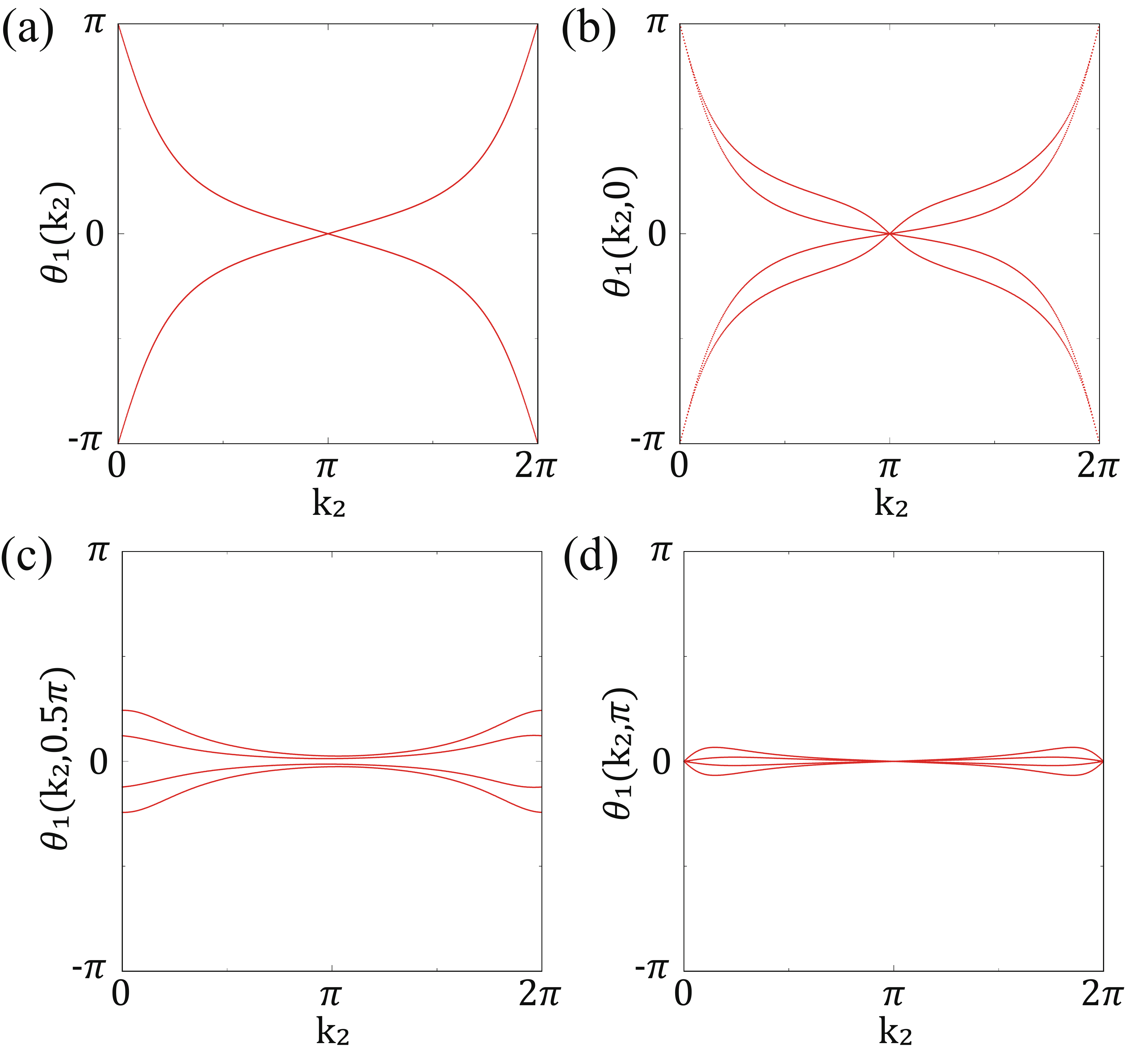}
	\caption{Wilson loop spectrum.
		(a) $2$D third-order TI.
		(b)-(d) $3$D fourth-order TI.
		The Wilson loop spectrum shows (a) $N_W=1$, (b) $N_W=2$ at $k_3=0$.
		The Wilson loop spectrum (c) at $k_3=0.5\pi$, (d) at $k_3=\pi$.
		The Wilson loop is defined along the $k_1$ direction.}
	\label{fig: WL_NP}
\end{figure}
%%%%%%%%%%%%%%%%%%%%%%%%%%%%%%%%%%%%%%%%%%%%%%%%%%%%%%
The fragile band topology of a $d$D $(d+1)$th-order TI is manifested by the winding of the (nested) Wilson loop spectrum.
The Wilson loop $W_{\bb{k}_0+\bb{G} \leftarrow \bb{k}_0}$ indicates the parallel transport of Bloch wave functions along a non-contractible loop in the Brillouin zone, from $\bb{k}_0$ to $\bb{k}_0+\bb{G}$ where $\bb{G}$ is a reciprocal lattice vector~\cite{yu2011equivalent,alexandradinata2014wilson,alexandradinata2016topological}.
In a tight-binding model, the Bloch wave function is replaced by the eigenstate $|n,\bb{k}\rangle$ of the tight-binding Hamiltonian $H(\bb{k})$, which satisfies $H(\bb{k}) |n, \bb{k} \rangle = E_{n}(\bb{k}) |n, \bb{k} \rangle$.
The Wilson loop is defined in terms of $|n, \bb{k} \rangle$ as
\ba
\label{eq: Wilson}
\Big[ W_{\bb{k}_0+\bb{G} \leftarrow \bb{k}_0} \Big]_{nm}
=& \lim_{N \rightarrow \infty} \, \bra{n, \bb{k}_0+\bb{G}} \Big[ \prod_{i=1}^{N-1} P(\bb{k}_i) \Big] \ket{m, \bb{k}_0},
\ea
where $\bb{k}_i=\bb{k}_0+\frac{i}{N} \bb{G}$ with an integer $i$, and $n_{\rm occ}$ denotes the number of occupied states.
And the projector into the occupied states $P(\bb{k})$ is defined by $P(\bb{k})=\sum_{n=1}^{n_{\rm occ}} \ket{n, \bb{k}}\bra{n, \bb{k}}$.
Note that the Wilson loop spectrum $\{{\theta}(\bb{k_0})\}$, $W_{\bb{k}_0+\bb{G} \leftarrow \bb{k}_0} \ket{\theta(\bb{k}_0)} = e^{i \theta(\bb{k}_0)} \ket{\theta(\bb{k}_0)}$, is independent of the start point $\bb{k}_0$.
Instead of the Wilson loop spectrum, we sometimes use the equivalent term, the Wilson bands in the following discussions.
Also, since we mostly consider the Wilson loop defined along the $k_1$ direction, we will denote $\{\theta_1(k_2,k_3,\dots)\}$ as the Wilson loop spectrum.
The definition and properties of the Wilson loop is summarized in Appendix~\ref{app: Wilson}.

%%%%%%%%%%%%%%%%%%%%%%%%%%%%%%%%%%%%%%%%%%%%%%%%%%%%%%%%
\begin{table*}[t!]
	\label{table: WilsonStability}
	\caption{Change of parity configuration and the Wilson loop spectrum after adding trivial orbitals.}
	\begin{tabular}{ccccccccccc}
		\hline
		\hline
		\multirow{2}{*}{Band rep.} & & \multicolumn{3}{c}{$\{n^W_{\theta}(k_2=0)\}$} & & \multicolumn{3}{c}{$\{n^W_{\theta}(k_2=\pi)\}$} & & \multirow{2}{*}{$N_W$} \\
		\cline{3-5} \cline{7-9} & & $n^W_0(0)$ & $n^W_\pi(0)$ & $n^W_c(0)$ & & $n^W_0(\pi)$ & $n^W_\pi(\pi)$ & $n^W_c(\pi)$ & & \\ \hline
		$\mathcal{B}_\gamma(d,2^{d-1}) \oplus p(A)$ & & $1$ & $2^{d-1}$ & $0$ & & $2^{d-1}-1$ & $0$ & $2$ & & $2^{d-2}-1$ \\
		$\mathcal{B}_\gamma(d,2^{d-1}) \oplus p(B)$ & & $0$ & $2^{d-1}+1$ & $0$ & & $2^{d-1}$ & $1$ & $0$ & & $2^{d-2}$ \\
		$\mathcal{B}_\gamma(d,2^{d-1}) \oplus p(C)$ & & $0$ & $2^{d-1}+1$ & $0$ & & $2^{d-1}$ & $1$ & $0$ & & $2^{d-2}$ \\
		$\mathcal{B}_\gamma(d,2^{d-1}) \oplus p(D)$ & & $1$ & $2^{d-1}$ & $0$ & & $2^{d-1}+1$ & $0$ & $0$ & & $2^{d-2}$ \\ \hline		
		\hline
	\end{tabular}
\end{table*}
%%%%%%%%%%%%%%%%%%%%%%%%%%%%%%%%%%%%%%%%%%%%%%%%%%%%%%%%%

Here, let us review the Wilson loop of inversion-symmetric systems studied comprehensively in Ref.~\cite{alexandradinata2014wilson}. For simplicity, let us consider a $2$D plane $(k_1,k_2)$ which is left invariant under the inversion.
The Wilson loop spectrum satisfies $\{\theta_1(k_2)\}=\{-\theta_1(-k_2)\}$ and can exhibit chiral and/or relative winding. We always assume that there is no chiral winding since it  represents the non-zero Chern number and we are interested in the insulators which are not stable TIs.
Figure~\ref{fig: WL_NP} shows the Wilson loop spectra for a $2$D third-order TI and a $3$D fourth-order TI.
In Fig.~\ref{fig: WL_NP}(a) relevant to the $2$D third-order TI, one Wilson band has the winding number $1$ while the other one has the winding number $-1$. Their winding number difference is called the relative winding number $N_W$, and $N_W=1$ in this case. Similarly, Fig.~\ref{fig: WL_NP}(b) has $N_W=2$ while Figs.~\ref{fig: WL_NP}(c) and~\ref{fig: WL_NP}(d) have $N_W=0$.

Due to the constraint $\{\theta_1(k_2)\}=\{-\theta_1(-k_2)\}$, the Wilson loop spectrum at $\bar{k}_2=0,\pi$ consists of $\theta_1(\bar{k}_2)=0$ or $\pi$ which are inversion symmetric by itself, or $(-\theta_0,\theta_0)$ which are inversion symmetric as a pair. Thus, we define $n^W_0(\bar{k}_2)$, $n^W_\pi(\bar{k}_2)$, and $n^W_{c}(\bar{k}_2)$, which indicate the number of Wilson loop spectrum with eigenvalues $\theta(\bar{k}_2)=0$, $\pi$, and $(-\theta_0,\theta_0)$, respectively.
For example, the parity configuration $\mathcal{B}_\gamma(2,2)$ shown in Fig.~\ref{fig: ParityConfig}(a) is mapped to $\{n^W_{\theta}(0)\}=(n^W_{0}(0),n^W_{\pi}(0),n^W_{c}(0))=(0,2,0)$ and $\{n^W_{\theta}(\pi)\}=(n^W_0(\pi),n^W_\pi(\pi),n^W_c(\pi))=(2,0,0)$.
Then the relative winding number $N_W$ is given by~\cite{alexandradinata2014wilson}
\ba
\label{eq: MapNW}
2N_W=\text{max}(n^W_{\pi}(0)-n^W_{\pi}(\pi)-n^W_{c}(\pi), \nn \\ 
n^W_{\pi}(\pi)-n^W_{\pi}(0)-n^W_{c}(0),0),
\ea
which is unambiguously determined by $n_{\xi}(\bb{k}^{{\rm inv}})$ where $n_{\xi}(\bb{k}^{{\rm inv}})$ denotes the number of the occupied bands at $\bb{k}^{\rm inv}=(0,0),(\pi,0),(0,\pi)$, or $(\pi,\pi)$ with inversion parity $\xi=\pm 1$.
The mapping between the inversion parities of occupied states $\{n_{\xi}(\bb{k}^{{\rm inv}})\}$ and the Wilson loop spectrum $\{n^W_{\theta}(\bar{k}_2)\}$ is studied for arbitrary number of occupied states $n_{\rm occ}$~\cite{alexandradinata2014wilson}, as summarized in Appendix.~\ref{app: Mapping}.

Equation~\eqref{eq: MapNW} implies that the relative winding $N_W$ is fragile, i.e., $N_W$ decreases as appropriately chosen trivial bands are added. To see this, let us consider the lattice model for a $d$D $(d+1)$th-order TI in Eq.~\eqref{eq: Htb} with $\lambda=1$, which has $2^{d-1}$ occupied eigenstates whose parity configuration corresponds to $\mathcal{B}_{\gamma}(d,N_{\mathcal{B}}=2^{d-1})$.
We focus on the $(k_1,k_2)$ plane with $k_{i>2}=0$.
It is straightforward to show that the Wilson loop spectrum of $\mathcal{B}_\gamma(d,2^{d-1})$ is given by $\{n^W_{\theta}(0)\}=\{n^W_{0}(0),n^W_{\pi}(0),n^W_{c}(0)\}=\{0,2^{d-1},0\}$ and $\{n^W_{\theta}(\pi)\}=\{n^W_0(\pi),n^W_\pi(\pi),n^W_c(\pi)\}=\{2^{d-1},0,0\}$.
Then, Eq.~\eqref{eq: MapNW} gives the relative winding number $N_W=2^{d-2}$.
Now, let us examine how $N_W$ changes as trivial bands are added.
In the $2$D unit cell in real space, which is reciprocal to the $(k_1,k_2)$ plane, there are four Wyckoff positions: $A=\bb{0}$, $B=\frac{1}{2}\bb{a}_1$, $C=\frac{1}{2}\bb{a}_1+\frac{1}{2}\bb{a}_2$, $D=\frac{1}{2}\bb{a}_2$ where $\bb{a}_{1,2}$ are relevant lattice vectors.
Then, adding a $p(B)$, $p(C)$, or $p(D)$ orbital does not change $N_W$. Here, we use the expression $p(W)$ to denote the $p$ orbital at the Wyckoff position $W$.
On the other hand, adding a $p(A)$ decreases $N_W$ by $1$.
In general, the relative winding of the Wilson loop can be unwound by adding $2^{d-2}$ $p(A)$ orbitals for a $d$D $(d+1)$th-order TI.
The above results are summarized in Table~\ref{table: WilsonStability}.

%%%%%%%%%%%%%%%%%%%%%%%%%%%%%%%%%%%%%%%%%%%%%%%%%%%%%
\begin{figure}[b!]
	\centering
	\includegraphics[width=0.49\textwidth]{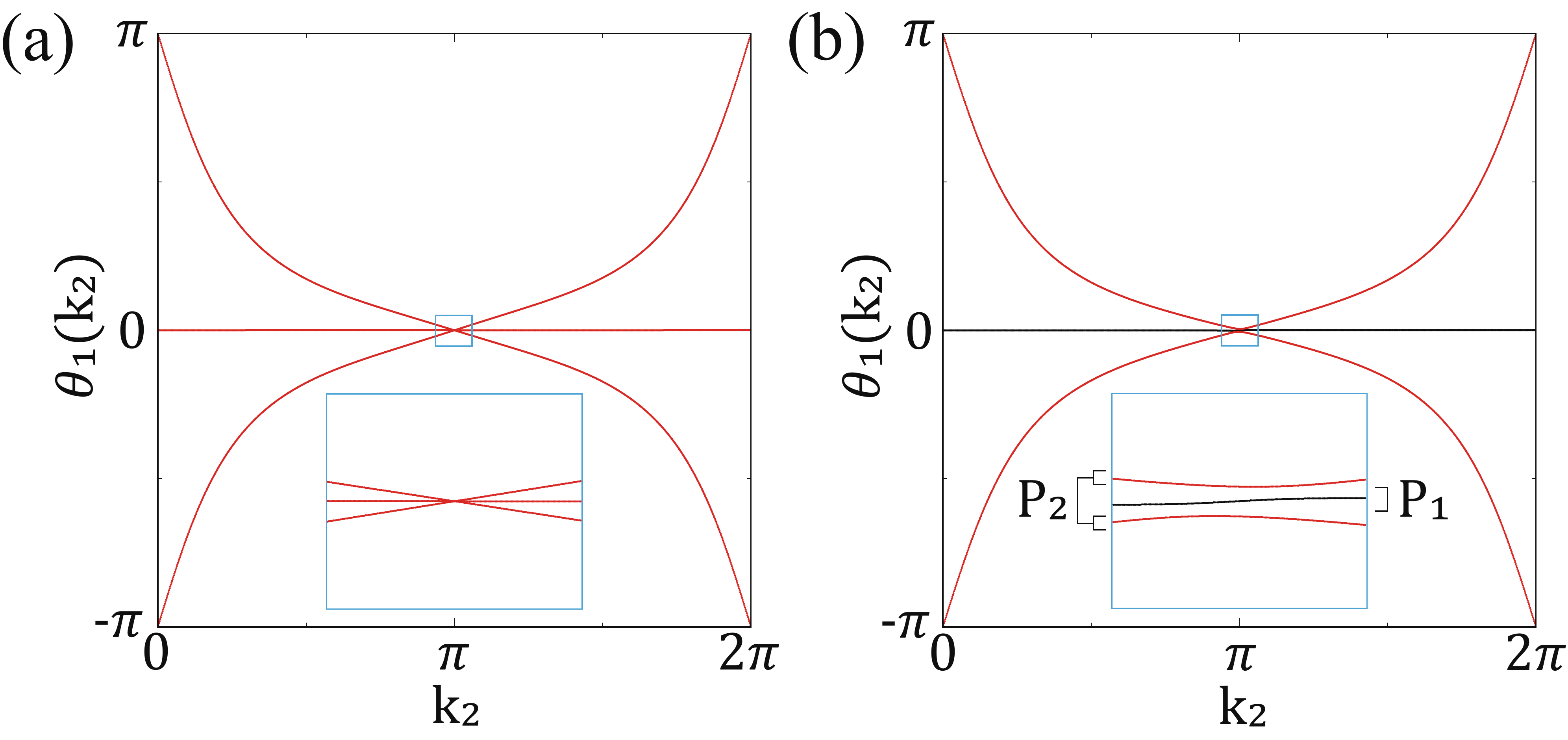}
	\caption{The Wilson loop spectrum of $2$D $3$rd order TI. (a) Adding $s(A)$ orbital cannot open the gap of Wilson spectrum. (b) Adding $p(A)$ orbital opens the gap of Wilson loop spectrum. Three bands are divided into $P_1$ and $P_2$. The black band (red bands) belongs (belong) to $P_1$ ($P_2$). $P_1$ band has the nested Berry phase $\pi$, and $P_2$ bands have the nested Berry phases ($0$, $\pi$).}
	\label{fig: WL_2d}
\end{figure}
%%%%%%%%%%%%%%%%%%%%%%%%%%%%%%%%%%%%%%%%%%%%%%%%%%%

We verify numerically the fragile nature of the relative winding number discussed above. Figure~\ref{fig: WL_2d} shows that the Wilson loop spectrum of $2$D third-order TIs changes as expected when $s(A)$ or $p(A)$ orbitals are added.
Namely, adding a $p(A)$ orbital opens a gap in the Wilson loop spectrum while adding a $s(A)$ orbital cannot.
The details of numerical calculations are summarized in Appendix.~\ref{app: Numerics}.
Since the Wilson bands are divided into two sectors ($P_{1,2}$) after we add a $p(A)$ orbital as shown in Fig.~\ref{fig: WL_2d}(b), the Berry phase $\tilde{\Phi}$ of each sector can be calculated.
$P_{1,2}$ regroup the Wilson bands in an inversion-symmetric manner, and thus have quantized Berry phases $\tilde{\Phi}[P_{1,2}]$ in the $k_2$ direction. We find $\tilde{\Phi}[P_1]=\pi$ and $\tilde{\Phi}[P_2]=(0,\pi)$.
On the other hand, let us note that the average $\Phi$ of the Wilson loop eigenvalues in $P_1$ and $P_2$ sectors along the $k_1$ direction are $0$ in both cases.
As the two Wilson bands in $P_2$ have a fixed symmetric value $\theta_1(k_2)=\pi$ at $k_2=0$, we interpret $\Phi[P_2]=0$ in a way that the Wannier centers of the two Wilson bands in the $\bb{a}_1$ direction are $\frac{1}{2}$ in both cases.
Thus, we conclude that the Wannier centers of three bands in $P_1$ and $P_2$ are $(0,\pi)$, $(\pi,0)$, and $(\pi,\pi)$, respectively.
These values support that the Wannier state representation for the $2$D third-order TI is $\ominus p(A) \oplus p(B) \oplus p(C) \oplus p(D)$~\cite{wieder2018axion}.
This is also consistent with Eq.~\eqref{eq: FrgWannier} and the corner charge accumulation~\cite{benalcazar2019quantization} in the square boundary shown in Fig.~\ref{fig: FiniteSize}(b).

%%%%%%%%%%%%%%%%%%%%%%%%%%%%%%%%%%%%%%%%%%%%%%%%%%%%%
\begin{figure*}[t!]
	\centering
	\includegraphics[width=0.75\textwidth]{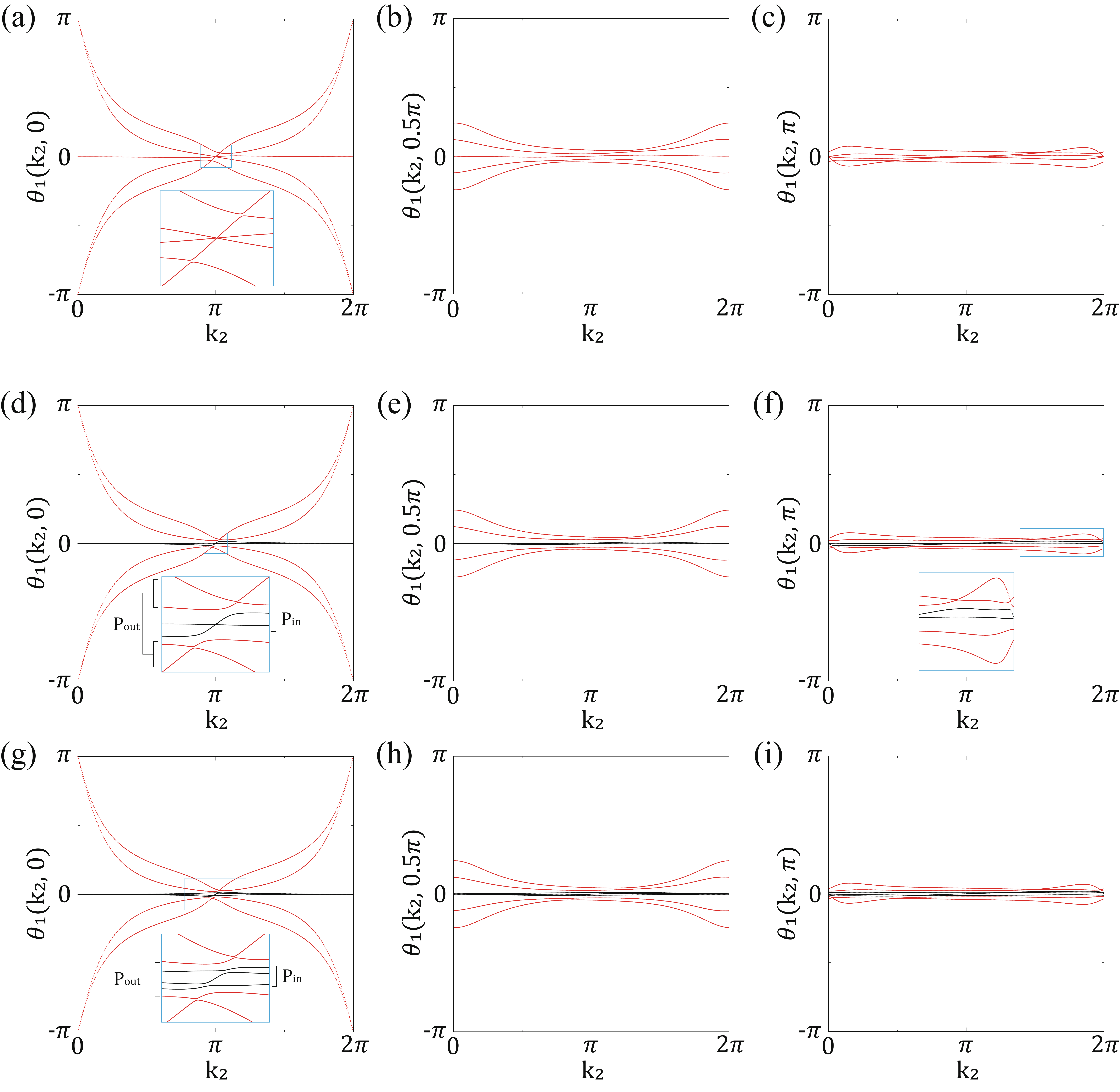}
	\caption{The Wilson loop spectrum of $3$D fourth-order TI (a)-(c) after adding one $p(A)$ orbital,
	(d)-(f) after adding two $p(A)$ orbitals, (g)-(i) after adding three $p(A)$ orbitals.
	(a) The Wilson loop spectrum at $k_3=0$. The $N_W=2$ relative winding is reduced to $N_W=1$.
	(b) The Wilson loop spectrum at $k_3=0.5\pi$. (c) The Wilson loop spectrum at $k_3=\pi$.
	(d) The Wilson loop spectrum at $k_3=0$. The $N_W=2$ relative winding is reduced to $N_W=0$.
	(e) The Wilson loop spectrum at $k_3=0.5\pi$. (f) The Wilson loop spectrum at $k_3=\pi$.
	(g) The Wilson loop spectrum at $k_3=0$. The $N_W=2$ relative winding is reduced to $N_W=0$.
	(h) The Wilson loop spectrum at $k_3=0.5\pi$. (i) The Wilson loop spectrum at $k_3=\pi$.
	As the Wilson spectra are gapped after adding two or three $p(A)$ orbitals, we can define two sectors of Wilson bands, $P_{\rm in}$ and $P_{\rm out}$.
	}
	\label{fig: WL_3d}
\end{figure*}
%%%%%%%%%%%%%%%%%%%%%%%%%%%%%%%%%%%%%%%%%%%%%%%%%%%

Let us note that the fragility of $N_W=1$ Wilson bands has already been studied in Ref.~\onlinecite{wang2018higher,wieder2018axion}. However, the fragility of $N_W=2$ Wilson bands in a $3$D fourth-order TI and in its higher-dimensional generalization has more involved structures and has not been studied yet. In the following, we show that, when $d>2$, adding $2^{d-2}$ $p$(A) orbitals are not sufficient to trivialize the fragile topology of $d$D $(d+1)$th-order TI even though it makes all the relative windings of the Wilson loops to be zero. This implies the nontrivial winding of the nested Wilson loops, which is addressed in the next section.

%%%%%%%%%%%%%%%%%%%%%%%%%%%%%%%%%%%%%%%%%%%
\begin{figure*}[t!]
	\centering
	\includegraphics[width=0.98\textwidth]{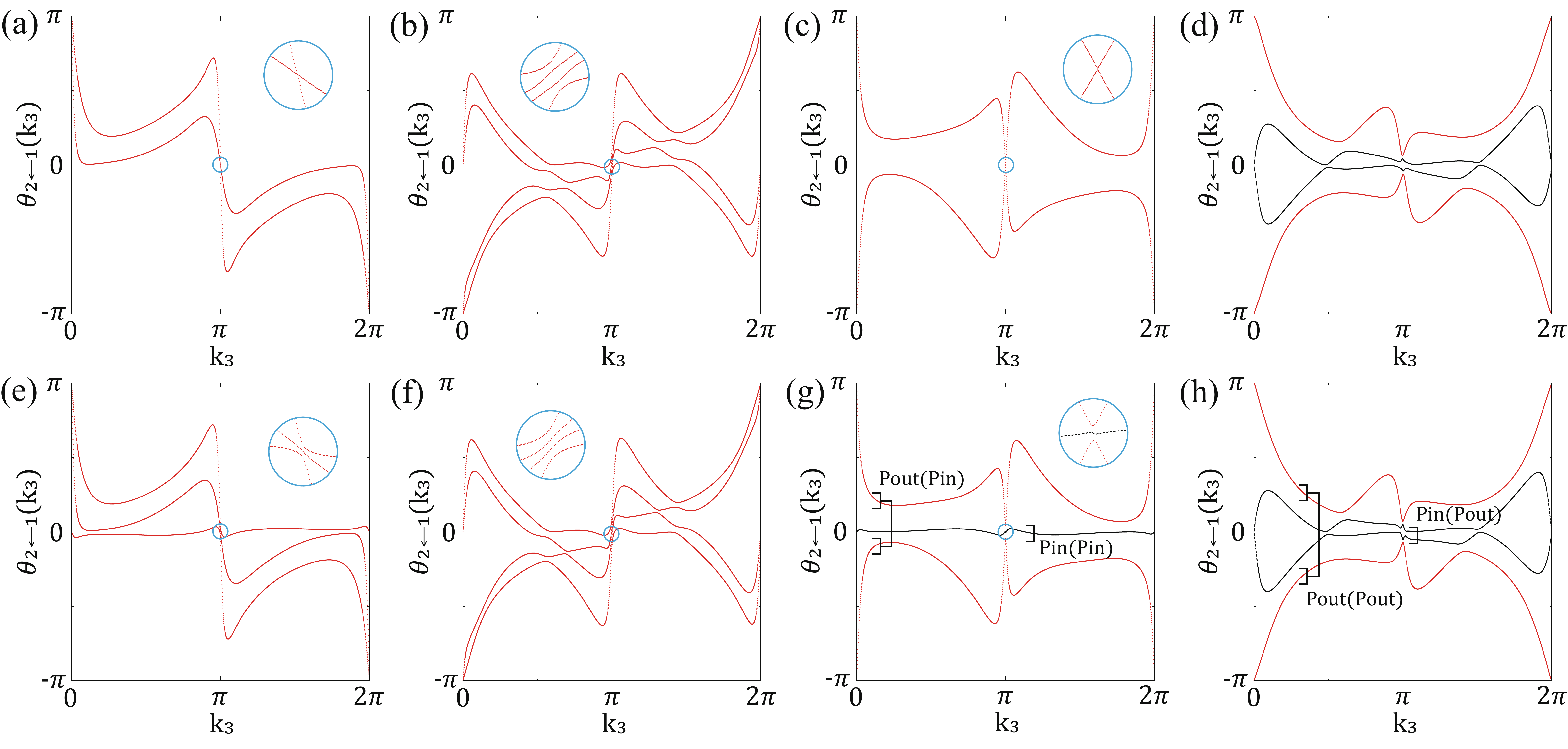}
	\caption{The nested Wilson loop spectra of $3$D fourth-order TI after adding two or three trivial $p(A)$ orbitals. (a)-(d) The nested Wilson loop spectrum of $3$D fourth-order TI with two additional $p(A)$ orbitals.
	The nested Wilson loop spectrum of (a) $P_{\rm in}$, (b) $P_{\rm out}$ before the Wilson band transition.
	The nested Wilson loop spectrum of (c) $P_{\rm in}$, (d) $P_{\rm out}$ after the Wilson band transition.
	(e)-(h) The nested Wilson loop spectrum of $3$D fourth-order TI with three additional $p(A)$ orbitals.
	The nested Wilson loop spectrum of (e) $P_{\rm in}$, (f) $P_{\rm out}$ before the Wilson band transition.
	The nested Wilson loop spectrum of (g) $P_{\rm in}$, (h) $P_{\rm out}$ after the Wilson band transition.}
	\label{fig: NestedWL}
\end{figure*}
%%%%%%%%%%%%%%%%%%%%%%%%%%%%%%%%%%%%%%%%%%%%%

\subsection{Nested Wilson loop and trivialization of fragile topology \label{subsec: NWilson}}
In this section, we discuss the fragility of the Wilson loop of the $3$D fourth-order TI.
Symmetry data analysis shows that the trivialization of the $3$D fourth-order TI requires at least three $p(A)$ orbitals although the relative winding of the Wilson bands can be gapped with two $p(A)$ orbitals.
To understand the origin of the mismatch in the number of required orbitals, we investigate the nested Wilson loop of the $3$D fourth-order TI with two or three additional $p(A)$ orbitals.

\subsubsection{Adding one $p(A)$ orbital \label{subsubsec: add1p}}
Adding one $p(A)$ orbital reduces the relative winding from $N_W=2$ to $N_W=1$ as shown in Figs.~\ref{fig: WL_3d}(a)-\ref{fig: WL_3d}(c).
Thus, the winding of the Wilson loop persists, and thus the nontrivial band topology still exists.

\subsubsection{Adding two $p(A)$ orbitals \label{subsubsec: add2p}}
Adding two $p(A)$ orbitals reduces the relative winding from $N_W=2$ to $N_W=0$ as shown in Figs.~\ref{fig: WL_3d}(d)-\ref{fig: WL_3d}(f).
Six Wilson bands are divided into two groups, $P_{\rm in}$ and $P_{\rm out}$, in an inversion-symmetric manner.
Then, we can compute the nested Wilson loops of $P_{\rm in}$ and $P_{\rm out}$, which show non-zero nested Chern numbers $C_{\rm nest}=-2$ and $C_{\rm nest}=2$, respectively, as shown in Figs.~\ref{fig: NestedWL}(a) and~\ref{fig: NestedWL}(b).
This can be explained by the fact that a $3$D fourth-order TI can be considered as two copies of the class A $3$D second order TI, i.e., the axion insulator, which is characterized by an odd nested Chern number~\cite{wieder2018axion}.
However, let us note that an even value of the nested Chern number itself does not guarantee the nontrivial bulk topology.
This is because only the parity of $C_{\rm nest}$ remains robust when a band crossing between Wilson bands in $P_{\rm in}$ and those in $P_{\rm out}$, or a Wilson band transition, happens while keeping the bulk gap~\cite{wieder2018axion}.
Since two $p(A)$ orbitals cannot trivialize the band topology according to the symmetry data analysis, the nontrivial bulk topology must be manifested in the nested Wilson loop.
We show that the nontrivial band topology appears in the form of the relative winding in the nested Wilson loop.
After the band crossing between Wilson bands, the nested Wilson loop of $P_{\rm in}$ shows the nested relative winding number $N_{W,{\rm nest}}=1$ while the nested Wilson loop of $P_{\rm out}$ is completely unwound as shown in Figs.~\ref{fig: NestedWL}(c) and~\ref{fig: NestedWL}(d).
In the nested Wilson loop spectra for $P_{\rm in}$ sector shown in Figs.~\ref{fig: NestedWL}(a) and~\ref{fig: NestedWL}(c), we have $|C_{\rm nest}| + 2 N_{W,{\rm nest}}=2$.
This means that the fragile band topology manifests as the nontrivial winding of the nested Wilson loop spectrum encoded in either $C_{\rm nest}$ or $N_{W,{\rm nest}}$.

By generalizing the above results, we propose an indicator $N_{d,{\rm nest}}$ which characterizes the fragile band topology.
First, let us define $N_{d,{\rm nest}}$ in a way similar to its counterpart $N_d$ in Eq.~\eqref{eq: MapNW} is defined for the Wilson loop:
\ba
N_{d,{\rm nest}} \equiv & \text{max}(n_\pi^{W_{\rm nest}}(0)-n_\pi^{W_{\rm nest}}(\pi)-n_c^{W_{\rm nest}}(\pi), \nn \\
& n_\pi^{W_{\rm nest}}(\pi)-n_\pi^{W_{\rm nest}}(0)-n_c^{W_{\rm nest}}(0))
\ea
where $n_\pi^{W_{\rm nest}}(\bar{k}_3)$ and $n_c^{W_{\rm nest}}(\bar{k}_3)$ are the number of eigenvalues equal to $\pi$ and that equal to a generic value, respectively, in the nested Wilson loop spectrum computed at $\bar{k}_3=0,~\pi$.
The indicator $N_{d,{\rm nest}}$ is invariant under a band crossing between the Wilson bands, and thus it determines the band topology of a given nested Wilson loop spectrum.
We note that there may be one or more gaps in the Wilson loop spectrum, and all possible nested Wilson loop spectra have to be considered. Let us distinguish the indicator $N_{d,{\rm nest}}$ into the following ways:
i) When $N_{d,{\rm nest}}$ is an odd integer, the nested Wilson loop spectrum shows an odd $C_{\rm nest}$. This implies the stable band topology.
ii) When $N_{d,{\rm nest}}$ is a positive even integer, $|C_{\rm nest}| + 2 N_{W,{\rm nest}}=N_{d,{\rm nest}}$ holds for $|C_{\rm nest}|<N_{d,{\rm nest}}$.
Thus, the nested Wilson loop shows the nested relative winding number equal to $N_{W,{\rm nest}}=\frac{1}{2} N_{d,{\rm nest}}$ even when the band crossing between the Wilson bands change $C_{\rm nest}$ to zero.
iii) When $N_{d,{\rm nest}}$ is a negative even integer or zero, the nested Wilson loop can be unwound by a band crossing between the Wilson bands.
The rules ii) and iii) can be applied to the nested Wilson loop spectrum of $P_{\rm in}$ and $P_{\rm out}$ which show $N_{d,{\rm nest}}=2$ and $-2$, respectively, as shown in Figs.~\ref{fig: NestedWL}(a)-\ref{fig: NestedWL}(d).
More details of our claim and the relevant proof are provided in Appendix.~\ref{app: FrgWL_deriv}.

\subsubsection{Adding three $p(A)$ orbitals \label{subsubsec: add3p}}
Adding the third $p(A)$ orbital to the case considered above makes little changes in the Wilson loop spectra. For instance, the third $p(A)$ orbital does not change the nested Chern number $C_{\rm nest}$ as shown in Figs.~\ref{fig: NestedWL}(e) and~\ref{fig: NestedWL}(f).
However, after the band crossing between Wilson bands in $P_{\rm in}$ and those in $P_{\rm out}$, the windings in the nested Wilson loops for both $P_{\rm in}$ and $P_{\rm out}$ have been unwound as shown in Figs.~\ref{fig: NestedWL}(g) and~\ref{fig: NestedWL}(h).
This can be inferred from the fact that the nested Wilson loop spectrum of $P_{\rm in}$ and $P_{\rm out}$ in Figs.~\ref{fig: NestedWL}(e)-\ref{fig: NestedWL}(h) show $N_{d,{\rm nest}}=0$ and $-2$, respectively.
Therefore, we conclude that the fragile TI is characterized by the nontrivial relative winding of the nested Wilson loop.

Since the Wannier obstruction is fully relaxed, we can find the Wannier state representations corresponding to the cases shown in Figs.~\ref{fig: NestedWL}(g) and~\ref{fig: NestedWL}(h).
One interesting feature of the nested Wilson loops in Fig.~\ref{fig: NestedWL}(g) is that even the nest Wilson bands from $P_{\rm in}$ can be divided further into two different groups of nested Wilson bands, $P_{\rm in}(P_{\rm in})$ and $P_{\rm out}(P_{\rm in})$, which is possible since there is no winding or relative winding in the spectrum. Similarly, Fig.~\ref{fig: NestedWL}(h) shows that the nested Wilson bands from $P_{\rm out}$ can be divided into $P_{\rm in}(P_{\rm out})$ and $P_{\rm out}(P_{\rm out})$ (see Appendix~\ref{app: Wilson} for more details).
Then, the corresponding averaged eigenvalues $\Phi$ and the Berry phases $\tilde{\Phi}$ are given by $\Phi[P_{\text{in,out}}]=\Phi[P_{\text{in,out}}(P_{\text{in,out}})]=0$, $\tilde{\Phi}[P_{\rm in}(P_{\rm in})]=\pi$, $\tilde{\Phi}[P_{\rm out}(P_{\rm in})]=(0,\pi)$, $\tilde{\Phi}[P_{\rm in}(P_{\rm out})]=(0,\pi)$, and $\tilde{\Phi}[P_{\rm out}(P_{\rm out})]=(0,\pi)$.
Here, $\Phi[P_i]$ ($\Phi[P_i(P_j)]$) indicates the average eigenvalues of the Wilson band $P_i$ [the nested Wilson band $P_i(P_j)$] while $\tilde{\Phi}[P_i]$ ($\tilde{\Phi}[P_i(P_j)]$) denotes the Berry phase of the Wilson band $P_i$ [the nested Wilson band $P_i(P_j)$].
The corresponding Wannier centers of the seven occupied states are $(0,\pi,0)$, $(\pi,0,0)$, $(\pi,\pi,0)$, $(0,0,\pi)$, $(0,\pi,\pi)$, $(\pi,0,\pi)$, and $(\pi,\pi,\pi)$, respectively.
This supports that the Wannier state representation for the $3$D fourth-order TI is $\bigoplus_{i=2}^8 \, p(W_i) \ominus 3p(W_1)$, which is consistent with Eq.~\eqref{eq: FrgWannier}. Also, the corner charge accumulation in the cubic geometry shown in Fig.~\ref{fig: FiniteSize}(e) can be related to the Wannier state representation in the same way as in $2$D.

\subsection{Generalization to higher dimensions and higher-order \label{subsec: WilsonGeneralization}}
The idea presented in the previous section can be generalized further to $d$D $k$th-order TI with $d \ge 3$ and $k>d$. 
The key observation is that only the inner Wilson bands in $P_{in}$ of the trivialized $l$th nested Wilson loop have the relative winding number in their $(l+1)$th nested Wilson loop spectrum.
We assume that the nested Chern numbers are always zero, which can always be achieved via the band crossing between Wilson bands.

Explicitly, according to the discussion in Sec.~\ref{sec: FragileReal}, a $d$D $k$th-order TI has an equivalent Wannier state representation given by
\ba
\label{eq: GeneralWannier1}
2^{n-d} \Big[ \bigoplus_{i=1}^{2^d} \, p(W_i) \ominus 2^{d-1} \, p(W_1) \Big],
\ea
where $d+k=2n+1$ and $n=d, d+1, \dots$ for class A systems.
Thus, the $d$D $k$th-order TI can be trivialized by adding $2^{n-d}(2^{d-1}-1)$ $p(W_1)$ orbitals according to Eq.~\eqref{eq: GeneralWannier1}.
Now, we give a heuristic argument suggesting that $2^{n-d} \times 2^{d-l-2}$ $p(W_1)$ orbitals trivialize the relative winding of $l$th nested Wilson loop for $l=0,1,\dots,d-2$. Here, the 0th nested Wilson loop means the conventional non-nested Wilson loop.
First, we consider the Wilson loop of $d$D $k$th-order TI defined along the $k_d$ direction. As shown in Sec.~\ref{sec: Wilsonloopwinding}, it has the relative winding number $2^{n-d} \times 2^{d-2}$.
Thus, adding $2^{n-d} \times 2^{d-2}$ $p(W_1)$ orbitals unwinds the Wilson loop completely.
Then the resulting Wannier state representation is
\ba
\label{eq: GeneralWannier2}
2^{n-d} \Big[ \bigoplus_{i=1}^{2^d} \, p(W_i) \ominus 2^{d-2} \, p(W_1) \Big].
\ea
Since there is no relative winding in the Wilson loop, the Wilson bands can be divided into two groups $P_{\rm in}$ and $P_{\rm out}$. Here, $P_{in}$ consist of the nearly flat Wilson bands near $\theta_1=0$ which come from the additional trivial orbitals.
This indicates that the Wannier centers of the Wilson bands in $P_{in}$ in the $\bb{a}_d$ direction are all zero. Then, the Wannier state representation of the Wilson bands in $P_{in}$ is equivalent to
\ba
\label{eq: GeneralWannier3}
2^{n-d} \Big[ \bigoplus_{i=1}^{2^{d-1}} \, p(W_i) \ominus 2^{d-2} \, p(W_1) \Big],
\ea
where we projected out the Wyckoff positions $W_i$ ($i=2^{d-1}+1,\dots,2^d$) whose coordinate in the $\bb{a}_d$ direction is $1/2$.
Since the Wannier state representation in Eq.~\eqref{eq: GeneralWannier3}, is equivalent to that for the $(d-1)$D $(k-1)$th-order TI, we can repeat a similar procedure to trivialize the first nested Wilson loop.
Repeating this procedure recursively, we conclude that the inner Wannier bands $P_{in}$ of the $l$th nested Wilson loop are equivalent to the $(d-l-1)$D $(k-l-1)$th-order TI which is characterized by the relative winding number $2^{n-d} \times 2^{d-l-3}$.
This completes our heuristic proof. Since adding $2^{n-d} \times 2^{d-l-2}$ $p(W_1)$ orbitals can unwind the $l$th nested Wilson loop for $l=0,1,\dots,d-2$, the total number of $p(W_1)$ orbitals required for complete trivialization is
\ba
\sum_{l=0}^{d-2} 2^{n-d} \times 2^{d-l-2} = 2^{n-d}(2^{d-1}-1).
\ea
In this way, we can understand the total number of additional orbitals required to trivialize the fragile topology in terms of the relative winding numbers in all possible nested Wilson loops.

\section{General pumping process \label{sec: Pumping}}
In this section, we describe the pumping process of inversion-symmetric TIs. The main claim is that the band topology of the ($d+1$)D $(k-1)$th-order TI can be understood in terms of the adiabatic pumping process between the $d$D $k$th-order TI and the trivial-parity insulator along the $(d+1)$th momentum direction. In particular, we discuss how the higher-order TIs with hinge states, i.e., $(d+1)$D $d$th-order TIs, can be constructed via the pumping process of $d$D $(d+1)$th-order TIs. This can be considered as the generalization of the idea proposed in Ref.~\onlinecite{wieder2018axion}.

\subsection{General pumping process}
Let us consider an adiabatic pumping process of the parity configurations of $d$D $k$th-order TIs along the $k_{d+1}$ direction between $k_{d+1}=0$ and $k_{d+1}=\pi$ planes.
We label the symmetry data in the $k_{d+1}=0$ plane as $\{n_{-}(\bb{k}^{{\rm inv}}_1)-n_{+}(\bb{k}^{{\rm inv}}_1),\dots,n_{-}(\bb{k}^{{\rm inv}}_{2^d})-n_{+}(\bb{k}^{{\rm inv}}_{2^d})\}$ and those in the $k_{d+1}=\pi$ plane as $\{n_{-}(\bb{k}^{{\rm inv}}_{2^d+1})-n_{+}(\bb{k}^{{\rm inv}}_{2^d+1}),\dots,n_{-}(\bb{k}^{{\rm inv}}_{2^{d+1}})-n_{+}(\bb{k}^{{\rm inv}}_{2^{d+1}})\}$, respectively.
At each $d$D subspace, the momentum-space symmetry data can be mapped to $2^d$ $\nu_W(k_{d+1})$ indices for $d$D inversion-symmetric insulator by using Eq.~\eqref{eq: Indices} as
\ba
\begin{pmatrix}
	\nu_{W_1}(0) \\
	\nu_{W_2}(0) \\
	\nu_{W_3}(0) \\
	\vdots \\
	\nu_{W_{2^d}}(0)
\end{pmatrix}
=A_d
\begin{pmatrix}
	n_-(\bb{k}^{{\rm inv}}_1) - n_+(\bb{k}^{{\rm inv}}_1) \\
	n_-(\bb{k}^{{\rm inv}}_2) - n_+(\bb{k}^{{\rm inv}}_2) \\
	n_-(\bb{k}^{{\rm inv}}_3) - n_+(\bb{k}^{{\rm inv}}_3) \\
	\vdots \\
	n_-(\bb{k}^{{\rm inv}}_{2^d}) - n_+(\bb{k}^{{\rm inv}}_{2^d})
\end{pmatrix},
\ea
and
\ba
\begin{pmatrix}
	\nu_{W_1}(\pi) \\
	\nu_{W_2}(\pi) \\
	\nu_{W_3}(\pi) \\
	\vdots \\
	\nu_{W_{2^d}}(\pi)
\end{pmatrix}
=A_d
\begin{pmatrix}
	n_-(\bb{k}^{{\rm inv}}_{2^d+1}) - n_+(\bb{k}^{{\rm inv}}_{2^d+1}) \\
	n_-(\bb{k}^{{\rm inv}}_{2^d+2}) - n_+(\bb{k}^{{\rm inv}}_{2^d+2}) \\
	n_-(\bb{k}^{{\rm inv}}_{2^d+3}) - n_+(\bb{k}^{{\rm inv}}_{2^d+3}) \\
	\vdots \\
	n_-(\bb{k}^{{\rm inv}}_{2^{d+1}}) - n_+(\bb{k}^{{\rm inv}}_{2^{d+1}})
\end{pmatrix}
.
\ea
Let us note that, thanks to the simple structure of inversion symmetry, the matrix $A_d$ in Eq.~\eqref{eq: Indices}, which connects the symmetry data in real and momentum spaces, satisfies the following simple recursion relation:
\ba
A_{d+1} = \frac{1}{2} \begin{pmatrix}
	A_d & A_d \\
	A_d & -A_d
\end{pmatrix},
\ea
which is proved in Appendix.~\ref{app: DerivProp}.
Then, $2^{d+1}$ $\nu_W$ indices for $(d+1)$D inversion-symmetric insulator are given by
\ba
\label{eq: nupump}
& \nu_{W_{i}} = \frac{1}{2}(\nu_{W_i}(0)+\nu_{W_i}(\pi)), \nn \\ 
& \nu_{W_{i+2^d}} = \frac{1}{2}(\nu_{W_i}(0)-\nu_{W_i}(\pi)),
\ea
where $i=1,\dots,2^d$.
Thus, the resulting phase can be diagnosed by using $\{\nu_W(0)\}$ and $\{\nu_W(\pi)\}$.

Equation~\eqref{eq: nupump} shows that any fragile TI or atomic insulators in $(d+1)$D can be understood as a pumping process between fragile TI or atomic insulators.
Since all $\nu_W$ indices for $(d+1)$D fragile TI or atomic insulator are integers, all $\nu_{W_i}(0)$ and $\nu_{W_i}(\pi)$ are also integers according to Eq.~\eqref{eq: nupump}. This means that the parity configurations at $k_{d+1}=0$ and $\pi$ correspond to either fragile TIs or atomic insulators.

In Sec.~\ref{subsec: Result2d3d}, it is mentioned that, when $n_{\rm occ}=2$, the fragile TI in higher dimensions can be obtained by a layer stacking of 2D fragile TI, and this process can also be described by a momentum-space pumping of fragile TIs.
To illustrate this idea, let us consider a trivial pumping process in which the parity configurations at $k_{d+1}=0$ and $\pi$ are identical to each other, i.e., $\nu_{W_i}(0)=\nu_{W_i}(\pi)$ for $i=1,\dots,2^d$.
In this situation, $\nu_W$ indices for $(d+1)$D inversion-symmetric insulator are given by $\nu_{W_i}=\nu_{W_i}(0)$ and $\nu_{W_{i+2^d}}=0$ for $i=1,\dots,2^d$.
This shows that a fragile TI with $n_{\rm occ}=2$ in $3$D, $\{\nu_W\}=\{-1,1,1,1,0,0,0,0\}$, for example, can be understood as a trivial pumping process of $2$D fragile TI with $\{\nu_W(0)\}=\{\nu_W(\pi)\}=\{-1,1,1,1\}$ in the $k_3$ direction, consistent with the previous description.

\subsection{Hamiltonian mapping}
Here we show that an adiabatic pumping between the $d$D $k$th-order TI and the trivial-parity insulator leads to the ($d+1$)D $(k-1)$th-order TI:
\begin{center}
\fbox{
	\centering
	\parbox{0.4\textwidth}{
		\centering
		($d+1$)D $(k-1)$th-order TI \\
		= \\
		Pumping between a $d$D $k$th-order TI and the trivial-parity insulator}
}.
\end{center}

The Hamiltonians for a class A $d$D $k$th-order TI and a $(d+1)$D $(k-1)$th-order TI are given by
\ba
\label{eq: Hpump1}
H_{{\rm A} (d,k)} = \sum_{i=1}^d \sin k_i \gamma_i - (d- \lambda + \sum_{i=1}^{d} \cos k_i) \mathcal{M},
\ea
and
\ba
H_{{\rm A} (d+1,k-1)} = \sum_{i=1}^{d+1} \sin k_i \gamma_i - (d+1 - \lambda + \sum_{i=1}^{d+1} \cos k_i) \mathcal{M},
\ea
}
where the number of $\gamma$ matrices is $q=d+k=2n+1$ and their dimension is $2^n \times 2^n$. The bulk mass term $\mathcal{M}$ and the inversion symmetry operator $I$ are given by $\mathcal{M}=I=\gamma_q$.
The parity configurations corresponding to $H_{{\rm A}(d,k)}$ and $H_{{\rm A}(d+1,k-1)}$ are $\mathcal{B}_{\gamma}(d,2^{n-1})$ and $\mathcal{B}_{\gamma}(d+1,2^{n-1})$, respectively.
The $\nu_W$ indices for $\mathcal{B}_{\gamma}(d,2^{n-1})$ are given by
\ba
2^{n-d} \{1-2^{d-1},1,\dots, 1\},
\ea
while those for the trivial parity configuration of $2^{n-1}$ bands with only positive parities are
\ba
\{-2^{n-1},0,\dots, 0\}.
\ea
Thus, the pumping between $\mathcal{B}_{\gamma}(d,2^{n-1})$ and the trivial parity configuration gives $2^{d+1}$ $\nu_W$ indices $2^{n-d-1} \{1-2^d,1,1,\dots, 1\}$, which is exactly the $\nu_W$ indices of $\mathcal{B}_{\gamma}(d+1,2^{n-1})$.

Also, one can explicitly construct a mapping between the Hamiltonians given by
\begin{gather}
\varphi_{\uparrow}[H_{{\rm A}(d,k)}](\bb{k},k_{d+1},\lambda) \nn \\
= H_{{\rm A}(d,k)}(\bb{k},\lambda-1+\cos k_{d+1}) + \sin k_{d+1} \gamma_{d+1},
\end{gather}
where $H_{{\rm A}(d,k)}(\bb{k},\lambda)$ is the Hamiltonian for a $d$D $k$th-order TI in Eq.~\eqref{eq: Hpump1}.
The map $\varphi_{\uparrow}[H]$ increases the dimension of the Hamiltonian $H$ by one, and describes the pumping process of the $d$D $k$th-order TI.
Here $\lambda$ is a parameter governing the topological phase transition between the trivial and topological phases.
$(d+k)$ should be an odd integer in order that the $d$D $k$th-order and $(d+1)$D $(k-1)$th-order TIs exist.
$H_{{\rm A}(d,k)}(\bb{k},\lambda)$ is in the trivial (topological) phase when $\lambda<0$ ($0<\lambda<2$).
For any $k_{d+1}$, the map $\varphi_{\uparrow}[H_{{\rm A}(d,k)}](\bb{k},k_{d+1},\lambda)=H_{{\rm A}(d+1,k-1)}$ preserves the bulk gap, since
\ba
\left\{\varphi_{\uparrow}[H_{{\rm A}(d,k)}](\bb{k},k_{d+1},\lambda)\right\}^2>0.
\ea
The nontrivial topology of $\varphi_{\uparrow}[H_{{\rm A}(d,k)}]$ originates from the band topology of $H_{{\rm A}(d,k)}$.
$\varphi_{\uparrow}[H_{{\rm A}(d,k)}]$ is in the topological (trivial) phase of $H_{A(d,k)}$ at $k_{d+1}=0$ ($k_{d+1}=\pi$) when $0<\lambda<2$.
Hence, the map $\varphi_{\uparrow}[H_{{\rm A}(d,k)}]$ describes the desired pumping process.

One can consider the following examples as an application of the pumping picture.
For instance, a pumping of a $1$D second-order TI described by the SSH model without chiral symmetry leads to the $2$D first-order TI, the Chern insulator.
A pumping of a $2$D third-order TI gives a $3$D second-order TI, the axion insulator~\cite{wieder2018axion}.
Finally, a pumping of a $3$D second-order TI gives a $4$D first-order TI, the $4$D Chern insulator.
In the case of other AZ classes, the Hamiltonian mapping is more restricted.
For instance, in class AI, a pumping of a $2$D third-order TI produces $\mathbb{Z}_2$ monopole nodal line semimetal~\cite{morimoto2014weyl, fang2015topological, zhao2016unified, zhao2017p, bzduvsek2017robust, song2018diagnosis, ahn2018band}.
In the case of class AII, a pumping of a $2$D third-order TI generates the doubled strong TI characterized by the $\mathbb{Z}_4$ symmetry indicator~\cite{po2017symmetry,fang2017rotation,khalaf2018symmetry}.
Finally, let us mention several related works discussing the pumping of a $2$D third-order TI. References~\cite{van2018higher,trifunovic2019geometric} discussed a charge pumping process of roto-inversion-symmetric systems. Reference~\cite{lee2019fractional} discussed a fractional charge bound to a vortex, which can be understood as a charge-pumping process, in the systems with various crystalline symmetries including inversion symmetry. Although the relevant symmetries are different from ours, all these pumping processes are essentially the pumping of a 2D third-order TI, or 2D second-order TI when the chiral symmetry is present, to a 3D second-order TI.

\section{Conclusion \label{sec: Discussions}}
In this work, we systematically study the bulk and boundary properties of fragile TIs protected by inversion symmetry.
In particular, we focus on the class A inversion-symmetric insulators which can be characterized by $(2^d+1)$ invariants, $n_{\rm occ}$ and $\{\nu_W\}$, defined in terms of symmetry data in momentum space.
Using these invariants, stable TIs, fragile TIs, and atomic insulators are distinguished. We propose an efficient method for diagnosing fragile topological phases protected by inversion symmetry, and estimate the number of parity configurations corresponding to atomic insulators, fragile TIs, and stable TIs, respectively.
Considering that the symmetry indicator cannot distinguish atomic insulators and fragile TIs, our diagnosis method goes beyond the recently proposed diagnosis schemes.

We have extended the notion of higher-order topology by generalizing the concept of $d$D $k$th-order TI to the cases with $k>d$, which is normally defined for $k \le d$. We show that $d$D $k$th-order TIs with $k>d$ have fragile band topology. In contrast to the case of stable TIs where the relative topology between a trivial and a topological insulators can be described by a Dirac Hamiltonian unambiguously, the relative topology between a fragile TI and an atomic insulator is more subtle. In particular, the Dirac Hamiltonian approach to the fragile TI is meaningful only when we specify trivial insulator taking into account the parity of occupied states at all inversion-invariant momenta. In our case, we define the trivial-parity insulator having positive-parity occupied states at all inversion-invariant momenta as a reference trivial insulator. Then, the band inversion described by the Dirac Hamiltonian gives a fragile TI unambiguously. In fact, once we have the full symmetry data at all inversion-invariant momenta, we can completely analyze the fragile band topology protected by inversion symmetry which we have demonstrated rigorously.

Moreover, we derive the bulk-boundary correspondence of $d$D $(d+1)$th-order TI, and identify a minimal fragile phase with the filling anomaly as the $d$D $(d+1)$th-order TI. In other words, a fragile TI corresponding to the $d$D $(d+1)$th-order phase can have nontrivial boundary charge. We also explain the topological origin of the charge accumulation on the boundary. We show how the mass winding, the induced current, the filling anomaly, and corner charges are related to each other in a unified way.
In the case of $d$D $k$th-order TIs with $k>(d+1)$ although there is no nontrivial charge accumulation, the net parity $\mathcal{I}$, serves as a bulk topological invariant for for any $k>d$.
In fact, the net parity characterizes the bulk topology of all inversion-symmetric insulators without gapless boundary states.
Also, from the careful study of Wilson loops and nested Wilson loops, we show that fragile topological phases are characterized by the relative winding in all possible Wilson loops.
Finally, we discuss the implication of the pumping process of inversion-symmetric insulators.

We expect that all the concepts developed in this paper, such as the mass winding, topological induced currents, the net parity, and the diagnosis of fragile topology based on symmetry data, can be applied to other topological crystalline insulators in general.
For example, the mass winding approach can be extended to fragile topological or atomic phases with rotation symmetries, considering the recent progress reported in Refs.~\onlinecite{benalcazar2017electric, benalcazar2017quantized, liu2019shift, benalcazar2019quantization}. We leave the extension of our theory to more general cases of fragile TIs protected by various space groups and magnetic space groups for future studies.

{\it Note added|.}
Recently, we noticed that a paper by Song {\it et al.}~\cite{song2019fragile} appeared on arXiv. They classified all fragile topological phases indicated by symmetry eigenvalues for the class AII and $230$ space groups. 
For inversion-symmetric cases, i.e., space group 2, linear inequalities for diagnosing fragile TIs are proposed in Ref.~\onlinecite{song2019fragile} which is equivalent to our nonlinear inequality in Fig.~\ref{fig: Classification}.

\acknowledgments
We thank Sungjoon Park for his participation in an early stage of this project.
Y.H. and J.A. were supported by IBS-R009-D1.
B.-J.Y. was supported by the Institute for Basic Science in Korea (Grant No. IBS-R009-D1) and Basic Science Research Program through the National Research Foundation of Korea (NRF) (Grant No. 0426-20190008), and the POSCO Science Fellowship of POSCO TJ Park Foundation (Grant No. 0426-20180002).
This work was supported in part by the U.S. Army Research Office under Grant No. W911NF-18-1-0137.

%\bibliography{reference.bib}

\appendix

\section{Derivation of diagnosis criterion \label{app: DerivDiag}}
In this appendix, we derive our criterion $\sum_W |\nu_W|>n_{\rm occ}$ that distinguishes the fragile topological and atomic phases. In the following, we adopt two assumptions:
(i) the band topology is not stable topological, i.e., it is either a fragile topological phase or an atomic phase,
and (ii) we consider only the band topology protected by inversion symmetry, such that inversion parity determines the band topology exactly.

Let us first prove that the phases with $\sum_{W} |\nu_W| > n_{\rm occ}$ cannot be atomic insulator.
Any atomic phase has $\mu_{W}$ integer greater than or equal to zero.
Then, the following inequality holds:
\ba
n_{\rm occ}&=\sum_{W} (\mu_{W,+}+\mu_{W,-}) \nn \\
&\ge \sum_{W} |\mu_{W,+}-\mu_{W,-}| = \sum_{W} |\nu_W|.
\ea
Thus, a phase which satisfies $\sum_{W} |\nu_W| > n_{\rm occ}$ cannot be atomic phase.

Now, let us prove that insulators satisfying $\sum_{W} |\nu_W| \le n_{\rm occ}$ are atomic insulators.
In other words, we can construct an equivalent Wannier state representation with $\mu_{W,\pm} \ge 0$.
From the $n_{\rm occ}$ occupied states, we first construct $\sum_W |\nu_W|$ Wannier states,
\ba
\label{eq: AtomicWann1}
\bigoplus_{W,\xi} \Big( \frac{s_W - \xi}{2} \Big) \nu_{W} w_{\xi}(W),
\ea
where $s_W=\text{sgn}(\nu_W)$.
By construction, we have $\mu_{W,\xi}=(\frac{s_W-\xi}{2}) \nu_W = (|\nu_W|-\xi \nu_W)/2 \ge 0$.
Then, we consider the remaining $n_{\rm occ}-\sum_W |\nu_W|$ states.
Since the Wannier state representation in Eq.~\eqref{eq: AtomicWann1} satisfies $\mu_{W,-}-\mu_{W,+}=\nu_W$ already, adding the remaining states should not change $\nu_W$ at each $W$.
It can be done by locating pairs of even- and odd-parity states at the Wyckoff positions.
It is possible to do so because $n_{\rm occ} - \sum_W |\nu_W| = \sum_{W,\xi} (1+ \xi s_W) \mu_{W,\xi}
$ and $(1\pm s_W)$ are even integers.
Thus, we can locate $\frac{1}{2}(n_{\rm occ}-\sum_W |\nu_W|)$ pairs of $s$ and $p$ orbitals at any $W$ without changing $\nu_W$.

In conclusion, an insulator satisfying $\sum_W |\nu_W|>n_{\rm occ}$ is fragile TI, and vice versa.

\section{Properties of $A_d$ and $\nu_W$ \label{app: DerivProp}}
\subsection{Representation of symmetric Wannier state}
Let us consider a Wannier state $w_{\xi}(W)$ centered at the Wyckoff position $W$ in the unit cell labeled by the lattice vector $\bb{R}$.
We denote such states as $\ket{w_{\xi,\bb{R}}(W)}$.
Inversion symmetry acts on $\ket{w_{\xi,\bb{R}}(W)}$ as
\ba
I \, \ket{w_{\xi,\bb{R}}(W)}
&= \xi \ket{w_{\xi,-\bb{R}-2W}(W)},
\ea
where $\xi=\pm 1$.
Then, the corresponding Bloch wave function $\ket{\psi_{\xi}^W (\bb{k})}$ is
\ba
\ket{\psi_{\xi}^W (\bb{k})}
=\frac{1}{\sqrt{N_{\rm cell}}} \sum_{\bb{R}} e^{i \bb{k} \cdot \bb{R}} \ket{w_{\xi,\bb{R}}(W)},
\ea
where $N_{\rm cell}$ is the number of unit cells.
Its transformation property under inversion is inherited from $\ket{w_{\xi,\bb{R}}(W)}$,
\ba
I \ket{\psi_{\xi}^W (\bb{k})}
&= \frac{1}{\sqrt{N_{\rm cell}}} \sum_{\bb{R}} e^{i \bb{k} \cdot \bb{R}} I \, \ket{w_{\xi,\bb{R}}(W)} \nn \\
&= \frac{1}{\sqrt{N_{\rm cell}}} \sum_{\bb{R}} e^{i \bb{k} \cdot \bb{R}} \xi \ket{w_{\xi,-\bb{R}-2W}(W)} \nn \\
&= \frac{1}{\sqrt{N_{\rm cell}}} \sum_{\bb{R}'} e^{-i \bb{k} \cdot (\bb{R}'+2W)} \xi \ket{w_{\xi,\bb{R}'}(W)} \nn \\
&= \xi e^{- 2i \bb{k} \cdot W} \ket{\psi_{\xi}^W (-\bb{k})}.
\ea

This means that the Bloch wave function $\ket{\psi^W_\xi(\bb{k})}$ constructed by $w_{\xi}(W_i)$ has parity $\xi \eta_{ij}$ at inversion-invariant momentum $\bb{k}^{{\rm inv}}_j$, where we define $\eta_{ij}\equiv \xi e^{-2 i \bb{k}^{{\rm inv}}_j \cdot W_i}$.
In other words, $|w_{\xi,\bb{R}}(W_i) \rangle$ generates $n_{\pm}(\bb{k}^{{\rm inv}}_j)=\frac{1}{2}(1 \pm \xi \eta_{ij})$ Bloch states with parity $\pm 1$ at $\bb{k}^{\rm inv}_j$.

\subsection{Properties of $\eta_{ij}$}
Since $\bb{k}^{{\rm inv}}_i = 2 \pi W_i$ and $\frac{2}{\pi}\bb{k}^{{\rm inv}}_i \cdot W_j \in \mathbb{Z}$, $\eta_{ij} = e^{-2 i \bb{k}^{\rm inv}_i \cdot W_j}$ satisfies
\ba
\label{eq: etaprop1}
\eta_{ij} = \eta_{ji} = \eta_{ij}^{-1},
\ea
and
\ba
\label{eq: etaprop2}
\sum_{i=1}^{2^d} \eta_{ij} = 2^d \delta_{1j}, \quad \sum_{l=1}^{2^d} \eta_{il} \eta_{lj} = 2^d \delta_{ij}.
\ea
Let first us prove $\sum_{i=1}^{2^d} \eta_{ij} = 2^d \delta_{1j}$.
When $j=1$, all $\eta_{i 1}$ are equal to $1$ because $W_1=0$ by definition.
Thus, $\sum_{i=1}^{2^d} \eta_{i 1} = \sum_{i=1}^{2^d} 1 = 2^{d}$.
When $j\ne 1$, $\eta_{ij}$ always has a partner $\eta_{i' j}=-\eta_{ij}$ for some $i'$.
Therefore, $\sum_{i=1}^{2^d} \eta_{ij}$ is zero.
Combining the above relations, we obtain $\sum_{i=1}^{2^d} \eta_{i j} = 2^{d} \delta_{j1}$.

In a similar manner, one can prove that $\sum_{l=1}^{2^d} \eta_{il} \eta_{lj}
= \sum_{l=1}^{2^d} e^{-2 i (\bb{k}^{{\rm inv}}_i-\bb{k}^{{\rm inv}}_j) \cdot W_l}
= 2^d \delta_{ij}$.
The proof is trivial when $i=j$.
When $i \ne j$, either $\bb{k}^{\rm inv}_i$ or $\bb{k}^{\rm inv}_j$ is non-zero.
Let us assume that $\bb{k}^{\rm inv}_i \ne 0$ without loss of generality.
Then, $\eta_{il} \eta_{lj}=e^{-2i(\bb{k}^{\rm inv}_i-\bb{k}^{\rm inv}_j)\cdot W_l}$ always has a partner $\eta_{il'} \eta_{l'j}=-\eta_{il} \eta_{lj}$ for some $l'$.
The summation of $\eta_{il} \eta_{lj}$ over $l$ thus vanishes.

\subsection{Derivation of the mapping $A_d$}

We derive the matrix $A_d$ that relates the number of the Wannier sates in real space to the number of occupied states of even and odd parity in momentum space.
As we show in the first subsection, $|w_{\xi,\bb{R}}(W_i) \rangle$ generates $n_{\pm}(\bb{k}^{{\rm inv}}_j)=\frac{1}{2}(1 \pm \xi \eta_{ij})$ Bloch states with parity $\pm 1$ at $\bb{k}^{\rm inv}_j$, a set of Wannier states $\sum_{W, \xi} \, \mu_{W, \xi} w_{\xi}(W)$ produces
\ba
\label{eq: numbersBZ}
n_{\pm}(\bb{k}^{{\rm inv}}_i)
&= \sum_{j=1}^{2^d} \sum_{\xi=\pm} \frac{1}{2}(1 \pm \xi \eta_{ij}) \mu_{W_j,\xi} \nn \\
&= \frac{1}{2} \sum_{j=1}^{2^d} \sum_{\xi=\pm} \mu_{W_j,\xi} \pm \frac{1}{2} \sum_{j=1}^{2^d} \eta_{ij} \sum_{\xi=\pm} \xi \mu_{W_j,\xi} \nn \\
&= \frac{1}{2} n_{\rm occ} \mp \frac{1}{2} \sum_{j=1}^{2^d} \eta_{ij} \nu_{W_j}.
\ea
Thus, we obtain
\ba
\label{eq: IndicesCov}
n_{-}(\bb{k}^{{\rm inv}}_i)-n_{+}(\bb{k}^{{\rm inv}}_i)
=\sum_{j=1}^{2^d} \eta_{ij} \nu_{W_j}.
\ea
Equivalently, $\nu_{W_i}=\sum_{j=1}^{2^d} 2^{-d} \eta_{ij} \left[n_{-}(\bb{k}^{{\rm inv}}_i)-n_{+}(\bb{k}^{{\rm inv}}_i)\right]$ because $\eta^{-1} = 2^{-d} \eta$ in the matrix notation as shown in Eqs.~\eqref{eq: etaprop1} and~\eqref{eq: etaprop2}.
We introduce a new notation $A_d$ by
\ba
\label{eq: IndicesApp}
\begin{pmatrix}
	\nu_{W_1} \\
	\nu_{W_2} \\
	\nu_{W_3} \\
	\vdots \\
	\nu_{W_{2^d}}
\end{pmatrix}
=A_d
\begin{pmatrix}
	n_{-}(\bb{k}^{{\rm inv}}_1) - n_{+}(\bb{k}^{{\rm inv}}_1) \\
	n_{-}(\bb{k}^{{\rm inv}}_2) - n_{+}(\bb{k}^{{\rm inv}}_2) \\
	n_{-}(\bb{k}^{{\rm inv}}_3) - n_{+}(\bb{k}^{{\rm inv}}_3) \\
	\vdots \\
	n_{-}(\bb{k}^{{\rm inv}}_{2^d}) - n_{+}(\bb{k}^{{\rm inv}}_{2^d})
\end{pmatrix},
\ea
where 
\ba
\label{eq: DefA}
\big[ A_d \big]_{ij}= \frac{1}{2^d} \eta_{ij} = \frac{1}{2^d} \exp(2 i \bb{k}^{{\rm inv}}_i \cdot W_j).
\ea

From Eqs.~\eqref{eq: etaprop1} and~\eqref{eq: etaprop2}, one can find that $A_d$ satisfies
\ba
\label{eq: PropA}
A_d^{-1} = 2^d A_d, \quad A_d^{T} = A_d, \quad \sum_{i=1}^{2^d} \big[ A_d \big]_{ij} = \delta_{j,1}.
\ea

For later use, we write Eq.~\eqref{eq: IndicesApp} in another form,
\ba
\label{eq: IndicesOther}
\begin{pmatrix}
	\nu_{W_1} + n_{\rm occ} \\
	\nu_{W_2} \\
	\nu_{W_3} \\
	\vdots \\
	\nu_{W_{2^d}}
\end{pmatrix}
=2 A_d
\begin{pmatrix}
	n_{-}(\bb{k}^{{\rm inv}}_1) \\
	n_{-}(\bb{k}^{{\rm inv}}_2) \\
	n_{-}(\bb{k}^{{\rm inv}}_3) \\
	\vdots \\
	n_{-}(\bb{k}^{{\rm inv}}_{2^d})
\end{pmatrix}
\ea
by using $n_{+}(\bb{k}^{{\rm inv}}_i) + n_{-}(\bb{k}^{{\rm inv}}_i) = n_{\rm occ}$.

Also, $A_{d+1}$ and $A_d$ have a recursion relation,
\ba
A_{d+1} = \frac{1}{2} \begin{pmatrix}
	A_d & A_d \\
	A_d & -A_d
\end{pmatrix}.
\ea
This can be easily proven by using $\frac{1}{\pi}(\bb{k}^{{\rm inv}}_{i+2^d} - \bb{k}^{{\rm inv}}_i)=2(W_{i+2^d} - W_{i})=(0,\dots,0,1)$ for $i=1,\dots,2^d$.

\subsection{Properties of $\nu_W$}
As $n_{\xi}(\bb{k}^{\rm inv})$ must be an integer which satisfies $0 \le n_{\xi}(\bb{k}^{{\rm inv}})\le n_{\rm occ}$, $\{\nu_W\}$ cannot take any values.
Most importantly, $\{\nu_W\}$ satisfy
\ba
\big| \sum_{j=1}^{2^d} [A_d]_{ij} \nu_{W_j} \big| = \frac{1}{2^d} n
\ea
where $n$ is an integer less than or equal to $n_{\rm occ}$, according to Eq.~\eqref{eq: numbersBZ}.
Also, $\nu_W$ indices have the following properties:
\ba
&a.
&2^{d-1} \nu_{W_{i}} \in \mathbb{Z}, \nn\\
&b.
& 2^{d-2} (\nu_{W_{i}} \pm \nu_{W_{j \ne i}}) \in \mathbb{Z},\nn\\
&c.
& |\nu_{W_i}| \le n_{\rm occ}, \nn\\
&d.
&|\nu_{W_i} \pm \nu_{W_j}| \le n_{\rm occ}, \nn\\
&e.
&\big|\sum_{W} \nu_{W}\big| \le n_{\rm occ}
\ea

\paragraph{Proof of $2^{d-1} \nu_{W_{i}} \in \mathbb{Z}$.}
As $n_{-}(\bb{k}^{{\rm inv}})$ is an integer greater than or equal to zero and $A_d$ satisfies $2^d [A_d]_{ij} = \pm 1$, $\nu_{W_i} = \sum_{j=1}^{2^d} 2 [A_d]_{ij} n_{-}(\bb{k}^{{\rm inv}}_j)$ (mod $1$) in Eq.~\eqref{eq: IndicesOther} implies
\ba
2^{d-1} \nu_{W_{i}} \in \mathbb{Z}.
\ea

\paragraph{Proof of $2^{d-2} (\nu_{W_{i}} \pm \nu_{W_{j}}) \in \mathbb{Z}$.}
From $2^{d-1} \left( [A_d]_{ij} \pm [A_d]_{i'j'} \right) = 0$ or $\pm 1$, it follows that $\nu_{W_i} \pm \nu_{W_{i'}} = \sum_{j=1}^{2^d} 2 \left( [A_d]_{ij} \pm [A_d]_{i'j} \right) n_{-}(\bb{k}^{{\rm inv}}_j)$ for $i,i' \ne 1$, which implies
\ba
2^{d-2} (\nu_{W_i} \pm \nu_{W_{i'}}) \in \mathbb{Z}
\ea
when $i,i'\ne 1$.
Similarly, $\nu_{W_1} \pm \nu_{W_{i}} = \sum_{j=1}^{2^d} 2 \left( [A_d]_{1j} \pm [A_d]_{ij} \right) n_{-}(\bb{k}^{{\rm inv}}_j) - n_{\rm occ}$ for $i \ne 1$ implies
\ba
2^{d-2} (\nu_{W_1} \pm \nu_{W_{i}}) \in \mathbb{Z}.
\ea
Thus, we obtain $2^{d-2} (\nu_{W_{i}} \pm \nu_{W_{j}}) \in \mathbb{Z}$ for $i \ne j$.
This completes the proof because when $i=i'$ the property $b$ reduces to the property $a$, which we prove above.

\paragraph{Proof of $|\nu_{W_i}| \le n_{\rm occ}$.}
Let us first prove $|\nu_{W_1}| \le n_{\rm occ}$.
From Eqs.~\eqref {eq: PropA} and~\eqref{eq: IndicesOther}, we obtain
\ba
|\nu_{W_1}|
&= \Big| 2 \sum_{j=1}^{2^d} [A_d]_{1j} n_{-}(\bb{k}^{{\rm inv}}_j)-n_{\rm occ} \Big| \nn \\
&= |2 n_{-}(\bb{k}^{{\rm inv}}_1) - n_{\rm occ}| \le n_{\rm occ}.
\ea
Thus, $|\nu_{W_1}| \le n_{\rm occ}$.
We can prove $|\nu_{W_i}| \le n_{\rm occ}$ for $i \ne 1$ in a similar way.
For $i \ne 1$, the number of $[A_d]_{ij}=\pm \frac{1}{2^d}$ is equal.
Let us relabel the $j$ indices such that $[A_d]_{i j} = \frac{1}{2^d}$ for $j=1,\dots,2^{d-1}$ and $-\frac{1}{2^d}$ for $j=2^{d-1}+1,\dots,2^d$.
Then, we obtain
\ba
|\nu_{W_i}| &= 2 \Big| \sum_{j=1}^{2^d} [A_d]_{ij} n_{-}(\bb{k}^{{\rm inv}}_j) \Big| \nn \\
&= 2^{1-d} \sum_{j=1}^{2^{d-1}} \Big| n_{-}(\bb{k}^{{\rm inv}}_j)-n_{-}(\bb{k}^{{\rm inv}}_{j+2^{d-1}}) \Big| \nn \\
&\le n_{\rm occ}.
\ea

\paragraph{Proof of $|\nu_{W_i} \pm \nu_{W_{j \ne i}}| \le n_{\rm occ}$.}
Let us first prove $|\nu_{W_1} \pm \nu_{W_{i \ne 1}}| \le n_{\rm occ}$.
We can relabel the $j$ indices for given $i$ such that $[A_d]_{i j} = \pm \frac{1}{2^d}$ for $j=1,\dots,2^{d-1}$ and $\mp \frac{1}{2^d}$ for $j=2^{d-1}+1,\dots,2^d$.
Then, $2^{d-1} \left( [A_d]_{1j} \pm [A_d]_{ij} \right)=1$ for $j=1,\dots,2^{d-1}$ and $0$ for $j=2^{d-1}+1,\dots,2^d$.
Thus,
\ba
|\nu_{W_1} \pm \nu_{W_i}|
&= \Big| 2 \sum_{j=1}^{2^d} \left( [A_d]_{1j} \pm [A_d]_{ij} \right) n_{-}(\bb{k}^{{\rm inv}}_j)-n_{\rm occ} \Big| \nn \\
&= \Big| 2^{2-d} \sum_{j=1}^{2^{d-1}} n_{-}(\bb{k}^{{\rm inv}}_j)-n_{\rm occ} \Big| \le n_{\rm occ}.
\ea

When $i,i' \ne1$ and $i\ne i'$, we can relabel the $j$ indices such that $2^{d-1} \Big( [A_d]_{ij} \pm [A_d]_{i'j} \Big)=1$ for $j=1,\dots,2^{d-2}$, $-1$ for $j=2^{d-2}+1,\dots,2^{d-1}$, and $0$ for $j=2^{d-1},\dots,2^d$.
The relabeling of $j$ indices is achieved as follows. Since both $\bb{k}^{{\rm inv}}_{i}$ and $\bb{k}^{{\rm inv}}_{i'}$ are inversion-invariant momenta, so is $\bb{k}^{{\rm inv}}_{i'}-\bb{k}^{{\rm inv}}_i$. Thus, we can write it as $\bb{k}^{{\rm inv}}_{i'}-\bb{k}^{{\rm inv}}_i=\frac{\Delta\bb{G}}{2}$ where $\Delta\bb{G}$ is a reciprocal lattice vector.
Then, $\eta_{ij} \pm \eta_{i'j}=\eta_{ij}(1 \pm \exp (i \Delta \bb{G} \cdot W_j) )$, thus $\eta_{ij} \pm \eta_{i'j}=0$ for $j=2^{d-1}+1,\dots,2^{d}$ after the relabeling of $j$ indices. From $\sum_{j} \eta_{ij} \pm \eta_{i'j}=0$, $\eta_{ij} \pm \eta_{i'j}=\pm 1$ are equally distributed, say $\eta_{ij} \pm \eta_{i'j}=1$ for $j=1,\dots,2^{d-2}$ and $-1$ for $j=2^{d-2}+1,\dots,2^{d-1}$.
Finally, 
\ba
|\nu_{W_i} \pm \nu_{W_{i'}}|
&= \Big| 2 \sum_{j=1}^{2^d} \Big( [A_d]_{ij} \pm [A_d]_{i'j} \Big) n_{-}(\bb{k}^{{\rm inv}}_j) \Big| \nn \\
&= \Big| 2^{2-d} \sum_{j=1}^{2^{d-2}} (n_{-}(\bb{k}^{{\rm inv}}_j)-n_{-}(\bb{k}^{{\rm inv}}_{j+2^{d-2}})) \Big| \nn \\
&\le n_{\rm occ}.
\ea

\paragraph{Proof of $|\sum_{W} \nu_{W}| \le n_{\rm occ}$.}
Using Eq.~\eqref{eq: PropA}, we obtain
\ba
|\sum_{W} \nu_{W}|
&= \Big| 2 \sum_{i,j=1}^{2^d} [A_d]_{ij} n_{-}(\bb{k}^{{\rm inv}}_j) - n_{\rm occ} \Big| \nn \\
&= | 2 n_{-}(\bb{k}^{{\rm inv}}_1) - n_{\rm occ}| \le n_{\rm occ}.
\ea

\section{Relations between $\nu_{W}$ indices and symmetry indicator \label{app: RelationtoSI}}
Symmetry indicators are constructed from symmetry eigenvalues in momentum space, and these distinguish stable topological phases from fragile topological and atomic phases~\cite{po2017symmetry}.
Here, we relate $\nu_W$ indices with inversion symmetry indicators.

In $d$ dimensions, one can define the $\mathbb{Z}_{2^{d-1}}$ symmetry indicator $z_{2^{d-1}}$ for inversion symmetry,
\ba
z_{2^{d-1}}
=\frac{1}{2} \sum_{i=1}^{2^d} \Big(n_{+}(\bb{k}^{{\rm inv}}_i)-n_{-}(\bb{k}^{{\rm inv}}_i)\Big) \quad (\text{mod }2^{d-1}).
\ea
To see why it is defined modulo $2^{d-1}$, we connect it to $\nu_W$ indices.
Let us recall that an atomic orbital localized at $W_i$ with inversion parity $\xi$ has parity $\xi \exp(2 i \bb{k}^{{\rm inv}}_j \cdot W_i)$ at inversion-invariant momentum $\bb{k}^{{\rm inv}}_j$.
Therefore, we have
\ba
z_{2^{d-1}}
&=\frac{1}{2} \sum_{i=1}^{2^d} \Big(n_{+}(\bb{k}^{{\rm inv}}_i)-n_{-}(\bb{k}^{{\rm inv}}_i)\Big) \nn \\
&= -\frac{1}{2} \sum_{i,j=1}^{2^d} \eta_{ij} \nu_{W_j} \nn \\ \nn
&= -2^{d-1} \nu_{W_1} \quad (\text{mod }2^{d-1}),
\ea
where Eqs.~\eqref{eq: IndicesCov} and~\eqref{eq: etaprop2} are used in the first and second lines.
Since all fragile topological and atomic phases have $\nu_W$ indices as integers so that they have trivial $z_{2^{d-1}}$ indicator, $z_{2^{d-1}}$=0, this indicator is nontrivial only for stable topological phases.

While only $\nu_{W_1}$ determines the strong index $z_{2^{d-1}}$, other $\nu_W$ indices are related to the weak indices $z^{(w)}$.
For example, in two dimensions, $\nu_{W_i} \pm \nu_{W_j}$ are constrained to be integers as shown in Appendix~\ref{app: DerivProp}.
Thus, all stable topological phases detected by $\nu_A=\frac{1}{2}$ modulo $1$ have $\nu_{B,C,D}$ as half integers.
Thus, $\nu_{B,C,D}$ do not form additional symmetry indicators.
This follows from the fact that there is no weak topology in two dimensions because there is no Wannier obstruction coming from the $1$D band topology: all $1$D insulators are atomic insulators.
On the other hand, we have three $\mathbb{Z}_2$ weak indices in three dimensions, $z^{(w)}_{i=1,2,3}$~\cite{fu2007topological}.
These generate seven distinct weak topological phases characterized by $(z^{(w)}_1,z^{(w)}_2,z^{(w)}_3)=(0,0,1),(0,1,0),(1,0,0),(1,1,0),(1,0,1),(0,1,1),(1,1,1)$.
The weak symmetry indicators,
\ba
z^{(w)}_1 
&= \frac{1}{2} \sum_{i=2,4,6,8} \left[(n_{+}(\bb{k}^{{\rm inv}}_i)-n_{-}(\bb{k}^{{\rm inv}}_i)\right]
\, (\text{mod } 2), \nn \\
z^{(w)}_2 
&=\frac{1}{2} \sum_{i=3,4,7,8} \left[(n_{+}(\bb{k}^{{\rm inv}}_i)-n_{-}(\bb{k}^{{\rm inv}}_i)\right]
\, (\text{mod } 2), \nn \\
z^{(w)}_3 
&=\frac{1}{2} \sum_{i=5,6,7,8} \left[(n_{+}(\bb{k}^{{\rm inv}}_i)-n_{-}(\bb{k}^{{\rm inv}}_i)\right]
\, (\text{mod } 2).
\ea
are expressed by $\nu_{W_1}$, $\nu_{W_2}$, $\nu_{W_3}$, and $\nu_{W_5}$ as follows:
\begin{align}
	z^{(w)}_1 
	&= 2(\nu_{W_1}-\nu_{W_2})\quad (\text{mod } 2), \nn \\
	z^{(w)}_2 
	&= 2(\nu_{W_1}-\nu_{W_3})\quad (\text{mod } 2), \nn \\
	z^{(w)}_3 
	&= 2(\nu_{W_1}-\nu_{W_5}) \quad (\text{mod } 2).
\end{align}
Note that symmetry-indicated stable topological phases can be either insulators or semimetals~\cite{po2017symmetry}. For example, in three dimensions, the phase is gapless when the strong indicator $z_4$ is odd~\cite{turner2012quantized, hughes2011inversion}.

\section{Filling anomaly from inversion of minimal number of bands corresponding to $d$D $(d+1)$th-order Dirac Hamiltonian \label{app: SumFormula}}
In this appendix, we discuss how an odd number of band inversions corresponding to $d$D $(d+1)$th-order Dirac Hamiltonian generates the filling anomaly.
Here, we mean a band inversion at inversion-invariant momentum $\bb{k}^{\rm inv}_i$ by a change of parity configuration by $\Delta n_-(\bb{k}^{\rm inv}_i)=\pm 2^{d-1}$.
A change of $\nu_W$ indices due to a band inversion at $\bb{k}^{\rm inv}_i$, $\bb{\Pi}(\bb{k}^{\rm inv}_i)$, is readily obtained from Eq.~\eqref{eq: Indices}:
\ba
\Pi_i(\bb{k}^{\rm inv}_j)=\Pi_j(\bb{k}^{\rm inv}_i)=\pm \eta_{ij}=\pm \exp(2 i \bb{k}^{\rm inv}_i \cdot W_j).
\ea
Here, $\pm$ corresponds to $\Delta n_-(\bb{k}^{\rm inv}_i)=\pm 2^{d-1}$.
For example, band inversions at $\bb{k}^{\rm inv}_1$ and $\bb{k}^{\rm inv}_2$ in $3$D correspond to $\bb{\Pi}(\bb{k}^{\rm inv}_1)=\pm \{1,1,1,1,1,1,1,1\}$ and $\bb{\Pi}(\bb{k}^{\rm inv}_2)=\pm \{1,-1,1,-1,1,-1,1,-1\}$, respectively.
Since $\Pi_i(\bb{k}^{\rm inv}_j)=\pm1$, changes in $\nu_{W_i}$ for $i=1,\dots,2^d$ are odd (even) integers simultaneously when the band inversions occur odd (even) number of times.

First, let us suppose that we have an insulator without the filling anomaly, i.e., $\nu_{W_i}$ are even integers for $i=2,\dots,2^d$.
After an odd (even) number of band inversions, $\nu_{W_i}$ become odd (even) integers for $i=2,\dots,2^d$ and this implies the presence (absence) of the filling anomaly.
Likewise, an odd (even) number of band inversions implies the absence (presence) of the filling anomaly when we start from an insulator with the filling anomaly, i.e., $\nu_{W_i}$ are odd integers for $i=2,\dots,2^d$.

Conversely, any inversion-symmetric insulators with (without) the filling anomaly become atomic insulator without (with) the filling anomaly after an odd (even) number of band inversions.
For example, when $n_{\rm occ}=6$, $\{\nu_W\}=\{-1,1,3,1,1,-1,1,1\}$ has the following parity configuration.
\ba
\begin{tabular}{cccccccc}
		$\bb{k}^{\rm inv}_1$ & $\bb{k}^{\rm inv}_2$ & $\bb{k}^{\rm inv}_3$ & $\bb{k}^{\rm inv}_4$ & $\bb{k}^{\rm inv}_5$ & $\bb{k}^{\rm inv}_6$ & $\bb{k}^{\rm inv}_7$ & $\bb{k}^{\rm inv}_8$ \\
		$-$ & $-$ & $+$ & $-$ & $-$ & $-$ & $-$ & $+$ \\
		$-$ & $-$ & $+$ & $-$ & $-$ & $-$ & $-$ & $+$ \\
		$-$ & $-$ & $+$ & $+$ & $-$ & $+$ & $+$ & $+$ \\
		$-$ & $-$ & $+$ & $+$ & $-$ & $+$ & $+$ & $+$ \\
		$-$ & $+$ & $+$ & $+$ & $+$ & $+$ & $+$ & $+$ \\
		$-$ & $+$ & $+$ & $+$ & $+$ & $+$ & $+$ & $+$
\end{tabular}.
\label{eq: parityex1}
\ea
We always re-order the energy levels so that odd-parity states are atop even-parity states.
After the band inversions at $\bb{k}^{\rm inv}_{1,2,5}$, $\nu_{W}$ indices changes by $-\sum_{i=1,2,5}\bb{\Pi}(\bb{k}^{\rm inv}_i)$, thus $\{\nu_W\}=\{-4,0,0,0,0,0,0,2\}$:
\ba
\begin{tabular}{cccccccc}
	$\bb{k}^{\rm inv}_1$ & $\bb{k}^{\rm inv}_2$ & $\bb{k}^{\rm inv}_3$ & $\bb{k}^{\rm inv}_4$ & $\bb{k}^{\rm inv}_5$ & $\bb{k}^{\rm inv}_6$ & $\bb{k}^{\rm inv}_7$ & $\bb{k}^{\rm inv}_8$ \\
	$-$ & $+$ & $+$ & $-$ & $+$ & $-$ & $-$ & $+$ \\
	$-$ & $+$ & $+$ & $-$ & $+$ & $-$ & $-$ & $+$ \\
	$+$ & $+$ & $+$ & $+$ & $+$ & $+$ & $+$ & $+$ \\
	$+$ & $+$ & $+$ & $+$ & $+$ & $+$ & $+$ & $+$ \\
	$+$ & $+$ & $+$ & $+$ & $+$ & $+$ & $+$ & $+$ \\
	$+$ & $+$ & $+$ & $+$ & $+$ & $+$ & $+$ & $+$
\end{tabular}.
\ea
Now the parity configuration is equivalent to $4 s(W_1) \oplus 2 p(W_8)$, which is an atomic insulator without the filling anomaly.
Hence, the number of band inversions modulo $2$, which is equal to $N_{\rm BI}=\sum_{i=1}^{2^d} \left\lfloor \frac{n_-(\bb{k}^{\rm inv}_i)}{2^{d-1}} \right\rfloor$ (mod $2$), determines the filling anomaly, i.e., $N_{\rm BI}=0$ ($N_{\rm BI}=1$) for the absence (presence) of the filling anomaly. Here, $\left\lfloor x \right\rfloor$ denotes the greatest integer less than or equal to $x$.

%%%%%%%%%%%%%%%%%%%%%%%%%%%%%%%%%%%%%%%%%%%%%%%%%%%%%%%%%%%%%%%%
\begin{table*}[t!]
	\caption{Classification of generic higher-order TIs/TSCs protected by inversion symmetry~\cite{khalaf2018higher}. Among the ten AZ classes, class A and AIII belong to complex AZ classes while other eight classes belong to real AZ classes. $s$ labels the AZ classes: $s=0,1$ for complex AZ classes and $s=0,1,\dots,7$ for real AZ classes. $T$, $P$, and $S$ denote time-reversal, particle-hole and chiral symmetries. $T^2$ are denoted as $0$ when the symmetry is absent. The same notation is adopted for $P$ and $S$. Inversion symmetry operator $I$ satisfies $[I,T]=\{I,P\}=0$. $\delta$ is given by dimension $d$ and order $k$, $\delta=d-k+1$ (mod $8$). Here, $k$ is an arbitrary positive integer. As for $\mathbb{Z}_2$ and $``\mathbb{Z}_2"$ classifications, $\mathbb{Z}_2$ and $``\mathbb{Z}_2"$ nontrivial classes exhibit gapless boundary states when $k \le d$. When $k=d+1$, only $``\mathbb{Z}_2"$ nontrivial classes exhibit the filling anomaly.}
	\begin{tabular}{c|cccc|cccccccc}
		\hline
		\hline
		AZ class & $s$ & $T^2$ & $P^2$ & $S^2$ & $\delta=0$ & $\delta=1$ & $\delta=2$ & $\delta=3$ & $\delta=4$ & $\delta=5$ & $\delta=6$ & $\delta=7$ \\ \hline
		A & $0$ & $0$ & $0$ & $0$ & $``\mathbb{Z}_2"$ & $0$ & $\mathbb{Z}_2$ & $0$ & $\mathbb{Z}_2$ & $0$ & $\mathbb{Z}_2$ & $0$ \\
		AIII & $1$ & $0$ & $0$ & $1$ & $0$ & $\mathbb{Z}_2$ & $0$ & $\mathbb{Z}_2$ & $0$ & $\mathbb{Z}_2$ & $0$ & $\mathbb{Z}_2$ \\
		\hline
		AI & $0$ & $1$ & $0$ & $0$ & $``\mathbb{Z}_2"$ & $0$ & $0$ & $0$ & $\mathbb{Z}_2$ & $0$ & $\mathbb{Z}_2$ & $\mathbb{Z}_2$ \\
		BDI & $1$ & $1$ & $1$ & $1$ & $\mathbb{Z}_2$ & $\mathbb{Z}_2$ & $0$ & $0$ & $0$ & $\mathbb{Z}_2$ & $0$ & $\mathbb{Z}_2$ \\
		D & $2$ & $0$ & $1$ & $0$ & $\mathbb{Z}_2$ & $\mathbb{Z}_2$ & $\mathbb{Z}_2$ & $0$ & $0$ & $0$ & $\mathbb{Z}_2$ & $0$ \\
		DIII & $3$ & $-1$ & $1$ & $1$ & $0$ & $\mathbb{Z}_2$ & $\mathbb{Z}_2$ & $\mathbb{Z}_2$ & $0$ & $0$ & $0$ & $\mathbb{Z}_2$ \\
		AII & $4$ & $-1$ & $0$ & $0$ & $``\mathbb{Z}_2"$ & $0$ & $\mathbb{Z}_2$ & $\mathbb{Z}_2$ & $\mathbb{Z}_2$ & $0$ & $0$ & $0$ \\
		CII & $5$ & $-1$ & $-1$ & $1$ & $0$ & $\mathbb{Z}_2$ & $0$ & $\mathbb{Z}_2$ & $\mathbb{Z}_2$ & $\mathbb{Z}_2$ & $0$ & $0$ \\
		C & $6$ & $0$ & $-1$ & $0$ & $0$ & $0$ & $\mathbb{Z}_2$ & $0$ & $\mathbb{Z}_2$ & $\mathbb{Z}_2$ & $\mathbb{Z}_2$ & $0$ \\
		CI & $7$ & $1$ & $-1$ & $1$ & $0$ & $0$ & $0$ & $\mathbb{Z}_2$ & $0$ & $\mathbb{Z}_2$ & $\mathbb{Z}_2$ & $\mathbb{Z}_2$ \\ \hline
		\hline
	\end{tabular}
	\label{table: AZinversionhigher}
\end{table*}
%%%%%%%%%%%%%%%%%%%%%%%%%%%%%%%%%%%%%%%%%%%%%%%%%%%%%%%%%%%%%%%%

\section{Layer construction of inversion-symmetric TIs and TSCs \label{app: LCandAZ}}

%%%%%%%%%%%%%%%%%%%%%%%%%%%%%%%%%%%%%%%%%%%%%%%%%%%%%%%%%%%%%%%%
\begin{figure}[b!]
	\centering
	\includegraphics[width=0.49\textwidth]{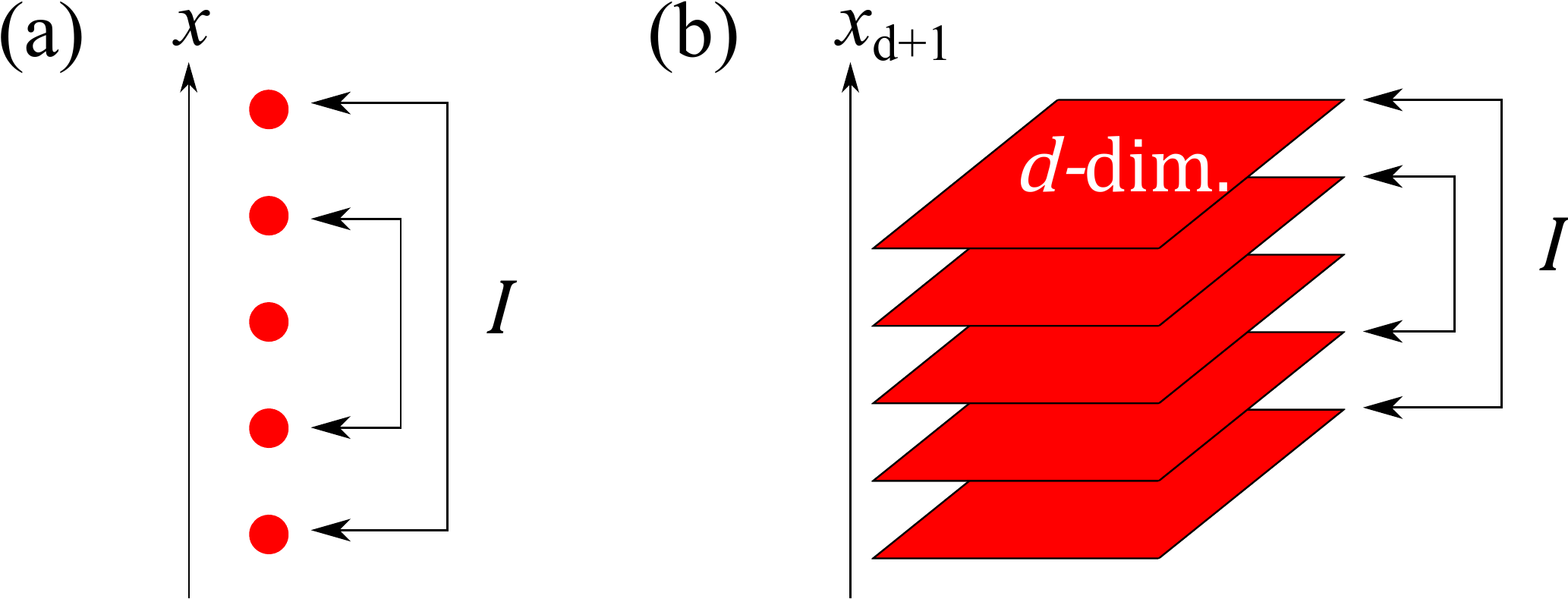}
	\caption{Layer construction.
		(a) Stacking $0$D systems gives a $1$D system. (b) Stacking $d$D systems gives a $(d+1)$D system.}
	\label{fig: layerconstruction}
\end{figure}
%%%%%%%%%%%%%%%%%%%%%%%%%%%%%%%%%%%%%%%%%%%%%%%%%%%%%%%%%%%%%%%%

Inversion-symmetric $k$th-order TIs/TSCs for $k\le d$ in $d$ dimensions can be constructed by staking first-order TIs/TSCs in $(d-k+1)$ dimensions~\cite{khalaf2018higher}.
Based on the layer construction, higher-order topological phases protected by inversion symmetry are neatly summarized in Table~\ref{table: AZinversionhigher}, which is reproduced from Ref.~\onlinecite{khalaf2018higher}.
On the other hand, it was pointed out in Refs.~\onlinecite{trifunovic2019higher,song2017topological,huang2017building} that topological phases without gapless boundary states are given by the layer construction starting from $0$D states.
In this way, any $d$D systems are constructed by stacking $0$D states (see Fig.~\ref{fig: layerconstruction}).
Here, we extended the layer construction in Ref.~\onlinecite{khalaf2018higher} to the case for $k>d$, where no gapless boundary states appear.
We focus on the case where inversion-symmetry operator $I$ satisfies the commutation relations with $T$ and $P$ in the same way as the Hamiltonian, i.e., $[I,T]=0$ and $\{I,P\}=0$.
When $I$ satisfies commutation relations different from that of $H$, only order of $k \le 2$ or 3 can be realized depending on the AZ classes, and we refer readers to \cite{khalaf2018higher, trifunovic2019higher} for discussions about such cases.

The layer construction can be employed to construct the Hamiltonian for generalized $k$th-order TIs/TSCs with $k>d$ as was done for $k\le d$ in Ref.~\onlinecite{khalaf2018higher}.
Before introducing a dimensional raising mapping of Hamiltonian based on the layer construction, we adopt the following convention:
\begin{enumerate}
	\item 
	$\mathcal{H}_{(d,k)}$ and $I_{(d,k)}$ denote the Dirac Hamiltonian and inversion symmetry operator of $d$D $k$th-order TI/TSC.
	\item
	$T_{(d,k)}$, $P_{(d,k)}$, and $S_{(d,k)}$ denote time-reversal, particle-hole, and chiral symmetry operators for $d$D $k$th-order TI/TSC.
	These non-spatial symmetries exist or not according to the AZ class.
	\item
	$\mathcal{M}_{(d,k)}$ denotes the bulk mass term and satisfies
	\ba
	[I_{(d,k)}, \mathcal{M}_{(d,k)}] =0, &\quad [T_{(d,k)}, \mathcal{M}_{(d,k)}]=0, \nn \\
	\{P_{(d,k)}, \mathcal{M}_{(d,k)}\}=0, &\quad \{S_{(d,k)}, \mathcal{M}_{(d,k)}\}=0.
	\ea
\end{enumerate}

Given a Hamiltonian $\mathcal{H}_{(d,k)}$, $\mathcal{H}_{(d+1,k+1)}$ is obtained by the dimensional raising mapping
\ba
\label{eq: LCMapping}
\mathcal{H}_{(d+1,k+1)}
= \mathcal{H}_{(d,k)} \otimes \tau_z + k_{d+1} \, \tau_x +m_{k,\bb{r}} \, \tau_y,
\ea
where $m_{k,\bb{r}}=-m_{k,-\bb{r}}$ denotes the boundary mass term.
Note that the boundary mass terms, $m_{a,\bb{r}}$, for $a=1,\dots,k-1$, in $\mathcal{H}_{(d,k)}$ are deformed to be odd under inversion symmetry in $(d+1)$D once the dimensional raising mapping Eq.~\eqref{eq: LCMapping} is done.
The symmetry operators of $(d+1)$D $(k+1)$th-order TI/TSC are given by
\ba
I_{(d+1,k+1)}
&=I_{(d,k)}\otimes \tau_z, \nn \\
T_{(d+1,k+1)}
&=T_{(d,k)}\otimes \tau_z, \nn \\
P_{(d+1,k+1)}
&=P_{(d,k)}\otimes \tau_0, \nn \\
S_{(d+1,k+1)}
&=S_{(d,k)}\otimes \tau_z,
\ea
thus, the dimensional raising mapping~\eqref{eq: LCMapping} does not change the AZ class.
Note that Eq.~\eqref{eq: LCMapping} is understood as a continuum limit of
\ba
H_{(d+1,k+1)}
=& H_{(d,k)} \otimes \tau_z - (1 - \cos k_{d+1}) \, \mathcal{M}_{(d+1,k+1)} \nn \\
&+ \sin k_{d+1} \, \tau_x,
\ea
where $\mathcal{M}_{(d+1,k+1)}=\mathcal{M}_{(d,k)}\otimes \tau_z$.
Accordingly, the Hamiltonian for $d$D $(d+n)$th-order TI/TSC, $\mathcal{H}_{(d,d+n)}$ for $n>0$ and $d\ge 1$, is constructed by a ``$0$D $n$th-order'' Hamiltonian, $\mathcal{H}_{(0,n)}$ by using the dimensional raising mapping repeatedly:
\ba
\mathcal{H}_{(d,d+n)}
&= \sum_{i=1}^d k_i \gamma_i + \lambda \mathcal{M}_{(d,d+n)} + \sum_{a=1}^{d+n-1} m_{a,\bb{r}} \gamma_{a+d}.
\ea
We define a $0$D $n$th-order Hamiltonian $\mathcal{H}_{(0,n)}$,
\ba
\mathcal{H}_{(0,n)} = \lambda \mathcal{M}_{(0,n)} + \sum_{a=1}^{n-1} m_{a,\bb{R}}\gamma_a,
\ea
where $\bb{R}$ is regarded as a set of auxiliary parameters that are promoted to $\bb{r}$ after the construction of $\mathcal{H}_{(d,d+n)}$ is done.
Thus, $m_{a,\bb{R}}$ transforms under the symmetry operators,
\ba
I:& \quad m_{a,\bb{R}} \rightarrow -m_{a,\bb{R}}, \nn \\
T:& \quad m_{a,\bb{R}} \rightarrow m_{a,\bb{R}}, \nn \\
P:& \quad m_{a,\bb{R}} \rightarrow m_{a,\bb{R}}, \nn \\
S:& \quad m_{a,\bb{R}} \rightarrow m_{a,\bb{R}}.
\ea
Now, let us regard $m_{a,\bb{R}}$ as momentum-like variables. Then, $T$ and $P$ can be regarded as space-time inversion symmetry and particle-hole symmetry combined with inversion symmetry.
Hence, the $0$D Hamiltonian $\mathcal{H}_{(0,n)}$ is identical to the inversion-symmetric $(n-1)$D Dirac Hamiltonian in the presence/absence of $IT$, $IP$, and $S$~\cite{bzduvsek2017robust}.
Therefore, $n$ is allowed when $n+s=1$ (mod $2$) in complex AZ classes, and $n+s=1,2,3,5$ (mod $8$) in real AZ classes where $s=0,1$ (mod $2$) and $s=0,\dots,7$ (mod $8$) label the complex and real AZ classes, respectively.
This is summarized in Table~\ref{table: AZinversionhigher}, and it is identical to the classification table for $k\ge d$ in Ref.~\onlinecite{khalaf2018higher}.
Therefore, our analysis generalizes Table~\ref{table: AZinversionhigher} to arbitrary non-negative integers $k$ and $d$.
Here, the inversion-protected $d$D $k$th-order topology has a $\mathbb{Z}_2$ classification, because a two-copy of $d$D $k$th-order TIs is not a $d$D $k$th-order TI anymore. Instead, it has a higher order. In our Dirac Hamiltonian construction in class A, a two-copy of $d$D $k$th-order TIs gives $d$D $(k+2)$th-order TI. For example in class AII, a two-copy of $d$D $(d+1)$th-order TIs gives $d$D $(d+5)$th-order TI.

\section{Boundary projection \label{app: Proj}}
In this appendix, we review the boundary projection introduced in Refs.~\onlinecite{khalaf2018higher, khalaf2018symmetry, geier2018second, trifunovic2019higher}.
Consider the Dirac Hamiltonian in $d$ dimensions, $\mathcal{H}(\bb{r})$:
\ba
\label{eq: DiracBP1}
\mathcal{H}(\bb{r}) = -i \sum_{i=1}^d \dr_i \gamma_i + \lambda(\bb{r}) \mathcal{M}.
\ea
Here, $\gamma_{i=1,\dots,d}$ are gamma matrices and $\mathcal{M}$ is bulk mass term.
gamma matrices $\gamma_i$ and the bulk mass term $\mathcal{M}$ satisfy $\{\gamma_i, \gamma_j\}=2\delta_{ij}$ and $\{\gamma_i,\mathcal{M}\}=0$.
We consider a system that is in the topological phase.
Let us take $\lambda>0$ in the topological phase and $\lambda<0$ in the trivial phase.
Then, $\lambda$ is positive inside the system while $\lambda$ is negative outside the system.
We set $\lambda(\bb{r})$ such that it approaches to $+1$ as $\bb{r}$ approaches the center of the system, $\bb{r}=0$, and it approaches to $-1$ as $\bb{r}$ goes to infinity.
On the boundary, $\lambda=0$ because $\lambda$ changes sign there.

For our analysis of the boundary theory of the Dirac Hamiltonian, we define $(d-1)$ orthonormal tangent basis vector $\bb{e}_{\bb{r},L=1,\dots,d-1}$ and the normal vectors $\bb{n}_{\bb{r}}$.
On the tangent space at $\bb{r}$, the Dirac Hamiltonian in Eq.~\eqref{eq: DiracBP1} is expressed as
\ba
\label{eq: DiracBP2}
\mathcal{H}_\perp(\bb{k}_S,\bb{r})
= \bb{k}_S \cdot \bb{\gamma} - i (\bb{n}_{\bb{r}} \cdot \bb{\gamma}) \dr_{x_\perp} + \lambda(x_\perp) \mathcal{M}.
\ea
Here, $\bb{k}_S$ denotes the boundary momentum normal to $\bb{n}_{\bb{r}}$, $\bb{k}_S=\bb{k}-(\bb{k} \cdot \bb{n}_{\bb{r}}) \bb{n}_{\bb{r}}$.
Also, we use the vector notation for gamma matrices, $\bb{\gamma}=(\gamma_1,\dots,\gamma_d)$, and $x_\perp$ denotes the normal coordinate at $\bb{r}$ on the boundary, $\delta x_\perp=\bb{n}_{\bb{r}} \cdot \delta \bb{r}$.
Note that the kinetic term, the first term in Eq.~\eqref{eq: DiracBP2}, can be expressed in terms of $\bb{e}_{\bb{r},L}$:
\ba
\bb{k}_S \cdot \bb{\gamma} = \sum_{L=1}^{d-1}(\bb{k}_S)_L (\bb{e}_{\bb{r},L} \cdot \bb{\gamma})
\ea
where $(\bb{k}_S)_L=\bb{k}_S \cdot \bb{e}_{\bb{r},L}$.

When the system is sufficiently large, $\mathcal{H}_\perp(\bb{k}_S,\bb{r})$ has eigenstate $\psi(\bb{k}_S,\bb{r})$ that is exponentially localized on the boundary.
One can show that the Schr\"{o}dinger equation,
\ba
\label{eq: BPEq}
\mathcal{H}_\perp(\bb{k}_S,\bb{r}) \psi(\bb{k}_S,\bb{r}) = E(\bb{k}_S) \psi(\bb{k}_S,\bb{r}),
\ea
is solved by the following ansatz:
\ba
\label{eq: BPSol}
\psi(\bb{k}_S,\bb{r}) = e^{\int^{x_\perp} dt \, \lambda(t)} \, P_+(\bb{r}) \tilde{\psi}(\bb{k}_S),
\ea
where we define projection operators $P_\pm(\bb{r})=\frac{1}{2}(\mathds{1} \mp i (\bb{n_r}\cdot \bb{\gamma}) \mathcal{M})$ is projection operator.
The choice of projection operator, $P_+(\bb{r})$ or $P_-(\bb{r})$, depends on the profile of $\lambda(\bb{r})$.
In our case, since $\lambda=-1$ outside the system, $P_+(\bb{r})$ is the proper choice.
$P_\pm(\bb{r})$ satisfies the following relation:
\begin{gather}
	P_\pm(\bb{r})^2=P_\pm(\bb{r}), \quad
	P_\pm(\bb{r}) P_\mp(\bb{r})=0, \nn \\
	[P_\pm(\bb{r}),\bb{e}_{\bb{r},L} \cdot \bb{\gamma}]=0, \quad
	P_\pm(\bb{r}) \mathcal{M} = \mathcal{M} P_\mp(\bb{r}), \nn \\
	P_\pm(\bb{r}) (\bb{n}_{\bb{r}} \cdot \bb{\gamma}) = (\bb{n}_{\bb{r}} \cdot \bb{\gamma}) P_\mp(\bb{r}).
	\label{eq: PJprop}
\end{gather}
With the help of the ansatz in Eqs.~\eqref{eq: BPSol} and~\eqref{eq: PJprop}, the eigenvalue problem in Eq.~\eqref{eq: BPEq} is reduced to the form,
\ba
P_{+}(\bb{r}) [\bb{k}_S \cdot \bb{\gamma} ] P_{+}(\bb{r}) \tilde{\psi}(\bb{k}_S) = E(\bb{k}_S) \tilde{\psi}(\bb{k}_S),
\ea
where we define the boundary Hamiltonian $h(\bb{k_S},\bb{r}) \equiv P_+(\bb{r}) \mathcal{H}_\perp(\bb{k}_S,\bb{r}) P_+(\bb{r})$.

Now, we consider the $k$th-order topology, which is defined by the presence of $(k-1)$ odd-parity mass terms $m_{a,\bb{r}} M_a$ with $a=1,\dots,k-1$, where $M_a$ satisfy $\{\gamma_i,M_a\}=0$, $\{\mathcal{M},M_a\}=0$, and $\{M_a,M_b\}=2\delta_{ab}$.
Inversion symmetry under $I=\mathcal{M}$ requires that $m_{a,\bb{r}}$ is odd under inversion, i.e., $m_{a,\bb{r}}=-m_{a,-\bb{r}}$.
Those mass terms break translation symmetry because of their $\bb{r}$ dependence, and we assume that they are finite only near the boundary.

After the boundary mass terms are included, the boundary Hamiltonian becomes
\ba
h(\bb{k_S},\bb{r})=P_+(\bb{r}) \Big[\bb{k}_S \cdot \bb{\gamma} + \sum_{a=1}^{k-1} m_{a,\bb{r}} M_a \Big] P_+(\bb{r}).
\ea
The eigenvalue problem for $h(\bb{k_S},\bb{r})$ can be solved in the following way.
The boundary-projected gamma matrices and mass matrices $\tilde{\gamma}_L$ and $\tilde{M}_a$, are defined by the projection,
\ba
\tilde{\gamma}_{L} 
&= P_{+}(\bb{r})(\bb{e}_{\bb{r},L} \cdot \bb{\gamma}) P_{+}(\bb{r})
=(\bb{e}_{\bb{r},L} \cdot \bb{\gamma}) P_{+}(\bb{r}),\nn \\
\tilde{M}_a 
&= P_{+}(\bb{r}) M_a P_+(\bb{r})
=M_a P_+(\bb{r}),
\ea
and satisfy $\{\tilde{\gamma}_L,\tilde{\gamma}_M\}=2\delta_{LM} P_+(\bb{r})$, $\{\tilde{M}_a,\tilde{M}_b\} = 2\delta_{ab} P_+(\bb{r})$ and $\{\tilde{\gamma}_L,\tilde{M}_a\}=0$.
Hence, energy eigenvalues are given by $E(\bb{k}_S,\bb{r})=\pm \sqrt{\bb{k}_S^2 + \sum_{a=1}^{k-1} m_{a,\bb{r}}^2}$.

So far, the boundary-projected mass terms $m_{a,\bb{r}} \tilde{M}_a$ are generated by the position-dependent mass terms $m_{a,\bb{r}} M_a$.
Alternatively, $m_{a,\bb{r}} \tilde{M}_a$ can be generated by a constant perturbation $\mathcal{H}_{\rm pert}$~\cite{geier2018second, trifunovic2019higher},
\ba
\mathcal{H}_{\rm pert} = i \sum_{i=1}^{d} \sum_{a=1}^{k-1} \Delta_{i,a} \, \gamma_i M_a \mathcal{M}.
\ea
After the boundary projection, $\mathcal{H}_{\rm pert}$ is reduced to
\ba
P_+(\bb{r}) \, \mathcal{H}_{\rm pert} \, P_+(\bb{r}) = \sum_{i=1}^{d} \sum_{a=1}^{k-1} \Delta_{i,a} \, \bb{n}_{\bb{r},i} \tilde{M}_a.
\ea
We use this method in numerical calculations.

\section{Induced current \label{app: Current}}
In this appendix, we derive the induced current $J^\mu_d$ in Eq.~\eqref{eq: top_current}.
Let us consider the Dirac Hamiltonian coupled to $(d+1)$ slowly varying background scalar fields $m_{a,\bb{r}}$,
\ba
\mathcal{H}(\bb{r}) = - i \sum_{i=1}^d \dr_{i} \gamma_i + \sum_{a=1}^{d+1} m_{a,\bb{r}} M_a,
\ea
where $\gamma$ denotes gamma matrix and $M_a=\gamma_{a+d}$.
gamma matrices satisfy $\{ \gamma_i, \gamma_j \} = 2 \delta_{ij}$ and are chosen to be $\gamma_1 \gamma_2 \dots \gamma_{2d+1}=\chi_d \mathds{1}_{2^d}$ where $\chi_d= i^d$.
Thus, $\text{Tr}[\gamma_{i_1} \gamma_{i_2} \dots \gamma_{i_{2d+1}}] = \epsilon_{i_1 i_2 \dots i_{2d+1}} \text{Tr}[\chi_d \mathds{1}_{2^d}]=\epsilon_{i_1 i_2 \dots i_{2d+1}} \chi_d 2^d$ where $\epsilon_{1 2 \dots 2d+1}=1$.
The Lagrangian is given by $\mathcal{L}=i \Psi^\dagger \dr_t \Psi - \bar{\Psi} \mathcal{H} \Psi$ where $\Psi$ denotes the Dirac spinor and $\bar{\Psi}=\Psi^\dagger \gamma^0$:
\ba
\label{eq: Lagrangian}
\mathcal{L} = \bar{\Psi} \Big[ i \slashed{\dr} - \sum_{a=1}^d m_{a,\bb{r}} \gamma^{a+d} - m_{d+1,\bb{r}} \mathds{1}_{2^d} \Big],
\ea
where we use the Feynman slash notation, $\slashed{A}=\sum_{\mu=0}^{d} A_\mu \gamma^\mu$, and define new gamma matrices $\gamma^0=\gamma_{2d+1}$ and $\gamma^{i \ne 0} = \gamma^0 \gamma_i$.
$\gamma^\mu$ satisfies $\{ \gamma^\mu, \gamma^\nu \}=-2 \eta^{\mu \nu}$ where $\eta_{\mu \nu}=\text{diag}(-1,1,\dots,1)$ for $\mu,\nu=0,1,\dots,2d$, and $\text{Tr}[\gamma^{\mu_1} \gamma^{\mu_2} \dots \gamma^{\mu_{2d+1}}] = \epsilon^{\mu_1 \mu_2 \dots \mu_{2d+1}} \chi_d (-1)^d 2^d$, where $\epsilon^{0 1 2 \dots 2d}=1$.

We calculate the current density $\braket{j^\mu_d(x)}$ by using
\ba
\braket{j^\mu_d(x)}
&= \braket{\bar{\Psi}(x) \gamma^\mu \Psi(x)} \nn \\
&= - \lim_{y \rightarrow x} \text{Tr} \Big[ \braket{T \Psi(x) \bar{\Psi}(y)} \gamma^\mu \Big] \nn \\
&= \lim_{y \rightarrow x} \text{Tr} \Big[ i G(x,y) \gamma^\mu \Big],
\ea
where $G(x,y) = i \braket{T \Psi(x) \bar{\Psi}(y)}$ is the propagator.
Here, we denote $x^\mu=(x^0,\bb{x})$ as $x$ for simplicity.
The propagator $G(x,y)$ satisfies
\ba
\Big[ - i \slashed{\dr}_x + \sum_{a=1}^d m_{a,\bb{x}} \gamma^{a+d} + m_{d+1,\bb{x}} \Big]G(x,y) = \delta^{(d+1)}(x-y).
\ea
Regarding $G(x,y)$ as a function of $x$ and $(x-y)$, we Fourier transform the propagator by~\cite{midorikawa1985fractional}
\ba
G(x,y) = \int \frac{d^{d+1} p}{(2\pi)^{d+1}} \, e^{i p \cdot (x-y)} \tilde{G}(x,p).
\ea
Note that translation symmetry is broken by $m_{a,\bb{r}}$, so $G(x,y)$ can not be a function of $(x-y)$ only.
Then, we expand $\tilde{G}(x,p)$ by its spatial derivative terms:
\ba
\label{eq: DerivExpansion}
\tilde{G}(x,p)
=& \frac{1}{\slashed{p} + \sum_{a=1}^d m_{a,\bb{x}} \gamma^{a+d} + m_{d+1,\bb{x}} - i \slashed{\dr}_x} \nn \\
=& \tilde{G}_0(x,p) + \tilde{G}_0(x,p) i \slashed{\dr}_x \tilde{G}_0(x,p) \nn \\
& + \tilde{G}_0(x,p) i \slashed{\dr}_x \tilde{G}_0(x,p) i \slashed{\dr}_x \tilde{G}_0(x,p) + \cdots \nn \\
=& \tilde{G}_0(x,p) \sum_{k=0}^{\infty} \big[ i \slashed{\dr} \, \tilde{G}_0(x,p) \big]^k.
\ea
Here, we define $\tilde{G}_0(x,p)$: 
\ba
\tilde{G}_0(x,p)
&= \frac{1}{\slashed{p} + \sum_{a=1}^d m_{a,\bb{x}} \gamma^{a+d} + m_{d+1,\bb{x}}} \nn \\
&= \frac{1}{p^2 + m_{\bb{x}}^2} (m_{d+1,\bb{x}} - \slashed{p}-\sum_{a=1}^d m_{a,\bb{x}} \gamma^{a+d}) \nn \\
&= \frac{1}{p^2 + m_{\bb{x}}^2} (M_{\bb{x}} - \slashed{p}),
\ea
where $m_{\bb{x}}^2 = \sum_{a=1}^{d+1} m_{a,\bb{x}}^2$ and $M_{\bb{x}} = m_{d+1,\bb{x}} - \sum_{a=1}^d m_{a,\bb{x}} \gamma^{a+d}$.
The lowest-order contribution comes from $k=d$ term in Eq.~\eqref{eq: DerivExpansion}.
Up to the lowest-order in the derivative, we have
\begin{widetext}
	\ba
	\braket{j^\mu_d(x)}
	&= i^{d+1} \int \frac{d^{d+1} p}{(2\pi)^{d+1}} \, \frac{1}{(p^2+m_{\bb{x}}^2)^{d+1}} \, \text{Tr} \Big[ (M_{\bb{x}} - \slashed{p}) \slashed{\dr} M_{\bb{x}} \slashed{\dr} M_{\bb{x}} \dots \slashed{\dr} M_{\bb{x}} \gamma^\mu \Big] \nn \\
	&= i^{d+1} \int \frac{d^{d+1} p}{(2\pi)^{d+1}} \, \frac{1}{(p^2+m_{\bb{x}}^2)^{d+1}} \, \text{Tr} \Big[ M_{\bb{x}} \slashed{\dr} M_{\bb{x}} \slashed{\dr} M_{\bb{x}} \dots \slashed{\dr} M_{\bb{x}} \gamma^\mu \Big].
	\ea
From
\begin{gather}
	\text{Tr} \Big[ M_{\bb{x}} \gamma^{\mu_1} \dr_{\mu_1} M_{\bb{x}} \gamma^{\mu_2} \dr_{\mu_2} M_{\bb{x}} \dots \gamma^{\mu_d} \dr_{\mu_d} M_{\bb{x}} \gamma^\mu \Big] \nn \\
	= (-1)^{d(d+1)/2} 2^d \chi_d \, \epsilon^{a_1 a_2 \dots a_{d+1}} \, \epsilon^{\mu \mu_1 \dots \mu_d} \, m_{a_1,\bb{x}} \dr_{\mu_1} m_{a_2,\bb{x}} \dots \dr_{\mu_d} m_{a_{d+1},\bb{x}},
	\end{gather}
we obtain
\ba
\braket{j^\mu_d(x)}
&= i^{d+1} (-1)^{d(d+1)/2} 2^d \chi_d \, \int \frac{d^{d+1} p}{(2\pi)^{d+1}} \, \frac{1}{(p^2+m_{\bb{x}}^2)^{d+1}} \, \epsilon^{a_1 a_2 \dots a_{d+1}} \, \epsilon^{\mu \mu_1 \dots \mu_d} \, m_{a_1,\bb{x}} \dr_{\mu_1} m_{a_2,\bb{x}} \dots \dr_{\mu_d} m_{a_{d+1},\bb{x}} \nn \\
&= - i^d (-1)^{d(d+1)/2} \frac{\chi_d}{d! {\rm Area} (S^d)} \epsilon^{a_1 a_2 \dots a_{d+1}} \, \epsilon^{\mu \mu_1 \dots \mu_d} \, \hat{m}_{a_1,\bb{x}} \dr_{\mu_1} \hat{m}_{a_2,\bb{x}} \dots \dr_{\mu_d} \hat{m}_{a_{d+1},\bb{x}} \nn \\
&= - (-1)^{d(d-1)/2} \frac{1}{d! {\rm Area} (S^d)} \epsilon^{a_1 a_2 \dots a_{d+1}} \, \epsilon^{\mu \mu_1 \dots \mu_d} \, \hat{m}_{a_1,\bb{x}} \dr_{\mu_1} \hat{m}_{a_2,\bb{x}} \dots \dr_{\mu_d} \hat{m}_{a_{d+1},\bb{x}},
\ea
\end{widetext}
where $\hat{m}_{a,\bb{x}} = \frac{m_{a,\bb{x}}}{m_{\bb{x}}}$ and ${\rm Area}(S^d)=\frac{2 \pi^{(d+1)/2}}{\gamma(\frac{d+1}{2})}$.
Hence, the induced current $\braket{j^\mu_d(x)}$ in $d$ dimensions (denoted as $J^\mu_d$ in the main text) is given by the winding number density of mass terms $m_{a,\bb{r}}$ where $a=1,\dots,d$.

The induced current density can be calculated either using the bulk Hamiltonian or the boundary Hamiltonian.
There is a caveat because the convention for the gamma matrices may change by the projection, which can lead a different choice of $\chi_d$ on the boundary while we take $\chi_d=i^d$ in the bulk.
Therefore, we need to carefully calculate $\tilde{\chi}_{d-1}\equiv 2^{-(d-1)}\text{Tr}[\tilde{\gamma}_1 \dots \tilde{\gamma}_{d-1} \tilde{M}_1 \dots \tilde{M}_d ] $ on the boundary, where $\tilde{\gamma}$ and $\tilde{M}$ are gamma matrices and mass matrices projected to boundary defined in Appendix.~\ref{app: Proj}.
It turns out that $\tilde{\chi}_{d-1}=-\chi_{d-1}$ as we show now.

Now, let us turn to the boundary theory when a domain between trivial and topological phases is formed.
The boundary Hamiltonian $\bb{h}(\bb{k}_S,\bb{r})$ is
\ba
h(\bb{k}_S,\bb{r})=P_+(\bb{r}) \Big[\bb{k}_S \cdot \bb{\gamma} + \sum_{a=1}^d m_{a,\bb{r}} M_a \Big] P_+(\bb{r}),
\ea
which is obtained by the projection to the boundary, introduced in Appendix~\ref{app: Proj}.
Since $h(\bb{k}_S,\bb{r})$ has $(d-1)$ kinetic terms and $d$ mass terms, the induced current $\braket{j^\mu_{d-1}(x)}$ on the $(d-1)$D boundary is given by repeating the above calculation.
The minus sign can be traced in the following way.
From the definition of $\tilde{\gamma}_{L=1,\dots,d-1}=(\bb{e}_{\bb{r},L} \cdot \bb{\gamma}) P_+(\bb{r})$ and $\tilde{M}_{a=1,\dots,d}=M_a P_+(\bb{r})$, where $\bb{e}_{\bb{r},L}$ are orthonormal bases at $\bb{r}$, we show $\text{Tr}[\tilde{\gamma}_1 \dots \tilde{\gamma}_{d-1} \tilde{M}_1 \dots \tilde{M}_d ] = -2^{d-1} \chi_{d-1}$ explicitly:
\begin{widetext}
	\ba
	\text{Tr}\left[\tilde{\gamma}_1 \dots \tilde{\gamma}_{d-1} \tilde{M}_1 \dots \tilde{M}_d \right]
	&=
	(\bb{e}_{\bb{r},1})_{\mu_1} \dots (\bb{e}_{\bb{r},d-1})_{\mu_{d-1}} \text{Tr}\left[\gamma_{\mu_1} P_+(\bb{r}) \dots \gamma_{\mu_{d-1}} P_+(\bb{r}) M_1 P_+(\bb{r}) \dots M_d P_+(\bb{r}) \right] \nn \\
	&=
	(\bb{e}_{\bb{r},1})_{\mu_1} \dots (\bb{e}_{\bb{r},d-1})_{\mu_{d-1}} \text{Tr}\left[\gamma_{\mu_1} \dots \gamma_{\mu_{d-1}} M_1 \dots M_d P_+(\bb{r}) \right] \nn \\
	&=-\frac{i}{2}
	(\bb{e}_{\bb{r},1})_{\mu_1} \dots (\bb{e}_{\bb{r},d-1})_{\mu_{d-1}} \text{Tr}\left[\gamma_{\mu_1} \dots \gamma_{\mu_{d-1}} M_1 \dots M_d (\bb{e}_{\bb{r},0} \cdot \bb{\gamma}) M_{d+1} \right] \nn \\
	&=\frac{i}{2}
	(\bb{e}_{\bb{r},0})_{\mu_0} \dots (\bb{e}_{\bb{r},d-1})_{\mu_{d-1}} \text{Tr}\left[ \gamma_1 \dots \gamma_d M_1 \dots M_{d+1} \right] \nn \\
	&=\frac{i}{2} \epsilon_{0 1 \dots d-1} \text{Det} \, (\bb{e}_{\bb{r}}) \, 2^d \chi_d \nn \\
	&=-2^{d-1} \chi_{d-1},
	\ea
	where we define $\bb{e}_{\bb{r},0} \equiv \bb{n}_{\bb{r}}$ in the third line, $\text{Det} (\bb{e}_{\bb{r}})$ is the determinant of $\bb{e}_{\bb{r}}$ in the matrix notation.
\end{widetext}

Finally, let us mention that $d$D $(d+1)$th-order phase is allowed for classes A, AI, AII, BDI, and D. However, when the chiral or particle-hole symmetry is exists, the induced current is zero. Thus, charge accumulation on boundary occurs only for classes A, AI, and AII.

\section{Review of the Wilson loop \label{app: Wilson}}
In this appendix, we review the Wilson loop and its symmetry transformation.
We closely follow Ref.~\onlinecite{alexandradinata2016topological}.
In addition, we introduce a more general method which exploits parallel transport and symmetry transformation of Berry connection.

\subsection{Tight-binding notation}
In the tight-binding model, Hamiltonian is spanned by a basis set $\ket{\bb{R} \, \alpha}$.
We denote $\bb{R}$ and $\bb{x}_\alpha$ as the unit cell position and the sublattice position of the orbital $\alpha$:
\ba
\hat{H}
&= \sum_{\bb{R}, \bb{R}'} \sum_{\alpha,\beta} \ket{\bb{R} \, \alpha} \, H(\bb{R} - \bb{R}')_{\alpha \beta} \, \bra{\bb{R}' \, \beta} \nn \\
&= \sum_{\bb{k}} \sum_{\alpha,\beta} \ket{\bb{k} \, \alpha} \, H(\bb{k})_{\alpha \beta} \, \bra{\bb{k} \, \beta}.
\ea
Here, $\ket{\bb{k} \, \alpha}$ is the Fourier transformation of $\ket{\bb{R} \, \alpha}$,
\ba
\label{eq: BasisTB}
&\ket{\bb{k} \, \alpha}
\equiv \frac{1}{\sqrt{N_{\rm cell}}} \sum_{\bb{R}} e^{i \bb{k} \cdot (\bb{R} + \bb{x}_\alpha)} \, \ket{\bb{R} \, \alpha},
\ea
where $N_{\rm cell}$ is the number of unit cells in the system.
These basis states are orthonormal: $\braket{\bb{R} \, \alpha | \bb{R}' \, \beta} = \delta_{\bb{R},\bb{R}'} \delta_{\alpha \beta}$ and $\braket{\bb{k} \, \alpha | \bb{k}' \, \beta} = \delta_{\bb{k},\bb{k}'} \delta_{\alpha \beta}$.

When the orbitals are located at $\bb{x}_\alpha \ne 0$, $\ket{\bb{k} \, \alpha}$ is not periodic in the Brillouin zone.
\ba
\ket{\bb{k} + \bb{G} \, \alpha}
= e^{i \bb{G} \cdot \bb{x}_\alpha} \ket{\bb{k} \, \alpha} = \ket{\bb{k} \, \beta} \, V(\bb{G})^{-1}_{\beta \alpha},
\ea
where $V(\bb{k})_{\alpha \beta} = e^{-i \bb{k} \cdot \bb{x}_\alpha} \delta_{\alpha \beta}$ (note that we take a convention
different from the one in Ref.~\onlinecite{alexandradinata2016topological}).
Diagonalizing $H(\bb{k})_{\alpha \beta}$, we obtain energy eigenstates $\ket{n, \bb{k}}_\alpha$, where
\ba
H(\bb{k})_{\alpha \beta} \, \ket{n, \bb{k}}_\beta
= E_n(\bb{k}) \, \ket{n, \bb{k}}_\alpha.
\ea
The energy eigenstates $\ket{n, \bb{k}}$ satisfy
\ba
\label{eq: PropEf}
\braket{n, \bb{k} | m, \bb{k}} = \delta_{n m}, \quad 
\sum_{n=1}^{n_{\rm tot}} \ket{n, \bb{k}}_\alpha \bra{n, \bb{k}}_\beta = \delta_{\alpha\beta},
\ea
where $n_{\rm tot}$ is the rank of Hamiltonian $H(\bb{k})_{\alpha \beta}$, which is also the number of orbitals: $\sum_\alpha 1 = n_{\rm tot}$.
Since $H(\bb{k} + \bb{G}) = V(\bb{G}) \, H (\bb{k}) \, V(\bb{G})^{-1}$, the ``periodic gauge'' is defined by
\ba
\label{eq: PGauge}
\ket{n, \bb{k} + \bb{G}} = V(\bb{G}) \, \ket{n, \bb{k}}.
\ea

\subsection{Symmetry transformation}
Consider a symmetry operation $\hat{\sigma}$: $\bb{r} \rightarrow D_{\sigma} \bb{r} + \bb{d}$.
$D_{\sigma}$ is representation of the point-group part of $\hat{\sigma}$, which is an element of orthogonal group, $O(d)$.
It acts on the basis $\ket{\bb{R} \, \alpha}$ as follows:
\ba
\hat{\sigma} \, \ket{\bb{R} \, \alpha}
&= \ket{\tilde{\bb{R}} \, \tilde{\alpha}} \, U(\sigma)_{\tilde{\alpha} \alpha},
\ea
where $\tilde{\bb{R}} = D_{\sigma} (\bb{R} + \bb{x}_{\alpha}) + \bb{d} - \bb{x}_{\tilde{\alpha}}$.
Since $U(\sigma)$ acts on sublattice degrees of freedom,
$U(\sigma)$ is bijective and independent of $\bb{R}$.
%Note that $\tilde{\bb{R}}$ depends on $\sigma$, $\bb{R}$, and $\alpha$.
Using the fact that $\bb{k} \cdot \bb{x} = D_{\sigma}\bb{k} \cdot D_{\sigma}\bb{x}$ which follows from $D_{\sigma} \in O(d)$, we obtain the symmetry transformation of $|\bb{k}, \alpha \rangle$ under $\hat{\sigma}$:
\ba
\label{eq: SymmUni}
\hat{\sigma} \, \ket{\bb{k} \, \alpha}
&= \frac{1}{\sqrt{N_{\rm cell}}} \sum_{\bb{R}} e^{i \bb{k} \cdot (\bb{R} + \bb{x}_{\alpha})} \, \hat{\sigma} \, \ket{\bb{R} \, \alpha} \nn \\
&= \frac{1}{\sqrt{N_{\rm cell}}} \sum_{\bb{R}} e^{i D_{\sigma}\bb{k} \cdot D_{\sigma}(\bb{R} + \bb{x}_{\alpha})} \, \ket{\tilde{\bb{R}} \, \beta} \, U(\sigma)_{\beta \alpha} \nn \\
&= e^{- i D_{\sigma}\bb{k} \cdot \bb{d}} \, \frac{1}{\sqrt{N_{\rm cell}}} \sum_{\tilde{\bb{R}}} e^{i D_{\sigma}\bb{k} \cdot (\tilde{\bb{R}} + \bb{x}_{\beta})} \, \ket{\tilde{\bb{R}} \, \beta} \, U(\sigma)_{\beta \alpha} \nn \\
&= e^{-i D_{\sigma}\bb{k} \cdot \bb{d}} \, \ket{D_{\sigma}\bb{k} \, \beta} \, U(\sigma)_{\beta \alpha} \nn \\
&\equiv \ket{D_{\sigma}\bb{k} \, \beta} \, U_{\sigma}(\bb{k})_{\beta \alpha},
\ea
where we define $U_{\sigma}(\bb{k}) = e^{- i D_{\sigma}\bb{k} \cdot \bb{d}} \, U(\sigma)$.
From the condition that $\hat{H}$ is symmetric under $\hat{\sigma}$, $\hat{H}=\hat{\sigma} \hat{H} \hat{\sigma}^{-1}$, we obtain
\ba
\label{eq: Huni}
H(D_{\sigma}\bb{k}) = U_{\sigma} (\bb{k}) \, H(\bb{k}) \, U_{\sigma}(\bb{k})^{-1}.
\ea
At high-symmetry momenta $\bb{k}^{\rm inv}_\sigma = D_{\sigma} \bb{k}^{\rm inv}_\sigma + \bb{G}(\bb{k}^{\rm inv}_\sigma)$,
$H(\bb{k}^{\rm inv}_\sigma)$ and $V(\bb{G}(\bb{k}^{\rm inv}_\sigma)) \, U(\bb{k}^{\rm inv}_\sigma)$ commute,
\ba
\label{eq: UniInv}
\big[ H(\bb{k}^{\rm inv}_\sigma), V(\bb{G}(\bb{k}^{\rm inv}_\sigma)) \, U(\bb{k}^{\rm inv}_\sigma) \big] =0.
\ea
For inversion symmetry, the inversion-invariant momenta are given by $\bb{k}^{\rm inv} = -\bb{k}^{\rm inv}$ (mod $\bb{G}$).

From Eq.~\eqref{eq: Huni}, we can find an energy eigenstate at $D_{\sigma}\bb{k}$, $|n, D_{\sigma}\bb{k}\rangle$, as a symmetry partner of $|n, \bb{k}\rangle$ up to unitary transformation $B_\sigma$, the sewing matrix:
\ba
\label{eq: UniPair}
\ket{n, D_{\sigma} \bb{k}} = U_{\sigma}(\bb{k}) \, \ket{m, \bb{k}} \, B_{\sigma}(\bb{k})_{m n}.
\ea
$B_\sigma$ appears here because a symmetry transformation can lead to a linear combination of energy eigenstates when they are degenerate.
It is straightforward to show that the sewing matrix $B_{\sigma}(\bb{k})$,
\ba
B_{\sigma}(\bb{k})_{nm} = \bra{n, \bb{k}} U_\sigma(\bb{k})^\dagger \ket{m, D_\sigma \bb{k}},
\ea
is a unitary matrix, $B_{\sigma}(\bb{k}) \in U(n_{\rm occ})$, and is periodic in the Brillouin zone:
\begin{gather}
	\sum_{p=1}^{n_{\rm occ}} B_{\sigma}(\bb{k})_{n p} [B_{\sigma}(\bb{k})^\dagger]_{p m} = \delta_{nm}, \\
	B_\sigma(\bb{k}+\bb{G}) = B_\sigma(\bb{k}).
\end{gather}

Now, we investigate the transformation of the non-Abelian Berry connection $[ a_{i}(\bb{k}) ]_{n m}$ defined by the occupied states $|n, \bb{k} \rangle$:
\ba
\label{eq: Berrya}
[ a_{i}(\bb{k}) ]_{n m}
&= \braket{n, \, \bb{k} | \, i \nabla_i \, | m, \, \bb{k}},
\ea
where $\nabla_i$ denotes $\frac{\dr}{\dr k_i}$.
Note that $a_i(\bb{k})$ is Hermitian.
\ba
[ a_i(\bb{k}) ]_{n m}
&= [ a_i(\bb{k}) ]^{*}_{m n}.
\ea
The Berry connection transforms under the gauge transformation $\ket{n,\bb{k}} \rightarrow \ket{n, \bb{k}}_G = \ket{m, \bb{k}} \, G(\bb{k})_{m n}$ of the occupied states as
\ba
\label{eq: BerryGauge}
a^{G}_i(\bb{k}) = G^{\dagger}(\bb{k}) \, a_i(\bb{k}) \, G(\bb{k}) + i G^{\dagger}(\bb{k}) \, \nabla_{i} G(\bb{k}).
\ea
From Eqs.~\eqref{eq: UniPair} and~\eqref{eq: Berrya},
we obtain the symmetry transformation of $a_i(\bb{k})$ under $\hat{\sigma}$,
\begin{gather}
	[ D_{\sigma}^{-1} ]_{i j} \, \left[a_{j}(D_{\sigma}\bb{k}) - \bb{d}_j \mathds{1} \right] \nn \\
	= [ B_{\sigma}(\bb{k}) ]^{\dagger} \, a_{i}(\bb{k}) \, B_{\sigma}(\bb{k})
	+ i \, [ B_{\sigma}(\bb{k}) ]^{\dagger} \, \nabla_i \, B_{\sigma}(\bb{k}),
	\label{eq: BerrySymmetry}
\end{gather}
where $\mathds{1}$ denotes an identity matrix in the occupied subspace.
The results Eqs.~\eqref{eq: BerryGauge} and~\eqref{eq: BerrySymmetry} are used to derive the symmetry transformations of the Wilson loop.

\subsection{Wilson line}
%%%%%%%%%%%%%%%%%%%%%%%%%%%%%%%%%%%%%%%%%%%%%%%%%%%%%%%%%%%%%%%%
\begin{figure}[b!]
	\centering
	\includegraphics[width=0.35\textwidth]{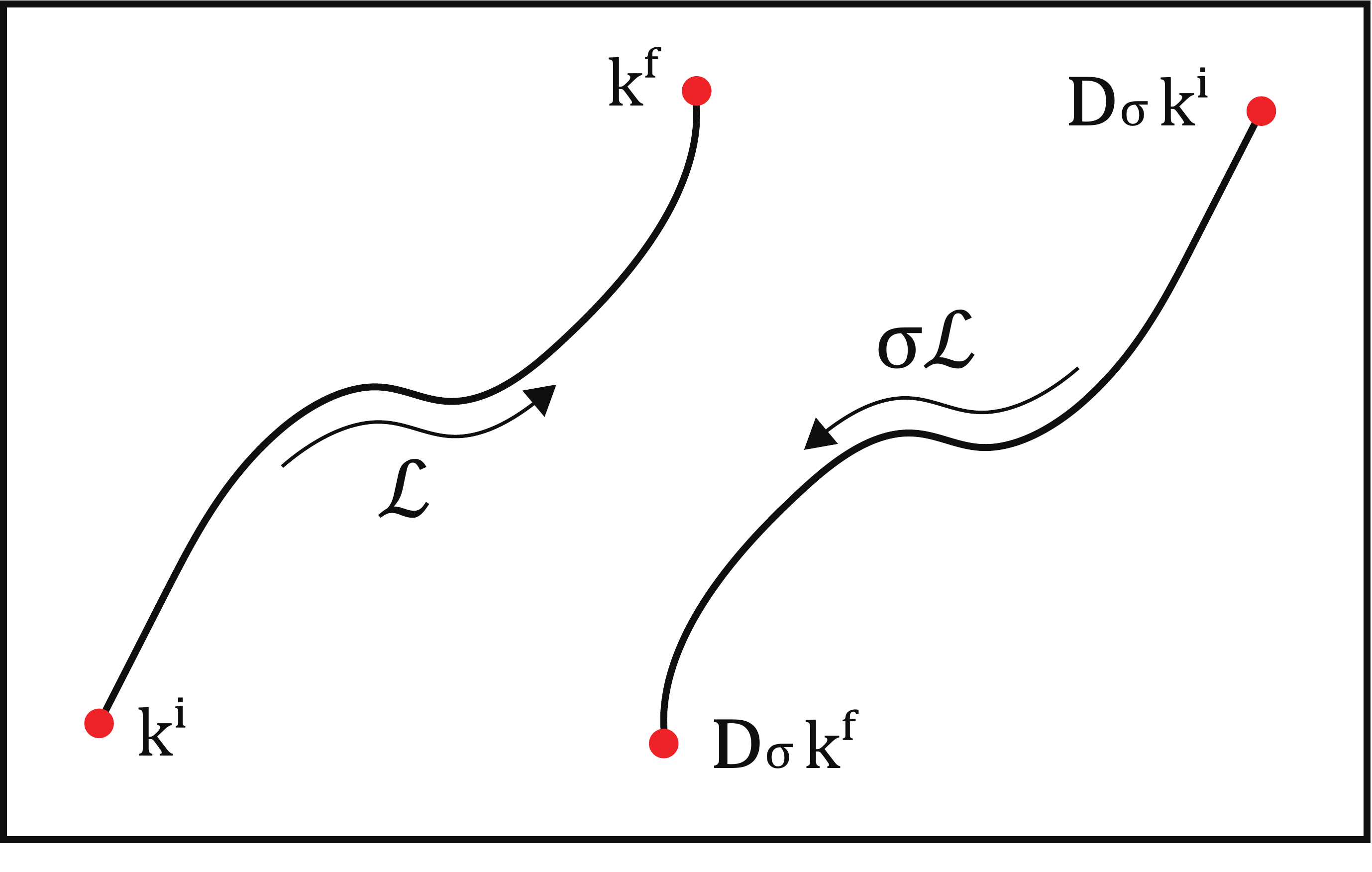}
	\caption{Paths $\mathcal{L}$ and $\sigma \mathcal{L}$ in which the Wilson lines $W_{\mathcal{L}}$ and $W_{\sigma \mathcal{L}}$ are defined, respectively. $\bb{k}^i$ ($\bb{k}^f$) is the start (end) point of $\mathcal{L}$. $\sigma\mathcal{L}$ is given by the symmetry transformation of $\mathcal{L}$ under $\hat{\sigma}$.}
	\label{fig: WilsonLine}
\end{figure}
%%%%%%%%%%%%%%%%%%%%%%%%%%%%%%%%%%%%%%%%%%%%%%%%%%%%%%%%%%%%%%%%

Let us consider a line $\mathcal{L}: \bb{k}^i \rightarrow \bb{k}^f$ which is parameterized by $\lambda$.
Start and end points $\bb{k}^{i,f}$ are given by $\bb{k}^i=\bb{k}(\lambda=0)$ and $\bb{k}^f=\bb{k}(\lambda=1)$ as shown in Fig.~\ref{fig: WilsonLine}.
Then, the Wilson line $[ W_{\mathcal{L}} ]_{n m}$ is defined by the path-ordered exponential of the Berry connection $[ a_{i}(\bb{k}) ]_{n m}$ along the line $\mathcal{L}$:
\ba
[ W_{\mathcal{L}} ]_{n m}
&= \Big[ P \, e^{i \int_{\mathcal{L}} dk^{i} a_{i}(\bb{k})} \Big]_{n m}.
\ea
For convenience, we choose a smooth gauge along $\mathcal{L}$.
Along the line $\mathcal{L}$, the covariant derivative $D_\lambda$ is defined as
\ba
\label{eq: CDeriv}
D_\lambda
&= \frac{d}{d \lambda} - i \frac{d k_i(\lambda)}{d \lambda} \, a_{i}(\bb{k}(\lambda)) \nn \\
&= \frac{d}{d \lambda} - i \dot{k}_i(\lambda) \, a_{i}(\bb{k}(\lambda)) \nn \\
&= \dot{k}_i(\lambda) \, \big[ \nabla_{i} - i a_{i}(\bb{k}(\lambda)) \big].
\ea
The Wilson line $W_{\mathcal{L}_\lambda}$ is the solution of
\ba
\label{eq: WilsonEq}
D_\lambda W_{\mathcal{L}_\lambda} = 0,
\ea
with the boundary conditions $W_{\mathcal{L}_\lambda=0}=\mathds{1}$ and $W_{\mathcal{L}_{\lambda=1}}=W_{\mathcal{L}}$.
Here, $\mathcal{L}_\lambda: \bb{k}^i \rightarrow \bb{k}(\lambda)$ and $\mathcal{L}_{\lambda=1}=\mathcal{L}$.
With the help of Eqs.~\eqref{eq: CDeriv} and~\eqref{eq: WilsonEq}, various properties of the Wilson line are derived.

First, $W_{\mathcal{L}}$ is unitary: $W_{\mathcal{L}}^{\dagger} = W_{\mathcal{L}}^{-1}$.
Also note that $W_{\mathcal{L}}^{\dagger} = W_{\mathcal{L}_{R}}$ where $W_{\mathcal{L}_{R}}$ is the ``reversed'' Wilson loop which is defined along the path-reversed line $\mathcal{L}_{R}: \bb{k}^f \rightarrow \bb{k}^i$.

Second, the gauge transformation of the Wilson line $W^{G}_{\mathcal{L}_{\lambda}}$ is derived from Eq.~\eqref{eq: BerryGauge}.
From Eq.~\eqref{eq: WilsonEq} for $W^{G}_{\mathcal{L}_{\lambda}}$,
\ba
\frac{d}{d \lambda} W^{G}_{\mathcal{L}_{\lambda}}
=& i \dot{k}_i(\lambda) \, a^{G}_{i}(\bb{k}(\lambda)) \, W^{G}_{\mathcal{L}_{\lambda}} \nn \\
=& i \dot{k}_i(\lambda) \, \big[ G^{\dagger}(\bb{k}(\lambda)) \, a_i(\bb{k}(\lambda)) \, G(\bb{k}(\lambda)) \nn \\
& + i G^{\dagger}(\bb{k}(\lambda)) \, \nabla_{i} G(\bb{k}(\lambda)) \big] \, W^{G}_{\mathcal{L}_{\lambda}}.
\ea
Hence,
\ba
\frac{d}{d \lambda} [ G(\bb{k}_{\lambda}) \, W^{G}_{\mathcal{L}_{\lambda}} ]
&= i \dot{k}^{\mu}_{\lambda} \, a_{\mu}(\bb{k}_{\lambda}) \, \big[ G(\bb{k}_{\lambda}) \, W^{G}_{\mathcal{L}_{\lambda}} \big].
\ea
The solution of the above equation is given by
\ba
\label{eq: WilsonLineGauge}
W^{G}_{\mathcal{L}_\lambda}
&= G^{\dagger}(\bb{k}(\lambda)) \, W_{\mathcal{L}_\lambda} \, G(\bb{k}_{1}),
\ea
with the boundary condition $W_{\mathcal{L}_{\lambda=0}}=\mathds{1}$.

Finally, we obtain the Wilson line $W_{\sigma \mathcal{L}_\lambda}$, which is the symmetry transformation of $W_{\mathcal{L}_\lambda}$ under $\hat{\sigma}$, where $\sigma \mathcal{L}_\lambda: D_{\sigma}\bb{k}^i \rightarrow D_{\sigma}\bb{k}^f$ as shown in Fig.~\ref{fig: WilsonLine}.
From Eqs.~\eqref{eq: BerrySymmetry} and~\eqref{eq: WilsonEq}, we derive
\begin{gather}
	W_{\sigma \mathcal{L}_\lambda} = e^{i \int_{\mathcal{L}_\lambda} d k_i \, [ D_{\sigma}^{-1}\bb{d} ]_i} \, [ B_{\sigma}(\bb{k}(\lambda)) ]^{\dagger} \, W_{\mathcal{L}_\lambda} \, B_{\sigma}(\bb{k}_1),
	\label{eq: WilsonLineUni}
\end{gather}
in a similar manner to the one used for deriving the gauge transformation.
For completeness, we derive the Wilson line $W_{\Theta \mathcal{L}}$, which is the symmetry transformation of the Wilson line $W_{\mathcal{L}_\lambda}$ under antiunitary symmetry $\hat{\Theta}=\hat{\sigma} \mathcal{K}$, where $\mathcal{K}$ denotes the complex-conjugation operator.
$\hat{\Theta}$ acts on $\bb{r}$ and $\bb{k}$ as $\hat{\Theta}: \bb{r} \rightarrow D_{\sigma}\bb{r} + \bb{d}$ and $\bb{k} \rightarrow -D_{\sigma}\bb{k}$.
Thus, $\Theta{\mathcal{L}}: -D_\sigma \bb{k}^i \rightarrow -D_\sigma \bb{k}^f$ is a line with a start point $-D_\sigma \bb{k}^i$ and an end point $-D_\sigma \bb{k}^f$.
The Wilson line $W_{\Theta \mathcal{L}}$ is given by
\begin{gather}
	W_{\Theta \mathcal{L}_\lambda} = e^{- i \int_{\mathcal{L}_\lambda} d k_i \, [ D_{\sigma}^{-1}\bb{d} ]_i} \, [ B_{\Theta}(\bb{k}(\lambda)) ]^{\dagger} \, W_{\mathcal{L}_\lambda}^* \, B_{\Theta}(\bb{k}_1),
	\label{eq: WilsonLineAntiUni}
\end{gather}
where $B_{\Theta}(\bb{k})$ is sewing matrix for $\hat{\Theta}$:
\ba
[B_{\Theta}(\bb{k})]^\dagger_{nm}=\braket{n, -D_\sigma \bb{k} | \Theta(\bb{k}) | m, \bb{k}} \, \mathcal{K},
\ea
and $\hat{\Theta} \ket{\bb{k} \, \alpha} = \ket{-D_\sigma \bb{k} \, \beta} \, \Theta(\bb{k})_{\beta \alpha}$.

\subsection{Wilson Loop}
From Eq.~\eqref{eq: WilsonLineGauge}, we conclude that the spectrum of $W_{\mathcal{L}}$ is gauge invariant when $\bb{k}^i=\bb{k}^f$ (mod $\bb{G}$).
In the Brillouin zone, consider a non-contractible loop $\mathcal{L}: \bb{k}_0 \rightarrow \bb{k}_0 + \bb{G}$.
For numerical calculations, we implement the discretized form,
\ba
\Big[ W_{\bb{k}_0+\bb{G} \leftarrow \bb{k}_0} \Big]_{nm}
=& \lim_{N \rightarrow \infty} \, \bra{n, \bb{k}_0+\bb{G}} \Big[ \prod_{i=1}^{N-1} P(\bb{k}_i) \Big] \ket{m, \bb{k}_0},
\ea
where $\bb{k}_i=\bb{k}_0+\frac{i}{N} \bb{G}$ and $P(\bb{k})=\sum_{n=1}^{n_{\rm occ}} \ket{n, \bb{k}} \bra{n, \bb{k}}$ is the projector for the occupied states.
However, we stick to formulate the Wilson loop in the continuum form
\ba
[ W_{\mathcal{L}} ]_{n m}
&= \Big[ P \, e^{i \int_{\mathcal{L}} dk^{i} a_{i}(\bb{k})} \Big]_{n m},
\ea
to exploit the covariant derivative $D_\lambda$ and symmetry transformation of the Berry connection $a(\bb{k})$.
The eigenstates $\ket{\theta(\bb{k}_0)}$ of the Wilson loop $W_{\bb{k}_0+\bb{G} \leftarrow \bb{k}_0}$ are given by $W_{\bb{k}_0+\bb{G} \leftarrow \bb{k}_0} \ket{\theta(\bb{k}_0)} = e^{i \theta(\bb{k}_0)} \ket{\theta(\bb{k}_0)}$.
Because of the periodic gauge, the Wilson loop is periodic,
\ba
\label{eq: WLPeri}
W_{\bb{k}_0+\bb{G}+\bb{G}' \leftarrow \bb{k}_0+\bb{G'}}=W_{\bb{k}_0+\bb{G} \leftarrow \bb{k}_0},
\ea
for any reciprocal vector $\bb{G}'$.
Also, the Wilson loop spectrum $\{\theta(\bb{k}_0)\}$ is independent of the start point $\bb{k}_0$,
\ba
\label{eq: WLInd}
W_{\bb{k}'_0+\bb{G} \leftarrow \bb{k}'_0}
&=W_{\bb{k}'_0+\bb{G} \leftarrow \bb{k}_0+\bb{G}} W_{\bb{k}_0+\bb{G} \leftarrow \bb{k}_0} W_{\bb{k}'_0 \leftarrow \bb{k}_0}^\dagger \nn \\
&=W_{\bb{k}'_0 \leftarrow \bb{k}_0} W_{\bb{k}_0+\bb{G} \leftarrow \bb{k}_0} W_{\bb{k}'_0 \leftarrow \bb{k}_0}^\dagger,
\ea
where $\bb{k}_0, \bb{k}'_0 \in \mathcal{L}: \bb{k}_0 \rightarrow \bb{k}_0 + \bb{G}$. 

Now, let us discuss a consequence of symmetry $\hat{\sigma}$ on the Wilson loop.
Although Eqs.~\eqref{eq: WilsonLineUni} and~\eqref{eq: WilsonLineAntiUni} applies to arbitrary symmetries and lines, we restrict our attention to inversion $I$ and $k_1$-directed loop $\mathcal{L}_1$ in three dimensions, $\mathcal{L}_1: (k_{1,0},k_2,k_3) \rightarrow (k_{1,0}+2\pi,k_2,k_3)$ for the discussion relevant to this paper.
Since inversion symmetry is symmorphic, one can find a coordinate system with $\bb{d}=0$.
Applying Eq.~\eqref{eq: WilsonLineUni},
\ba
\label{eq: WLInversion}
W_{I\mathcal{L}_1} = [ B_I(\bb{k}^{(1)}_0) ]^{\dagger} \, W_{\mathcal{L}_1} \, B_I(\bb{k}^{(1)}_0),
\ea
where $\bb{k}^{(1)}_0=(k_{1,0},k_2,k_3)$.
With help of Eqs.~\eqref{eq: WLPeri} and~\eqref{eq: WLInd}, $W_{I\mathcal{L}_1}$ satisfies $W_{I\mathcal{L}_1}=W_{\mathcal{L}_1}^{-1}$ and its spectrum is $\{-\theta^a_1(-\bb{k}^{(1)}_0)\}$.
Also, Eq.~\eqref{eq: WLInversion} means that the spectrum of $W_{I\mathcal{L}_1}$, $\{-\theta^a_1(-\bb{k}^{(1)}_0)\}$, is identical to that of $W_{\mathcal{L}_1}$, $\{\theta^a_1(\bb{k}^{(1)}_0)\}$.
Thus, $\ket{\theta^a_1(\bb{k}^{(1)}_0)}$ and $\ket{-\theta^{a'}_1(-\bb{k}^{(1)}_0)}$ are paired by inversion symmetry, where $\ket{\theta^a_1(\bb{k}^{(1)}_0)}$ is an eigenstate of $W_{\mathcal{L}_1}$ with an eigenvalue $e^{i\theta^a_1(\bb{k}^{(1)}_0)}$:
\ba
\label{eq: WSInversion}
\ket{-\theta^{a'}_{1}(-\bb{k}^{(1)}_0)} = [B_I(\bb{k}^{(1)}_0)]^\dagger \ket{\theta^a_{1}(\bb{k}^{(1)}_0)}.
\ea
This can be viewed as a unitary analogy of particle-hole symmetry.
We define the sewing matrix $B^{(1)}_{I}(\bb{k}^{(1)}_0)$ as
\ba
\label{eq: SewingWL}
B^{(1)}_{I}(\bb{k}^{(1)}_0)_{ab} = \braket{-\theta^a_{1}(-\bb{k}^{(1)}_0) | [B_I(\bb{k}^{(1)}_0)]^\dagger |\theta^b_{1}(\bb{k}^{(1)}_0)}.
\ea

Now, we define eigenstates $\ket{u^a_1(\bb{k}^{(1)}_0)}$ of $\hat{W}_{\mathcal{L}_1}=\ket{n, \bb{k}^{(1)}_0} W_{\mathcal{L}_1} \bra{n, \bb{k}^{(1)}_0}$ with eigenvalue $e^{i \theta^a_1(\bb{k}^{(1)}_0)}$, where
\ba
\label{eq: WilsonWannier1}
\ket{u^a_1(\bb{k}^{(1)}_0)} = \sum_{n=1}^{n_{\rm occ}} \ket{\theta^a_1(\bb{k}^{(1)}_0)}_n \ket{n, \bb{k}^{(1)}_0}.
\ea

%%%%%%%%%%%%%%%%%%%%%%%%%%%%%%%%%%%%%%%%%%%%%%%%%%%%%%%%%%%%%%%%
\begin{figure}[t!]
	\centering
	\includegraphics[width=0.35\textwidth]{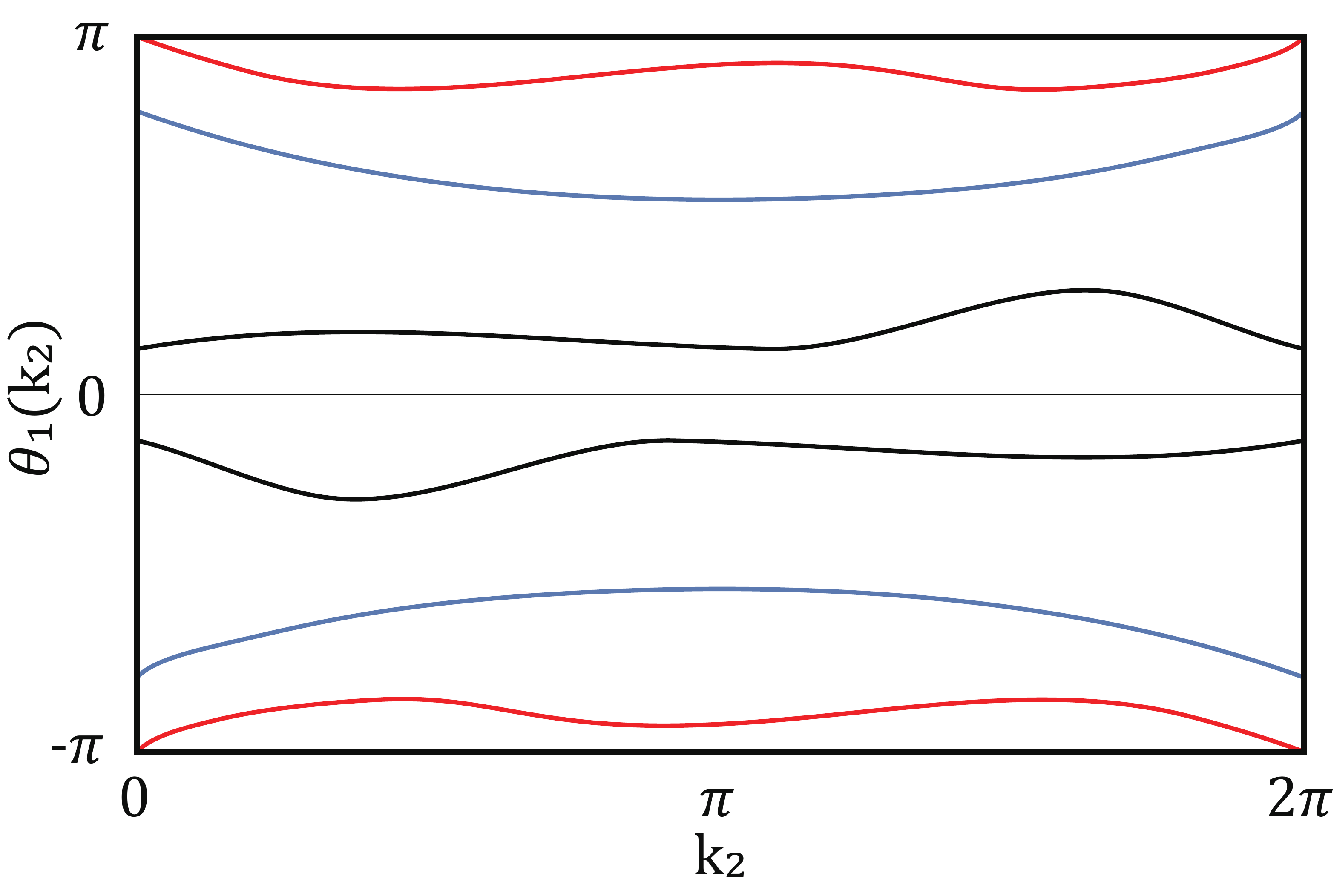}
	\caption{Inversion-symmetric Wilson loop spectrum. The spectrum is particle-hole symmetric, i.e., $\{\theta_1(-\bb{k}^{(1)}_0)\}=\{-\theta_1(\bb{k}^{(1)}_0)\}$. Hence, the projector $\tilde{P}^{(1)}$ for the nested Wilson loop must include pairs of particle-hole symmetric Wilson bands marked by the same colors.}
	\label{fig: WL_Inversion}
\end{figure}
%%%%%%%%%%%%%%%%%%%%%%%%%%%%%%%%%%%%%%%%%%%%%%%%%%%%%%%%%%%%%%%%

\subsection{Nested Wilson loop}
When the Wilson bands are gapped, a nested Wilson loop can be defined~\cite{benalcazar2017quantized,benalcazar2017electric}.
We first choose an inversion-symmetric subset of the Wilson bands by defining a projector $\tilde{P}^{(1)}$ preserving inversion symmetry as shown in Fig.~\ref{fig: WL_Inversion}.
The Wilson bands $\ket{\theta^a_1(\bb{k}^{(1)}_0)}$ selected by $\tilde{P}^{(1)}$ are labeled by $a=1,\dots,n_{\tilde{P}^{(1)}}$.
To define the nested Wilson loop, we choose the loop $\mathcal{L}_2$ along the $k_2$ direction: $\mathcal{L}_2: (k_{1,0},k_{2,0},k_3) \rightarrow (k_{1,0},k_{2,0}+2\pi,k_3)$.
In terms of the Berry connection $[a^{(1)}_i(\bb{k})]_{ab}$,
\ba
[a^{(1)}_i(\bb{k})]_{ab}
&= \braket{u^a_1(\bb{k}) | \, i \nabla_i \, | u^b_1(\bb{k})},
\ea
the nested Wilson loop is defined by
\ba
[ W^{(1)}_{\mathcal{L}_2} ]_{a b}
&= \Big[ P \, e^{i \int_{\mathcal{L}_2} dk^{i} a^{(1)}_{i}(\bb{k})} \Big]_{a b}.
\ea
Repeating the derivation of the symmetry transformation of the Wilson loop, we obtain
\ba
W^{(1)}_{I\mathcal{L}_2} = [ B^{(1)}_I(\bb{k}^{(2)}_0) ]^{\dagger} \, W^{(1)}_{\mathcal{L}_2} \, B^{(1)}_{I}(\bb{k}^{(2)}_0),
\ea
where $\bb{k}^{(2)}_0=(k_{1,0},k_{2,0},k_3)$.
Thus, $\ket{\theta^A_{2\leftarrow1}(\bb{k}^{(2)}_0)}$ and $\ket{-\theta^{A'}_{2\leftarrow1}(-\bb{k}^{(2)}_0)}$ are paired by inversion symmetry where $\ket{\theta^A_{2\leftarrow1}(\bb{k}^{(2)}_0)}$ is eigenstate of $W^{(1)}_{\mathcal{L}_2}$ with eigenvalue $e^{i\theta^A_{2\leftarrow1}(\bb{k}^{(2)}_0)}$:
\ba
\label{eq: NWLInversion}
\ket{-\theta^{A'}_{2\leftarrow1}(-\bb{k}^{(2)}_0)} = [B^{(2)}_I(\bb{k}^{(2)}_0)]^\dagger \ket{\theta^A_{2\leftarrow1}(\bb{k}^{(2)}_0)}
\ea
Thus, sewing matrix is defined as
\ba
B^{(2)}_{I}(\bb{k}^{(2)}_0)_{AB} = \braket{-\theta^A_{2\leftarrow1}(-\bb{k}^{(2)}_0) | [B^{(1)}_I(\bb{k}^{(2)}_0)]^\dagger |\theta^B_{2\leftarrow1}(\bb{k}^{(2)}_0)}.
\ea

As for the nested Wilson loop, we define eigenstates $\ket{u^A_{2\leftarrow1}(\bb{k}^{(2)}_0)}$ of $\hat{W}^{(1)}_{\mathcal{L}_2}=\ket{n, \bb{k}^{(2)}_0} W^{(1)}_{\mathcal{L}_2} \bra{n, \bb{k}^{(2)}_0}$ with eigenvalue $e^{i \theta^A_{2\leftarrow1}(\bb{k}^{(2)}_0)}$, where
\ba
\label{eq: WilsonWannier2}
\ket{u^A_{2\leftarrow1}(\bb{k}^{(2)}_0)} = \sum_{n=1}^{n_{\rm occ}} \sum_{a=1}^{n_{\tilde{P}^{(1)}}} \ket{\theta^A_{2\leftarrow1}(\bb{k}^{(2)}_0)}_a \ket{\theta^a_1(\bb{k}^{(1)}_0)}_n \ket{n, \bb{k}^{(2)}_0}.
\ea

When the nested Wilson bands are gapped, the doubly nested Wilson loop can be defined after we project the nested Wilson bands with an inversion-symmetric projector $\tilde{P}^{(2)}$. The nested Wilson bands $\ket{\theta^A_{2\leftarrow1}(\bb{k}^{(2)}_0)}$ selected by $\tilde{P}^{(2)}$ are labeled by $A=1,\dots,n_{\tilde{P}^{(2)}}$.
For the doubly nested Wilson loop, the loop $\mathcal{L}_3$ is chosen to be $k_3$-directed line, $\mathcal{L}_3: (k_{1,0},k_{2,0},k_{3,0}) \rightarrow (k_{1,0},k_{2,0},k_{3,0}+2\pi)$.
If we define a Berry connection $[a^{(2)}_i(\bb{k}^{(2)}_0)]_{AB}$,
\ba
[a^{(2)}_i(\bb{k}^{(2)}_0)]_{AB}
&= \braket{u^A_{2\leftarrow1}(\bb{k}^{(2)}_0) | \, i \nabla_i \, | u^B_{2\leftarrow1}(\bb{k}^{(2)}_0)},
\ea
the doubly nested Wilson loop can be defined by
\ba
[ W^{(2)}_{\mathcal{L}_3} ]_{AB}
&= \Big[ P \, e^{i \int_{\mathcal{L}_3} dk^{i} a^{(2)}_{i}(\bb{k}^{(2)})} \Big]_{AB}.
\ea

Repeating this procedure, the $l$th nested Wilson loop $W^{(l)}_\mathcal{L}$ can be defined for an integer.

\subsection{Mapping between inversion parities and Wilson loop spectrum \label{app: Mapping}}
Here, we review the mapping between parity configuration of the occupied states and the Wilson loop spectrum.
We recommend to refer~\cite{alexandradinata2014wilson} for the comprehensive study of the mapping.
The mapping takes two steps.

\paragraph{Step 1: Mapping from parity configuration to Wilson loop spectrum.}
Without loss of generality, we can focus on the $(k_1, k_2)$ plane in $d$D Brillouin zone.
In this plane, inversion symmetry in two dimensions, $(k_1,k_2)\rightarrow(-k_1,-k_2)$, is induced by inversion symmetry in $d$ dimensions.
Then, we can define the number of occupied states at $(\bar{k}_1,\bar{k}_2)=-(\bar{k}_1,\bar{k}_2)$ with parity $\xi$, $n_\xi(\bar{k}_1,\bar{k}_2)$.
At $\bar{k}_2=0$ or $\pi$, one can define the fewest number $n_s(\bar{k}_2)$ among $\{n_\xi(0,\bar{k}_2),n_\xi(\pi,\bar{k}_2)\}$.
When $n_s(k_s)=n_{\xi_s}(k_s,\bar{k}_2)$, the mapping from $\{n_\xi(\bar{k}_1,\bar{k}_2)\}$ to the Wilson loop spectrum $\{\theta_1(\bar{k}_2)\}$ is given by the following rules:
\ba
n^W_{\frac{\pi}{2}(1+\xi_s)}(\bar{k}_2) &= n_{+}(k_s+\pi,\bar{k}_2)-n_s(\bar{k}_2), \nn \\
n^W_{\frac{\pi}{2}(1-\xi_s)}(\bar{k}_2) &= n_{-}(k_s+\pi,\bar{k}_2)-n_s(\bar{k}_2), \nn \\
n^W_{c}(\bar{k}_2) & = 2 n_s(\bar{k}_2),
\ea
where $n^W_0(\bar{k}_2)$, $n^W_\pi(\bar{k}_2)$ and $n^W_{c}(\bar{k}_2)$ are the number of Wilson loop spectrum $\{\theta_1(\bar{k}_2)\}$ with eigenvalues $\theta_1(\bar{k}_2)=0$, $\pi$ and $(-\theta_0,\theta_0)$, respectively. Here, we denote $\theta_0$ as a generic value not equal to $0$ or $\pi$.

\paragraph{Step 2: Mapping from Wilson loop spectrum to relative winding number.}
With $\{n^W_\theta(\bar{k}_2)\}=(n^W_{\xi_s}(\bar{k}_2),n^W_{-\xi_s}(\bar{k}_2), n^W_{c}(\bar{k}_2))$ for a given Wilson loop spectrum, $N_d$ is defined as
\ba
N_d={\rm max}(n^W_{\pi}(0)-n^W_{\pi}(\pi)-n^W_{c}(\pi), \nn \\
n^W_{\pi}(\pi)-n^W_{\pi}(0)-n^W_{c}(0)).
\ea
When $|C|<N_d$ where $C$ is the Chern number defined in the $(k_1,k_2)$ plane, $N_d$ is given by $N_d=2 N_W+|C|$.
If $|C| \ge N_d$, the relative winding number $N_W$ is zero.
Also note that the Chern number $C$ and $N_d$ satisfies
\ba
\label{eq: Chernparity}
C=N_d=n^W_\pi(\pi)-n^W_\pi(0) \quad \text{(mod 2)}.
\ea
Since a non-zero Chern number implies the stable band topology, let us assume that $C=0$.
Then, the Wilson loop is unwound if $N_d \le 0$. 
Otherwise, $N_d > 0$, the relative winding number $N_W$ is given by $N_W=\frac{1}{2}N_d$.

\section{Characterization of fragile band topology in the nested Wilson loop spectrum \label{app: FrgWL_deriv}}
In this Appendix, we provide the proof for the claim that the nested Wilson loop spectrum characterized by a positive even integer $N_{d,{\rm nest}}$, which is defined as
\ba
\label{eq: Ndnest}
N_{d,{\rm nest}} = & \text{max}(n_\pi^{W_{\rm nest}}(0)-n_\pi^{W_{\rm nest}}(\pi)-n_c^{W_{\rm nest}}(\pi), \nn \\
& n_\pi^{W_{\rm nest}}(\pi)-n_\pi^{W_{\rm nest}}(0)-n_c^{W_{\rm nest}}(0)),
\ea
implies the fragile band topology.
Here, $n^{W_{\rm nest}}_\pi(\bar{k}_3)$ and $n^{W_{\rm nest}}_{c}(\bar{k}_3)$ are the number of eigenvalues $\theta_{2 \leftarrow 1}(\bar{k}_3)$, which is equal to $\pi$ and takes a generic value, respectively, in the nested Wilson loop spectrum at $\bar{k}_3=0$ and $\pi$.

Before proving the claim, we note that the nested Wilson loop spectrum satisfies similar relations discussed in Appendix~\ref{app: Mapping} since the Wilson loop and the nested Wilson loop spectra are constrained by the inversion symmetry in the same way.
First, we can assign an inversion parity to each Wilson loop eigenstate at inversion-invariant momenta $\bb{k}^{\rm inv}$. At $\bb{k}^{\rm inv}$, $| \theta_1(\bb{k}^{\rm inv}) \rangle$ is an inversion eigenstate with definite parity when $\theta_1(\bb{k}^{\rm inv})=0$ and $\pi$.
On the other hand, $|\theta_1(\bb{k}^{\rm inv})\rangle$ and $|-\theta_1(\bb{k}^{\rm inv})\rangle$ are paired by inversion symmetry when $\theta_1(\bb{k}^{\rm inv})$ is a generic value. This pair forms inversion-eigenstates with even and odd parities. The parities of the Wilson loop eigenstates at $\bb{k}^{\rm inv}$ are encoded in the sewing matrix $B^{(1)}_{I}(\bb{k}^{\rm inv})$ in Eq.~\eqref{eq: SewingWL}.
By repeating Step 1 in Appendix~\ref{app: Mapping}, we obtain an eigenvalue configuration in the nested Wilson loop spectrum $\{n_\theta^{W_{\rm nest}}(\bar{k}_3)\}=(n_0^{W_{\rm nest}}(\bar{k}_3),n_\pi^{W_{\rm nest}}(\bar{k}_3),n_c^{W_{\rm nest}}(\bar{k}_3))$ at $\bar{k}_3=0$ and $\pi$. This configuration uniquely determines $N_{d,{\rm nest}}$ in Eq.~\eqref{eq: Ndnest}.
Second, the nested Chern number $C_{\rm nest}$, the nested relative winding number $N_{W,{\rm nest}}$ and $N_{d,{\rm nest}}$ satisfy relations similar to ones for the Chern number $C$, the relative winding number $N_W$ and $N_d$ presented in Step 2 in Appendix~\ref{app: Mapping}.
\\

We begin our proof by showing that a band crossing between the Wilson bands does not change $N_{d,{\rm nest}}$ while it changes the nested Chern number $C_{\rm nest}$ only by an even integer.
When a gap closing and reopening in the Wilson loop spectrum happens at a generic momentum, it does not change the parity configuration of the Wilson loop eigenstates. As the parity configuration determines an eigenvalue configuration in the nested Wilson loop spectrum, $N_{d,{\rm nest}}$ is invariant.
If the Wilson bands are inverted at an inversion-invariant momentum $\bb{k}^{\rm inv}$, only pairs of Wilson bands $|\theta_1(\bb{k}^{\rm inv})\rangle$ and $|-\theta_1(\bb{k}^{\rm inv})\rangle$ are involved in the Wilson band inversion. Thus, there is no net exchange of inversion-parities, and $N_{d,{\rm nest}}$ is invariant.
Consequently, the Wilson band crossing does not change $N_{d,{\rm nest}}$.
On the other hand, the Wilson band crossing can change $C_{\rm nest}$ by an even integer since it always happens pairwise due to inversion symmetry~\cite{wieder2018axion}.

For an integer-valued $N_{d,{\rm nest}}$, there are three cases to be considered: $N_{d,{\rm nest}}$ is equal to i) an odd integer, ii) a positive even integer and iii) a negative even integer or 0.
\begin{enumerate}[i)]
 	\item 
 	When $N_{d,{\rm nest}}$ is an odd integer, Eq.~\eqref{eq: Chernparity} implies that $C_{\rm nest}$ is also an odd integer. Since a Wilson band crossing cannot change an odd $C_{\rm nest}$ to zero, the nested Wilson loop spectrum indicates the stable band topology.
 	\item
 	When $N_{d,{\rm nest}}$ is a positive even integer, Eq.~\eqref{eq: Chernparity} implies that $C_{\rm nest}$ is also an even integer. For even $C_{\rm nest}$, a Wilson band crossing can change $C_{\rm nest}$ to zero. However, the nested Wilson loop spectrum cannot be gapped and it shows the relative winding since $N_{d,{\rm nest}}=2 N_{W,{\rm nest}} + |C_{\rm nest}|$ holds when $|C_{\rm nest}| < N_{d,{\rm nest}}$.
 	\item
	When $N_{d,{\rm nest}}$ is a negative even integer or zero, the nested relative winding number $N_{W,{\rm nest}}$ is always zero. This is because the relative winding exists only when $|C_{\rm nest}| < N_{d,{\rm nest}}$. Thus, the nested Wilson loop spectrum can be gapped when a Wilson band crossing happens.
\end{enumerate}

\section{Numerical details of the Wilson loop calculation \label{app: Numerics}}
In this appendix, we summarized the numerical details in Sec.~\ref{sec: FragileMomentum}.

\begin{figure*}[t!]
	\centering
	\includegraphics[width=0.98\textwidth]{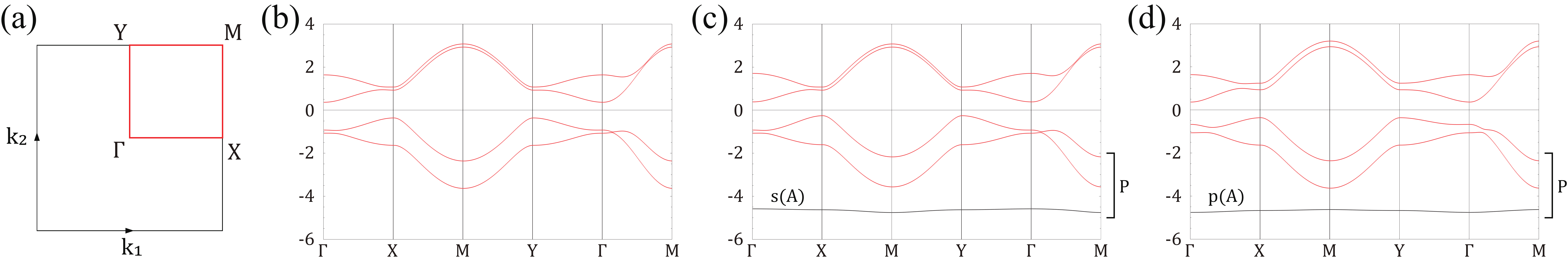}
	\caption{Band structure of the tight-binding model in Eq.~\eqref{eq: H(2,3)}. (a) $2$D Brillouin zone. High-symmetry lines are shown as red lines. (b) Energy bands along the high-symmetry lines. (c) Energy bands after adding a $s(A)$ orbital. (d) Energy bands after adding a $p(A)$ orbital.}
	\label{fig: 2DBand}
\end{figure*}

\begin{figure*}[t!]
	\centering
	\includegraphics[width=0.98\textwidth]{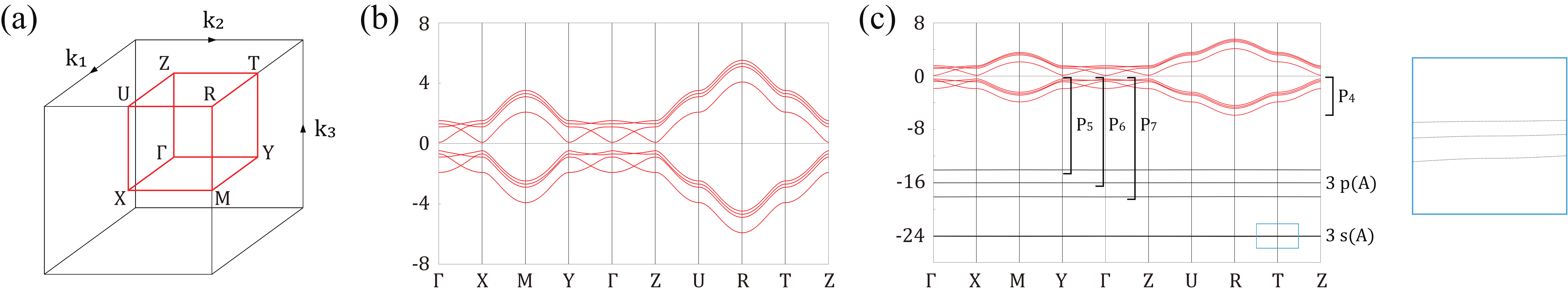}
	\caption{Band structure of the tight-binding model in Eq.~\eqref{eq: H(3,4)}. (a) $3$D Brillouin zone. High-symmetry lines are shown as red lines. (b) Energy bands along the high-symmetry lines. (c) Energy bands after adding three $s(A)$ and three $p(A)$ orbitals. (c) $P_{4,5,6,7}$ denote the projector for the Wilson loop.}
	\label{fig: 3DBand}
\end{figure*}

\subsection{$2$D third-order TI}
The tight-binding Hamiltonian of $2$D third-order TI is given by
\ba
\label{eq: H(2,3)}
H_{{\rm A} (2,3)}(\bb{k}) =& \sum_{i=1}^2 \sin k_i \gamma_i - (2 - \lambda - \sum_{i=1}^2 \cos k_i ) \gamma_5 \nn \\
& + i \sum_{i=1}^2 \sum_{a=3}^4 \Delta_{i,a} \, \gamma_i \gamma_a \gamma_5,
\ea
where $\{\gamma_1, \dots, \gamma_5\}=\{\tau_x\sigma_x, \tau_x\sigma_y, \tau_x\sigma_z, \tau_y\sigma_0, \tau_z\sigma_0\}$ and $\gamma_5=-\gamma_1 \dots \gamma_4$.
For numerical calculations, $\Delta_{1,3}=0.240$, $\Delta_{1,4}=-0.196$, $\Delta_{2,3}=-0.259$, and $\Delta_{2,4}=-0.204$ are used.
$H_{{\rm A} (2,3)}(\bb{k})$ is inversion symmetric, i.e., $I_0 \, H_{{\rm A} (2,3)}(\bb{k}) {I_0}^{-1} = H_{{\rm A} (2,3)}(-\bb{k})$, with inversion symmetry operator $I_0$:
\ba
I_0 = \tau_z \sigma_0 =
\begin{pmatrix}
	\mathds{1}_2 & \\
	& -\mathds{1}_2
\end{pmatrix}.
\ea
The energy bands of $H_{{\rm A} (2,3)}(\bb{k})$ are shown in Fig.~\ref{fig: 2DBand}(b).

\subsubsection{$2$D third-order TI $+$ $s(A)$}
Now, we add an $s$ orbital localized at the Wyckoff position $A=(0,0)$.
On-site energy $h_1$ and an interaction Hamiltonian between the original four bands and $s$ orbital are chosen as
\ba
h_1=-4.5, \quad h_2=0.1 \times (4,6,0,0)^T.
\ea
Thus, the perturbation Hamiltonian $H_c$ is given by
\ba
H_c=
\begin{pmatrix}
	0_4 & h_2 \\
	h_2^T & h_1
\end{pmatrix},
\ea
where $0_4$ denotes $4 \times 4$ null matrix.

Hence, Hamiltonian $H_{+ s(A)}$ and inversion symmetry operator $I$ for the five bands are given by
\begin{gather}
	H_{+ s(A)} = H_{{\rm A} (2,3)} + H_c, \nn \\
	I=
	\begin{pmatrix}
		I_0 & 0 \\
		0 & 1
	\end{pmatrix},
\end{gather}
and $I \, H_{+ s(A)}(\bb{k}) \, I^{-1} = H_{+ s(A)}(-\bb{k})$.
The energy bands of $H_{+ s(A)}(\bb{k})$ are shown in Fig.~\ref{fig: 2DBand}(c).

\subsubsection{$2$D third-order TI $+$ $p(A)$}
We add a $p$ orbital localized at the Wyckoff position $A=(0,0)$.
On-site energy $h_1$ and an interaction Hamiltonian between the original four bands and $p$ orbital are chosen as
\ba
h_1=-4.5, \quad h_2=0.1 \times (0,0,6,-8)^T.
\ea
Thus, the perturbation Hamiltonian $H_c$ is given by
\ba
H_c=
\begin{pmatrix}
	0_4 & h_2 \\
	h_2^T & h_1
\end{pmatrix}.
\ea

Hence, Hamiltonian $H_{+ p(A)}$ and inversion symmetry operator $I$ for the five bands are given by
\begin{gather}
	H_{+ p(A)}=H_{{\rm A} (2,3)} + H_c, \nn \\
	I=
	\begin{pmatrix}
		I_0 & 0 \\
		0 & -1
	\end{pmatrix},
\end{gather}
and $I \, H_{+ p(A)}(\bb{k}) \, I^{-1} = H_{+ p(A)}(-\bb{k})$.
The energy bands of $H_{+ p(A)}(\bb{k})$ are shown in Fig.~\ref{fig: 2DBand}(d).

\subsection{$3$D fourth-order TI}
The tight-binding Hamiltonian of $3$D fourth-order TI is given by
\ba
\label{eq: H(3,4)}
H_{{\rm A} (3,4)}(\bb{k}) =& \sum_{i=1}^3 \sin k_i \check{\gamma}_i - (3 - \lambda - \sum_{i=1}^3 \cos k_i ) \check{\gamma}_7 \nn \\
& + i \sum_{i=1}^3 \sum_{a=4}^6 \check{\Delta}_{i,a} \, \check{\gamma}_i \check{\gamma}_a \check{\gamma}_7.
\ea
where $\{\check{\gamma}_1, \dots, \check{\gamma}_7\}=$ $\{\mu_x\tau_x\sigma_x, \mu_x\tau_x\sigma_y, \mu_x\tau_x\sigma_z, \mu_x\tau_y\sigma_0,$ $\mu_x\tau_z\sigma_0, \mu_y\tau_0\sigma_0, \mu_z\tau_0\sigma_0\}$ and $\check{\gamma}_7=i \check{\gamma}_1 \dots \check{\gamma}_6$.
For numerical calculations, $\check{\Delta}_{1,4}=-0.25$, $\check{\Delta}_{1,5}=0.05$, $\check{\Delta}_{1,6}=0.10$, $\check{\Delta}_{2,4}=-0.05$, $\check{\Delta}_{2,5}=0.30$, $\check{\Delta}_{2,6}=-0.10$, $\check{\Delta}_{3,4}=0.05$, $\check{\Delta}_{3,5}=-0.05$, and $\check{\Delta}_{3,6}=0.35$ are used.
$H_{{\rm A} (3,4)}(\bb{k})$ is inversion symmetric, i.e., $I_0 \, H_{{\rm A} (3,4)}(\bb{k}) {I_0}^{-1} = H_{{\rm A} (3,4)}(-\bb{k})$, with inversion symmetry operator $I_0$,
\ba
I_0 = \mu_z \tau_0 \sigma_0 =
\begin{pmatrix}
	\mathds{1}_4 & \\
	& -\mathds{1}_4
\end{pmatrix}.
\ea
Energy bands of $H_{{\rm A} (3,4)}(\bb{k})$ are shown in Fig.~\ref{fig: 3DBand}(b).

\begin{figure*}[t!]
	\centering
	\includegraphics[width=0.98\textwidth]{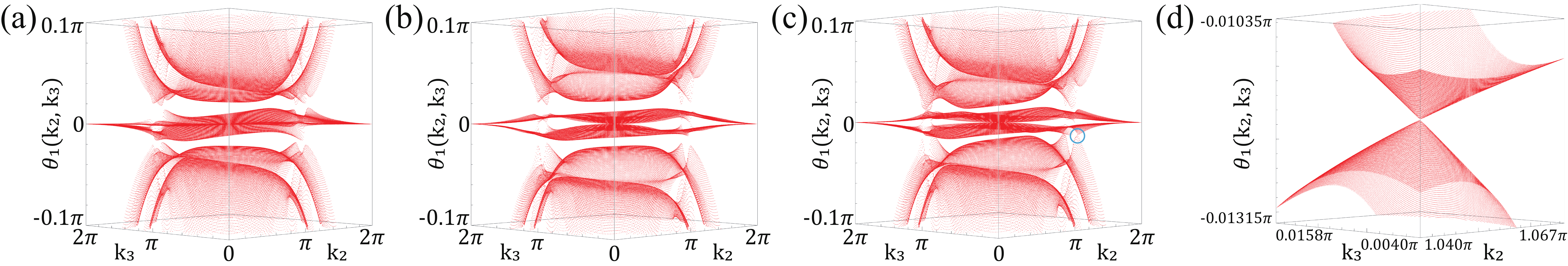}
	\caption{Wilson bands with a projector $P_6$ for: (a) $H_{c1}=H_c(q=1)$. (b) $H_{c2}=H_c(q=0)$. (c) $H_c(q=0.49)$. The Wilson band transition occurs at $(k_2,k_3,\theta_1(k_2,k_3))=(1.054 \pi, 0.0099 \pi, -0.0117 \pi)$ (blue circle) and $-(1.054 \pi, 0.0099 \pi, -0.0117 \pi)$. (d) The band crossing between Wilson bands at $(1.054 \pi, 0.0099 \pi, -0.0117 \pi)$ (blue circle in (c)).}
	\label{fig: WilsonBand}
\end{figure*}

\subsubsection{$3$D fourth-order TI $+$ $3p(A)$}
Now, we add three $p$ orbitals localized at the Wyckoff position $A=(0,0,0)$.
For this, we add six orbitals at generic Wyckoff positions $\bb{x}_{9,\dots,14}$.
\ba
&\bb{x}_9=-\bb{x}_{10}=(-0.4,-0.4,-0.4), \nn \\
&\bb{x}_{11}=-\bb{x}_{12}=(-0.2,-0.2,-0.2), \nn \\
&\bb{x}_{13}=-\bb{x}_{14}=(-0.1,0.1,-0.1).
\ea
Note that the original eight orbitals are localized at $\bb{x}_{1,\dots,8}=(0,0,0)$.
The Hamiltonian for the additional six bands is chosen to be
\ba
h_1=
\begin{pmatrix}
	-19 & -5 & 0 & 0 & 0 & 0 \\
	-5 & -19 & 0 & 0 & 0 & 0 \\
	0 & 0 & -20 & -4 & 0 & 0 \\
	0 & 0 & -4 & -20 & 0 & 0 \\
	0 & 0 & 0 & 0 & -21 & -3 \\
	0 & 0 & 0 & 0 & -3 & -21 \\
\end{pmatrix},
\ea
and it is invariant under the inversion symmetry operator $I_c$:
\ba
I_c=
\begin{pmatrix}
	0 & 1 & 0 & 0 & 0 & 0 \\
	1 & 0 & 0 & 0 & 0 & 0 \\
	0 & 0 & 0 & 1 & 0 & 0 \\
	0 & 0 & 1 & 0 & 0 & 0 \\
	0 & 0 & 0 & 0 & 0 & 1 \\
	0 & 0 & 0 & 0 & 1 & 0
\end{pmatrix}.
\ea
Diagonalizing $h_1$ we obtain three $s$ and three $p$ orbitals localized at $A$.
Three $s$ orbitals are degenerate with an energy eigenvalue $-24$, and three $p$ orbitals have energy eigenvalues $-14$, $-16$, and $-18$, respectively.
Now, we couple the additional six bands to the original eight bands according to $h_c$:
\ba
h_2=
0.05 \times
\begin{pmatrix}
	11 & 11 & -7 & -7 & -2 & -2 \\
	-3 & -3 & -1 & -1 & -1 & -1 \\
	-9 & -9 & -10 & -10 & 7 & 7 \\
	2 & 2 & 10 & 10 & 17 & 17 \\
	-1 & 1 & -9 & 9 & 7 & -7 \\
	-3 & 3 & -2 & 2 & -17 & 17 \\
	-8 & 8 & 4 & -4 & -5 & 5 \\
	-15 & 15 & 3 & -3 & 13 & -13
\end{pmatrix}.
\ea
Thus, the perturbation Hamiltonian $H_{c1}$ is given by
\ba
H_{c1}=
\begin{pmatrix}
	0_{8} & h_2 \\
	h_2^T & h_1 \\
\end{pmatrix}.
\ea

To sum up, the Hamiltonian $H$ and inversion symmetry operator $I$ for the 14 bands are given by
\begin{gather}
	H = H_{{\rm A} (3,4)} + H_{c1}, \nn \\
	I=
	\begin{pmatrix}
		I_0 & 0 \\
		0 & I_c
	\end{pmatrix},
\end{gather}
and $I \, H(\bb{k}) \, I^{-1} = H(-\bb{k})$.
The energy bands of $H(\bb{k})$ are shown in Fig.~\ref{fig: 3DBand}(c).
In order to study the effect of adding $p$ orbitals, we define projectors $P_5$, $P_6$, and $P_7$ as the projections to the occupied states with an additional $p(A)$ orbital, additional two $p(A)$ orbitals, and additional three $p(A)$ orbitals, respectively [see Fig.~\ref{fig: 3DBand}(c) for the definition of $P_{5,6,7}$].

\subsubsection{Wilson band transition}
The nested Wilson loop spectrum depends on the topology of Wilson bands.
Thus, a bulk-gap-preserving perturbation can change the topology of the nested Wilson loop through a band crossing between Wilson bands, which we call a Wilson band transition. 
To investigate how the nested Wilson loop spectrum changes after Wilson band transition systematically, we choose a new coupling Hamiltonian ${h_2}'$ as
\ba
{h_2}'=
0.05 \times
\begin{pmatrix}
	18 & 18 & 4 & 4 & 19 & 19 \\
	-5 & -5 & -1 & -1 & 5 & 5 \\
	-10 & -10 & 6 & 6 & 4 & 4 \\
	6 & 6 & 15 & 15 & -4 & -4 \\
	-8 & 8 & -15 & 15 & -4 & 4 \\
	-9 & 9 & 3 & -3 & -1 & 1 \\
	-4 & 4 & 6 & -6 & 3 & -3 \\
	-2 & 2 & 4 & -4 & 5 & -5
\end{pmatrix}.
\ea
Thus, new perturbation Hamiltonian $H_{c2}$ is given by
\ba
H_{c2}=
\begin{pmatrix}
	0_{8} & {h_2}' \\
	({h_2}')^T & h_1 \\
\end{pmatrix}.
\ea
Now, we define an interpolating Hamiltonian $H_c(q)$ and $H(\bb{k},q)$ as
\ba
H_c(q) &= q H_{c1} + (1-q) H_{c2}, \nn \\
H(\bb{k},q) &= H_{{\rm A} (3,4)} + H_c(q),
\ea
and find critical values $q_c$ where the Wilson bands close the gap.
The Wilson bands with the projector $P_6$ close the gap at $q_c=0.49$ and the band crossing points are $(k_2,k_3,\theta_1(k_2,k_3)) \simeq \pm (1.054 \pi, 0.0099 \pi, -0.0117 \pi)$ as shown in Figs.~\ref{fig: WilsonBand}(c) and~\ref{fig: WilsonBand}(d).
For Wilson bands with the projector $P_7$, the Wilson band transition occurs at $q_c=0.52$ and the gap closes at $(k_2,k_3,\theta_1(k_2,k_3)) \simeq \pm (1.040 \pi, 0.0083 \pi, -0.0114 \pi)$.
After the Wilson band transition, the nested Chern number changes from $\pm2$ to $0$ as shown in Fig.~\ref{fig: NestedWL}.

%\bibliography{reference.bib}
%merlin.mbs apsrev4-1.bst 2010-07-25 4.21a (PWD, AO, DPC) hacked
%Control: key (0)
%Control: author (0) dotless jnrlst
%Control: editor formatted (1) identically to author
%Control: production of article title (0) allowed
%Control: page (1) range
%Control: year (0) verbatim
%Control: production of eprint (0) enabled
%

\end{document}